# *Planck* 2013 results. XXIII. Isotropy and statistics of the CMB

Planck Collaboration: P. A. R. Ade[92], N. Aghanim[63], C. Armitage-Caplan[97], M. Arnaud[76], M. Ashdown[73,6], F. Atrio-Barandela[18], J. Aumont[63], C. Baccigalupi[91], A. J. Banday[100,9], R. B. Barreiro[70], J. G. Bartlett[1,71], N. Bartolo[35], E. Battaner[101], R. Battye[72], K. Benabed[61], A. Benoît[26,64,99], J.-P. Bernard[100,9], M. Bersanelli[39,53], P. Bielewicz[100,9,91], J. Bobin[76], J. J. Bock[71,10], A. Bonaldi[72], L. Bonavera[70], J. R. Bond[8], J. Borrill[13,94], F. R. Bouchet[64,99], M. Bridges[73,6,67], M. Bucher[1], C. Burigana[52,37], R. C. Butler[52], J.-F. Cardoso[77,1,64], A. Catalano[78,75], A. Challinor[67,73,11], A. Chamballu[76,15,63], R.-R. Chary[60], H. C. Chiang[29,7], L.-Y Chiang[66], P. R. Christensen[87,43], S. Church[96], D. L. Clements[59], S. Colombi[64,99], L. P. L. Colombo[25,71], F. Couchot[74], A. Coulais[75], B. P. Crill[71,88], M. Cruz[20], A. Curto[6,70], F. Cuttaia[52], L. Danese[91], R. D. Davies[72], R. J. Davis[72], P. de Bernardis[38], A. de Rosa[52], G. de Zotti[48,91], J. Delabrouille[1], J.-M. Delouis[64,99], F.-X. Désert[56], J. M. Diego[70], H. Dole[63,62], S. Donzelli[53], O. Doré[71,10], M. Douspis[63], A. Ducout[64], X. Dupac[45], G. Efstathiou[67], F. Elsner[64,99], T. A. Enßlin[81], H. K. Eriksen[8], Y. Fantaye[68], J. Fergusson[11], F. Finelli[52,54], O. Forni[100,9], M. Frailis[50], E. Franceschi[52], M. Frommert[17], S. Galeotta[50], K. Ganga[1], M. Giard[100,9], G. Giardino[46], Y. Giraud-Héraud[1], J. González-Nuevo[70,91], K. M. Górski[71,102], S. Gratton[73,67], A. Gregorio[40,50], A. Gruppuso[52], F. K. Hansen[8], M. Hansen[87], D. Hanson[83,71,8], D. L. Harrison[67,73], G. Helou[10], S. Henrot-Versillé[74], C. Hernández-Monteagudo[12,81], D. Herranz[70], S. R. Hildebrandt[10], E. Hivon[64,99], M. Hobson[6], W. A. Holmes[71], A. Hornstrup[16], W. Hovest[81], K. M. Huffenberger[27], A. H. Jaffe[59], T. R. Jaffe[100,9], W. C. Jones[29], M. Juvela[28], E. Keihänen[28], R. Keskitalo[23,13], J. Kim[87], T. S. Kisner[80], J. Knoche[81], L. Knox[32], M. Kunz[17,63,3], H. Kurki-Suonio[28,47], G. Lagache[63], A. Lähteenmäki[2,47], J.-M. Lamarre[75], A. Lasenby[6,73], R. J. Laureijs[46], C. R. Lawrence[71], J. P. Leahy[72], R. Leonardi[45], C. Leroy[63,100,9], J. Lesgourgues[98,90], M. Liguori[35], P. B. Lilje[68], M. Linden-Vørnle[16], M. López-Caniego[70], P. M. Lubin[33], J. F. Macías-Pérez[78], B. Maffei[72], D. Maino[39,53], N. Mandolesi[52,5,37], A. Mangilli[64], D. Marinucci[42], M. Maris[50], D. J. Marshall[76], P. G. Martin[8], E. Martínez-González[70], S. Masi[38], M. Massardi[51], S. Matarrese[35], F. Matthai[81], P. Mazzotta[41], J. D. McEwen[26,85], P. R. Meinhold[33], A. Melchiorri[38,55], L. Mendes[45], A. Mennella[39,53], M. Migliaccio[67,73], K. Mikkelsen[68], S. Mitra[58,71], M.-A. Miville-Deschênes[63,8], D. Molinari[36,52], A. Moneti[64], L. Montier[100,9], G. Morgante[52], D. Mortlock[59], A. Moss[93], D. Munshi[92], J. A. Murphy[86], P. Naselsky[87,43], F. Nati[38], P. Natoli[37,4,52], C. B. Netterfield[21], H. U. Nørgaard-Nielsen[16], F. Noviello[72], D. Novikov[59], I. Novikov[87], S. Osborne[96], C. A. Oxborrow[16], F. Paci[91], L. Pagano[38,55], F. Pajot[63], D. Paoletti[52,54], F. Pasian[50], G. Patanchon[1], H. V. Peiris[26], O. Perdereau[74], L. Perotto[78], F. Perrotta[91], F. Piacentini[38], M. Piat[1], E. Pierpaoli[25], D. Pietrobon[71], S. Plaszczynski[74], D. Pogosyan[30], E. Pointecouteau[100,9], G. Polenta[4,49], N. Ponthieu[63,56], L. Popa[65], T. Poutanen[47,28,2], G. W. Pratt[76], G. Prézeau[10,71], S. Prunet[64,99], J.-L. Puget[63], J. P. Rachen[22,81], B. Racine[1], C. Räth[82], R. Rebolo[69,14,44], M. Reinecke[81], M. Remazeilles[72,63,1], C. Renault[78], A. Renzi[31], S. Ricciardi[52], T. Riller[81], I. Ristorcelli[100,9], G. Rocha[71,10], C. Rosset[1], A. Rotti[58], G. Roudier[1,75,71], J. A. Rubiño-Martín[69,44], B. Ruiz-Granados[101], B. Rusholme[60], M. Sandri[52], D. Santos[78], G. Savini[89], D. Scott[24], M. D. Seiffert[71,10], E. P. S. Shellard[11], T. Souradeep[58], L. D. Spencer[92], J.-L. Starck[76], V. Stolyarov[6,73,95], R. Stompor[1], R. Sudiwala[92], F. Sureau[76], P. Sutter[64], D. Sutton[67,73], A.-S. Suur-Uski[28,47], J.-F. Sygnet[64], J. A. Tauber[46], D. Tavagnacco[50,40], L. Terenzi[52], L. Toffolatti[19,70], M. Tomasi[53], M. Tristram[74], M. Tucci[17,74], J. Tuovinen[84], M. Türler[57], L. Valenziano[52], J. Valiviita[47,28,68], B. Van Tent[79], J. Varis[84], P. Vielva[70], F. Villa[52], N. Vittorio[41], L. A. Wade[71], B. D. Wandelt[64,99,34], I. K. Wehus[71], M. White[31], A. Wilkinson[72], D. Yvon[15], A. Zacchei[50], and A. Zonca[33]

*(Affiliations can be found after the references)*

Received XX, 2013; accepted XX, 2013

## ABSTRACT

The two fundamental assumptions of the standard cosmological model — that the initial fluctuations are statistically isotropic and Gaussian — are rigorously tested using maps of the cosmic microwave background (CMB) anisotropy from the *Planck* satellite. The detailed results are based on studies of four independent estimates of the CMB that are compared to simulations using a fiducial ΛCDM model and incorporating essential aspects of the *Planck* measurement process. Deviations from isotropy have been found and demonstrated to be robust against component separation algorithm, mask choice, and frequency dependence. Many of these anomalies were previously observed in the *WMAP* data, and are now confirmed at similar levels of significance (about 3 σ). However, we find little evidence of non-Gaussianity, with the exception of a few statistical signatures that seem to be associated with specific anomalies. In particular, we find that the quadrupole-octopole alignment is also connected to a low observed variance in the CMB signal. A power asymmetry is now found to persist on scales corresponding to about ℓ = 600 and can be described in the low-ℓ regime by a phenomenological dipole modulation model. However, any primordial power asymmetry is strongly scale-dependent and does not extend to arbitrarily small angular scales. Finally, it is plausible that some of these features may be reflected in the angular power spectrum of the data, which shows a deficit of power on similar scales. Indeed, when the power spectra of two hemispheres defined by a preferred direction are considered separately, one shows evidence of a deficit in power, while its opposite contains oscillations between odd and even modes that may be related to the parity violation and phase correlations also detected in the data. Although these analyses represent a step forward in building an understanding of the anomalies, a satisfactory explanation based on physically motivated models is still lacking.

## 1. Introduction

This paper, one of a set associated with the 2013 release of data from the *Planck*[1] mission (Planck Collaboration I 2014)

---
[1] *Planck* (http://www.esa.int/Planck) is a project of the European Space Agency (ESA) with instruments provided by two sci-

describes a set of studies undertaken to determine the statistical properties of the cosmic microwave background (CMB).

The standard cosmological model is described well by the *Friedmann-Lemaître-Robertson-Walker* (FLRW) solution of the Einstein field equations. This model is characterized by a homogeneous and isotropic metric and a scale factor of the expanding Universe. At very early times it is hypothesized that the Universe went through a period of accelerated expansion, the so-called *cosmological inflation*, driven by a hypothetical scalar field, the *inflaton*. During inflation the universe behaves as a de Sitter space, providing the conditions in which some of the present properties of the universe can be realized and specifically relaxing the problem of initial conditions. In particular, the seeds that gave rise to the present large-scale matter distribution via gravitational instability originated as quantum fluctuations of the inflaton about its vacuum state. These fluctuations in the inflaton produce energy perturbations that are distributed as a homogeneous and isotropic Gaussian random field. Linear theory relates those energy fluctuations to the CMB anisotropies, implying a distribution for the anisotropies very close to that of an isotropic Gaussian random field.

The scope of this work is to use *Planck* data to test the Gaussianity and near isotropy of the CMB in intensity, as expected in the standard cosmology paradigm. Testing these fundamental properties is crucial for the validation of the standard cosmological scenario, and has profound implications for our understanding of the physical nature of the Universe and the initial conditions of structure formation. Moreover, the confirmation of the isotropic and Gaussian nature of the CMB is essential for justifying the corresponding assumptions usually made when estimating the CMB power spectra and other quantities to be obtained from the *Planck* data. Conversely, the detection of significant deviations from these assumptions that are not consistent with known systematic effects or foreground residuals would necessitate major revision of current methodological approaches for derivating the mission's many science results.

Significant deviations from Gaussianity are expected from non-linear processes that lead to secondary anisotropies, e.g., the integrated Sachs-Wolfe (ISW) effect and lensing. Indeed, these effects are the subject of two companion *Planck* papers (Planck Collaboration XVII 2014; Planck Collaboration XIX 2014, respectively). However, remarkably, a number of anomalies, by which we mean features of the observed sky that are not statistically consistent with the best-fit ΛCDM model, have been found in the *WMAP* data. Indeed, the *WMAP* team (Spergel et al. 2003) themselves initially proposed some intriguing discrepancies in the form of a lack of power on large angular scales, which was subsequently reconfirmed for the three-year data in Copi et al. (2007). Further examples include an alignment of the low-order multipoles (Tegmark et al. 2003; Schwarz et al. 2004; Bielewicz et al. 2005; Land & Magueijo 2005a), a north-south asymmetry in both power spectra (Eriksen et al. 2004a; Hansen et al. 2009) and various measures of non-Gaussianity (Eriksen et al. 2004c, 2005; Räth et al. 2007a), parity asymmetry in the power spectrum corresponding to large angular scales (Kim & Naselsky 2010a), and a region of significant temperature decrement — the so-called *cold spot* (Vielva et al. 2004; Cruz et al. 2005).

Whilst *WMAP* have presented refutations of these anomalies, either by criticizing the robustness of the statistical meth-ods employed (Bennett et al. 2011) or by associating them with systematic artefacts of the data processing that have been corrected in the nine-year data release (Bennett et al. 2012), *Planck* represents a unique opportunity to independently assess their existence. Its higher angular resolution and sensitivity and wider frequency range enable a better understanding and removal of the Galactic and extragalactic foregrounds thus allowing a larger fraction of the sky to be useful for performing isotropy and Gaussianity analysis and to confirm and interpret those anomalies.

Throughout this paper, we quantify the significance of the test statistic in terms of the *p*-value. This is the probability of obtaining a test statistic at least as extreme as the observed one, under the assumption that the null hypothesis (i.e., Gaussianity and isotropy of the CMB) is true. In some tests, where it is very justified to only use a one-tailed probability, the *p*-value is replaced by the corresponding upper or lower-tail probability. A low *p*-value is indicative of a tension between the data and the assumed statistical model (the null hypothesis). This can arise either when the assumed cosmological model is incorrect, if unknown or unmodelled aspects of the foreground emission or the measurement process exist, or as a result of a natural statistical fluctuation. The most interesting possibility, of course, is that a low *p*-value is an indication of new physics.

From the theoretical point of view, there are many variants of inflation that predict high levels of non-Gaussianity and new scenarios motivated by string and M-theory. In addition, there are many physical effects that might give rise to a deviation from isotropy or the presence of non-Gaussianity. Those deviations may be classified according to their physical nature and origin as follows: non-standard inflationary models, geometry and topology of the Universe, and topological defects. The main results from these areas, as well as the detailed descriptions of methodologies and of specific theoretically-motivated model constraints, are provided in the companion papers (Planck Collaboration XXIV 2014; Planck Collaboration XXVI 2014; Planck Collaboration XXV 2014).

This paper covers all relevant aspects related to the phenomenological study of the statistical isotropy and Gaussian nature of the CMB measured by the *Planck* satellite. It is organized as follows. Section 2 describes the *Planck* data used for the analyses. Section 3 explains the main characteristics of the simulations that constitute our reference set of Gaussian sky maps representative of the null hypothesis. In Sect. 4 the null hypothesis is tested with a number of standard tests that probe different aspects of non-Gaussianity. The *WMAP* anomalies are revisited in the light of the *Planck* data in Sect. 5. In Sect. 6 the implications of the found deviations of the null hypothesis on $C_\ell$ and cosmological parameters estimations are discussed. Finally, Sect. 7 provides the main conclusions of the paper.

## 2. Data description

In this paper, we utilize data from the *Planck*-2013 data release corresponding to the nominal period of the *Planck* mission. In part, this comprises sky maps at nine frequencies, with corresponding "half-ring" maps that are generated by separating the data for a given pointing period into two halves, plus maps generated from data within the first and second survey periods. The maps are provided in HEALPix[2] format, with a pixel size defined by the $N_{side}$ parameter. This set of maps allows a variety of consistency checks to be made, together with estimates of the entific consortia funded by ESA member states (in particular the lead countries France and Italy), with contributions from NASA, (USA) and telescope reflectors provided by a collaboration between ESA and a scientific consortium led and funded by Denmark instrumental noise contributions to analyses and limits on time-varying systematic artefacts. Full details are provided in a series of companion papers (Planck Collaboration II 2014; Planck Collaboration VI 2014).

[2] http://healpix.sourceforge.net

**Table 1.** List of the masks that have been used for the analyses described in this paper. All of them have been generated at $N_{side} = 2048$, and when needed, they have been degraded to a lower resolution, as explained in the text. The CL masks have been constructed following the procedure described in Planck Collaboration XV (2014) but for different sky coverages.

| Mask name | Sky coverage [% of unmasked pixels] |
|---|---|
| CS-SMICA89[1] ...... | 89.0 |
| U73[1] ............. | 73.0 |
| CG90[1] ............ | 90.0 |
| CG80[1] ............ | 80.0 |
| CG70[1] ............ | 70.0 |
| CG60[1] ............ | 60.0 |
| CL65[2] ............ | 65.1 |
| CL58[2] ............ | 57.8 |
| CL48[2] ............ | 48.0 |
| CL37[2] ............ | 37.3 |
| CL25[2] ............ | 24.7 |

[1] Planck Collaboration XII (2014)
[2] Planck Collaboration XV (2014)

Our main results are based on the CMB maps resulting from sophisticated component separation algorithms applied to the frequency maps, as detailed in Planck Collaboration XII (2014). The four methods — `Commander-Ruler`, `NILC`, `SEVEM` and `SMICA` — are used to generate estimates of the CMB sky with an effective angular resolution of around $7'$ or better, with accompanying symmetrized beam profiles, analysis masks, and half-ring maps.

In Table 1 we list the different masks that have been used in the analyses described in this paper. These masks have been produced at full resolution ($N_{side}$=2048) and are described in papers (Planck Collaboration XII 2014; Planck Collaboration XV 2014). In general, we make use of a standardized common mask that merges those associated with the individual methods (listed in Table 1 as U73). Where appropriate, we manipulate the masks for use at lower resolution, and, in particular, they have been degraded to lower resolutions ($N_{side}$ =1024, 512, 256, 128, 64, 32 and 16). The masks with resolutions $N_{side}$=128–1024 have been degraded using the following procedure: first, the masks are degraded to their final resolution using the `ud_degrade HEALPix` routine, and then, a conservative approach is followed setting to zero any pixels with a value lower than 0.8. If the masks have to be degraded to even lower resolutions, $N_{side}$ =16–64, the procedure that has been used is different. First, the full-resolution mask is smoothed with a Gaussian kernel with a FWHM equal to three times the pixel size of the low resolution mask that we want to produce. Then the mask is degraded using `ud_degrade` to their final resolution. Finally, those pixels with a value lower or equal than 0.5 have been set to zero and the rest have been set to 1. This criterion, less conservative than the one used for the higher resolution masks, is a compromise between minimizing the amount of sky that is being masked and the level of contamination left unmasked. Note that in some cases the more conservative criterion of a 0.8 threshold has been used for the lower resolutions, as stated in the corresponding analyses.

## 3. Simulations

The derivation of results to be presented in this paper requires extensive simulations, essential aspects of which include: 1) modelling the *Planck* instrumental effects that affect the quality of the data, including instrumental noise and identified systematic effects; 2) replicating the foreground removal approach and estimating the extent of foreground residuals; and 3) modelling the intrinsic statistical properties, Gaussian or otherwise, of the CMB signals expected from specific models of the Universe.

The full focal plane (FFP6) simulations described in Planck Collaboration (2013) provide a complete realization of the *Planck* mission capturing all characteristics of the scanning strategy, telescope, detector responses, and data reduction pipeline over the nominal mission period of 15.5 months. The Planck Sky Model (PSM) is used as input, encompassing the best current estimate of the microwave sky at *Planck* wavelengths including Galactic and extragalactic astrophysical foreground emission. The outputs include a complete set of maps for all detectors with accompanying half-ring and survey splits generated for a reference CMB sky. These have been used to test and validate various analysis tools, employed in turn to evaluate the CMB component separation algorithms as applied to the data set. This also allows an FFP6-based estimate of the foreground residuals remaining in the CMB sky after component separation to be evaluated, and their impact on various statistical estimators quantified.

An accompanying set of Monte Carlo simulations, that are allowed to vary within the cosmic variance, provides us with the reference set of Gaussian sky maps used for the null tests we employ. These simulations include `FEBeCoP` (Mitra et al. 2011) beam convolution at each of the *Planck* frequencies, which are then propagated through the various component separation pipelines using the same weights as derived from the *Planck* nominal mission data analysis. A fiducial CMB power spectrum has been adopted (and varied in the simulations within the cosmic variance) based on an analysis of the *Planck* data at an advanced, but not final stage of processing. Only small changes relative to the final *Planck* power spectrum presented in (Planck Collaboration XV 2014; Planck Collaboration XVI 2014) are observed.

## 4. Are the primordial fluctuations Gaussian?

As has been previously established, it is of major interest to determine whether the statistical properties of the primordial CMB anisotropies correspond to an isotropic Gaussian random field. Recent attempts to test this hypothesis have mainly relied on the *WMAP* data that have less sensitivity and cover a narrower frequency interval. *Planck* represents a unique opportunity to probe fundamental statistical properties of the Universe with cosmic variance limited sensitivity up to $\ell \approx 2000$ and minimum foreground contamination.

There is no unique signature of non-Gaussianity, however, different tests can allow us to probe different types of non-Gaussianity. As a consequence, it is important to subject the data to a variety of tests, and we do so in this section using a number of non-parametric tools. Specific signatures of non-Gaussianity are sought in three companion papers (Planck Collaboration XXIV 2014; Planck Collaboration XXV 2014; Planck Collaboration XXVI 2014).

Any signal on the sphere, $T(\boldsymbol{x})$, can be written in terms of the following spectral representation:

$$T(\boldsymbol{x}) = \sum_{\ell=0}^{\infty} \sum_{m=-\ell}^{\ell} a_{\ell m} Y_{\ell m}(\boldsymbol{x}), \qquad (1)$$

where $\boldsymbol{x}$ is a unit direction vector, $Y_{\ell m}$ the spherical harmonics and

$$a_{\ell m} = \int \mathrm{d}\boldsymbol{x} T(\boldsymbol{x}) Y_{\ell m}^*(\boldsymbol{x}), \qquad (2)$$

$m = 0, \pm 1, \ldots, \pm \ell, \quad \ell = 1, 2, \ldots$

For an isotropic random field the $a_{\ell m}$ coefficients are uncorrelated:

$$\langle a_{\ell m} a_{\ell' m'}^* \rangle = \delta_{\ell'}^{\ell} \delta_{m'}^{m} C_{\ell}, \qquad (3)$$

where $\delta$ is the Kronecker delta and $C_{\ell}$ is the angular power spectrum. Note that for a Gaussian and isotropic random field, the $a_{\ell m}$ coefficients are independent and, if the mean is zero, the angular power spectrum provides a complete characterization of its statistical distribution.

In this paper, we examine the goodness-of-fit of the data to the *Planck* best-fit fiducial CMB model, which constitutes our null hypothesis. The methods adopted constitute a broad range of statistical tools that allow the study of complementary statistical properties of the null hypothesis in both the real and harmonic space data representations. Claims of either consistency with the fiducial *Planck* cosmological model or of evidence for non-Gaussianity must be demonstrably robust to data selection and specifics of the data analysis. Residuals from the diffuse Galactic foreground are likely to be non-Gaussian in nature, and point-sources can be a source of non-Gaussianity on small angular scales. In addition, the analysis of multifrequency data must be considered in order to confirm that any claimed non-Gaussianity has a thermal (cosmological) spectrum. Moreover, the combined ISW-lensing effect produces secondary anisotropies that significantly deviate from Gaussianity and whose effect has been detected in the *Planck* data (Planck Collaboration XIX 2014). This non-Gaussian effect has to be considered when testing the null hypothesis.

We address these issues by analysing the cosmologically interesting subset of *Planck* frequency channels. Specifically, we analyse the uncorrected sky maps at 70, 100, 143 and 217 GHz, as a function of Galactic mask, to assess the likely contamination due to Galactic foregrounds. These tests have direct relevance for the *Planck* likelihood approach described in Planck Collaboration XV (2014) and the parameter estimation results presented in Planck Collaboration XVI (2014). We then consider the foreground cleaned versions of these maps generated by the SEVEM algorithm (Planck Collaboration XII 2014). Such a comparison also allows a semi-independent cross-check of the cosmological signal seen by *Planck* LFI (70 GHz) and HFI (100, 143, 217 GHz). Although the cosmological content of the cleaned LFI and HFI data sets are independent, the cleaning makes use of difference maps generated from the remaining *Planck* frequency bands. Nevertheless, since the calibration and beam responses of the data are well understood over the ful range of frequencies, there will be no leakage of cosmological signal between the instrument specific frequencies.

We then continue with analyses of the CMB sky estimates provided by four component separation approaches (Commander-Ruler, NILC, SEVEM, and SMICA) described in

Planck Collaboration XII (2014), together with the corresponding mask appropriate for these methods. The largest sky area possible should be used for definitive statements about Gaussianity, since, in the absence of foreground residuals or systematic artefacts, it represents a superior sample of the Universe. Conversely, overly conservative sky cuts suffer from a loss of information.

**Table 2.** Lower tail probablity for the variance, skewness, and kurtosis estimators at $N_{\mathrm{side}} = 2048$, using the U73 mask and four different component separation methods.

| Method | Probability | | |
|---|---|---|---|
| | Variance | Skewness | Kurtosis |
| C-R........... | 0.021 | 0.189 | 0.416 |
| NILC.......... | 0.020 | 0.191 | 0.392 |
| SEVEM......... | 0.014 | 0.206 | 0.419 |
| SMICA......... | 0.017 | 0.189 | 0.419 |

### 4.1. One dimensional moments

In this section we perform some of the simplest Gaussianity tests, such as comparing the sample skewness and kurtosis of the data with simulations. The skewness, $\gamma$, and kurtosis, $\kappa$, of a random variable, $X$, are defined as

$$\gamma(X) = \frac{\langle X - \langle X \rangle \rangle^3}{(\sigma^2(X))^{3/2}}, \qquad (4)$$

$$\kappa(X) = \frac{\langle X - \langle X \rangle \rangle^4}{(\sigma^2(X))^2} - 3, \qquad (5)$$

where $\sigma^2$, the variance is:

$$\sigma^2(X) = \langle X^2 \rangle - \langle X \rangle^2. \qquad (6)$$

Here $X$ represents the pixel temperature at some $N_{\mathrm{side}}$ and the averages are performed over the sky. The skewness is a measure of the asymmetry of the probability distribution of a real-valued random variable. Qualitatively, a positive (negative) skew indicates that the tail on the right (left) side of the probability density function is longer than the left (right) side. A zero value indicates that the values are relatively evenly distributed on both sides of the mean, typically but not necessarily, implying a symmetric distribution. The kurtosis is a measure of the peakedness of the distribution and the heaviness of its tail. A distribution with positive (negative) excess kurtosis indicates that the distribution has a more acute (wider) peak around the mean and fatter (thinner) tails. Normal random variables have zero skewness and kurtosis.

The sample variance, $\sigma^2$ is also considered in this section as a further consistency test, although it is not a normality test statistic. We begin by analysing the full resolution combined maps, applying the U73 mask for the four different component separation methods. The results for the variance, skewness, and kurtosis estimators are shown in Table 2. All four methods show similar results. The data are consistent with simulations for the skewness and kurtosis estimators, whereas the variance is anomalously low. This inconsistency was already reported for the *WMAP* data in Monteserín et al. (2008) and Cruz et al. (2011) at resolution $N_{\mathrm{side}} = 256$ for a mask allowing slightly less sky coverage.

The mask dependence of our results is studied by recalculating the estimators using the CL58 and CL37 masks, which allow sky fractions of $f_{\mathrm{sky}} = 58\%$ and $f_{\mathrm{sky}} = 37\%$, respectively. The `SMICA` cleaned maps at full resolution are considered. The most significant lower tail probability is obtained for the CL58, mask as can be seen in Table 3. The lower tail probabilities show a small dependence on the mask used, which could indicate either the presence of Galactic foreground residuals with larger sky coverage, or the increase of the sampling variance, and consequently a less significant probability, when a smaller fraction of the sky is considered.

**Table 3.** Lower tail probablity for the variance, skewness, and kurtosis estimators at $N_{\mathrm{side}} = 2048$, for the `SMICA` method, using different masks.

| | Probability | | |
|---|---|---|---|
| Mask | Variance | Skewness | Kurtosis |
| U73, $f_{\mathrm{sky}} = 73\%$ . . . . . . . . . . . | 0.017 | 0.189 | 0.419 |
| CL58, $f_{\mathrm{sky}} = 58\%$ . . . . . . . . . . | 0.003 | 0.170 | 0.363 |
| CL37, $f_{\mathrm{sky}} = 37\%$ . . . . . . . . . . | 0.030 | 0.314 | 0.266 |
| Ecliptic North, $f_{\mathrm{sky}} = 36\%$ . . . . | 0.001 | 0.553 | 0.413 |
| Ecliptic South, $f_{\mathrm{sky}} = 37\%$ . . . . | 0.483 | 0.077 | 0.556 |
| Galactic North, $f_{\mathrm{sky}} = 37\%$ . . . . | 0.001 | 0.788 | 0.177 |
| Galactic South, $f_{\mathrm{sky}} = 36\%$ . . . . | 0.592 | 0.145 | 0.428 |

**Table 4.** Lower tail probablity for the variance, skewness, and kurtosis estimators at $N_{\mathrm{side}} = 2048$, for the `SEVEM` cleaned maps at different frequencies.

| | Probability | | |
|---|---|---|---|
| Map | Variance | Skewness | Kurtosis |
| 100 GHz . . . . . . . . . . . . . . | 0.023 | 0.195 | 0.488 |
| 143 GHz . . . . . . . . . . . . . . | 0.014 | 0.221 | 0.460 |
| 217 GHz . . . . . . . . . . . . . . | 0.025 | 0.196 | 0.481 |

In order to identify any foreground contamination, the frequency dependence of our estimators is analysed. We use the `SEVEM` cleaned maps and the U73 mask. Note that as the 70 GHz full resolution noise is high we do not consider 70 GHz in this comparison. As the 100 GHz noise is not negligible we estimate the variance taking into account the noise dispersion as described in Cruz et al. (2011). The results are similar to those found for the combined map, as can be seen in Table 4. There is a small frequency dependence since the 100 GHz and 217 GHz maps show slightly higher variance and kurtosis than the 143 GHz map. However the 143 GHz map has a dominant contribution to the combined map, hence the foreground residuals in the combined map are likely to be small. The lower tail probabilities for the variance at 100 GHz, 143 GHz, and 217 GHZ are respectively 0.023, 0.014, 0.025, whereas the skewness and kurtosis are compatible with simulations.

In order to study the hemispherical asymmetry reported in Cruz et al. (2011) we reanalyse the `SMICA` data and simulations considering independently the northern and southern ecliptic hemispheres outside the U73 mask. A clear asymmetry is found in the variance, with an anomalously low value found in the northern hemisphere, as seen in Table 3. Hemispherical asymmetries are further analysed in Sect. 5.3.

The results for different resolutions using the U73 mask are shown in Table 5. Note that the $N_{\mathrm{side}} = 2048$ and 512 U73 masks have $f_{\mathrm{sky}} = 73\%$, while the low resolution masks at $N_{\mathrm{side}} = 64$, 32, and 16 have $f_{\mathrm{sky}} = 78\%$. The variance is anomalously low at all the considered resolutions, whereas at low resolutions, the skewness is anomalously low and the kurtosis anomalously high. These results will be further analysed in Sect. 5.2. However, it is clear that, except on the largest angular scales, there is no evidence for non-Gaussian behaviour in the data using these simple statistical measures.

**Table 5.** Lower tail probablity for the variance, skewness, and kurtosis estimators at different resolutions, for the four component separation methods, using the U73 mask.

| | Probability | | |
|---|---|---|---|
| Method | Variance | Skewness | Kurtosis |
| $N_{\mathrm{side}} = 2048$ | | | |
| C-R . . . . . . . . | 0.021 | 0.189 | 0.416 |
| NILC . . . . . . . | 0.020 | 0.191 | 0.392 |
| SEVEM . . . . . . | 0.014 | 0.206 | 0.419 |
| SMICA . . . . . . | 0.017 | 0.189 | 0.419 |
| $N_{\mathrm{side}} = 512$ | | | |
| C-R . . . . . . . . | 0.017 | 0.207 | 0.368 |
| NILC . . . . . . . | 0.017 | 0.198 | 0.390 |
| SEVEM . . . . . . | 0.013 | 0.218 | 0.408 |
| SMICA . . . . . . | 0.014 | 0.196 | 0.390 |
| $N_{\mathrm{side}} = 64$ | | | |
| C-R . . . . . . . . | 0.011 | 0.041 | 0.935 |
| NILC . . . . . . . | 0.011 | 0.041 | 0.935 |
| SEVEM . . . . . . | 0.008 | 0.058 | 0.900 |
| SMICA . . . . . . | 0.011 | 0.041 | 0.943 |
| $N_{\mathrm{side}} = 32$ | | | |
| C-R . . . . . . . . | 0.020 | 0.015 | 0.968 |
| NILC . . . . . . . | 0.019 | 0.016 | 0.960 |
| SEVEM . . . . . . | 0.012 | 0.026 | 0.932 |
| SMICA . . . . . . | 0.019 | 0.016 | 0.967 |
| $N_{\mathrm{side}} = 16$ | | | |
| C-R . . . . . . . . | 0.023 | 0.013 | 0.974 |
| NILC . . . . . . . | 0.022 | 0.022 | 0.972 |
| SEVEM . . . . . . | 0.019 | 0.022 | 0.964 |
| SMICA . . . . . . | 0.027 | 0.021 | 0.982 |

### 4.2. N-pdf analysis

Under the assumption of Gaussianity, the $N$-probability density function ($N$-pdf) is given by a multivariate Gaussian function:

$$f(\boldsymbol{T}) = \frac{1}{(2\pi)^{N_{\mathrm{pix}}/2} \det \mathsf{C}^{1/2}} \exp{-\frac{1}{2} \left( \boldsymbol{T} \mathsf{C}^{-1} \boldsymbol{T}^{\mathrm{T}} \right)}, \qquad (7)$$

where $\boldsymbol{T}$ is a vector formed from the measured temperatures $T(\boldsymbol{x})$ over all positions allowed by the applied mask, $N_{\mathrm{pix}}$ is the number of pixels in the vector, $\mathsf{C}$ is the covariance of the Gaussian field (of size $N_{\mathrm{pix}} \times N_{\mathrm{pix}}$).

Unfortunately, the calculation of $\boldsymbol{T}\mathsf{C}^{-1}\boldsymbol{T}^{\mathrm{T}}$ is computationally unfeasible for the full *Planck* resolution at `HEALPix` $N_{\mathrm{side}} = 2048$. At a lower resolution, the problem is tractable, and the noise level can also be considered negligible compared to the CMB signal. That implies that under the assumption of isotropy the covariance matrix $\mathsf{C}$ is fully defined by the *Planck* angular power spectrum ($C_\ell$):

$$\mathsf{C}_{ij} = \sum_{\ell=0}^{\ell_{\mathrm{max}}} \frac{2\ell+1}{4\pi} C_\ell b_\ell^2 P_\ell \left( \cos \theta_{ij} \right), \qquad (8)$$

where $\mathsf{C}_{ij}$ is the covariance between pixels $i$ and $j$, and $\theta_{ij}$ is angle between them, $P_\ell$ are the Legendre polynomials, $b_\ell$ is an effective window function associated with the $N_{\mathrm{side}}$ resolution, and $\ell_{\max}$ is the maximum multipole probed.

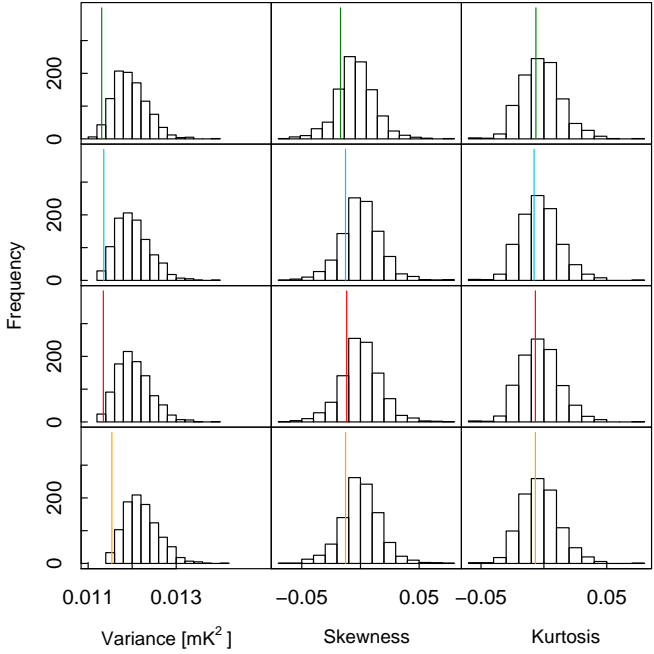

**Fig. 1.** Variance, skewness, and kurtosis for the combined map of the four different component separation methods. From top row to bottom row C-R, NILC, SEVEM, SMICA.

Under the multivariate Gaussian hypothesis, the argument on the exponential in equation 7 should follow a $\chi^2$ distribution with $N_{\mathrm{pix}}$ degrees of freedom, or, equivalently (for $N_{\mathrm{pix}} \gg 1$) a normal distribution $\mathcal{N}\left(N_{\mathrm{pix}}, \sqrt{N_{\mathrm{pix}}}\right)$.

We begin by analysing the $\chi^2$ quantity for low resolution maps at $N_{\mathrm{side}} = 32$ and filtering with a $5°$ FWHM Gaussian. $1\,\mu\mathrm{K}$ uncorrelated regularization noise is added to the covariance matrix before inverting it. Regularization noise realizations are added to the data and simulations for consistency (see Eriksen et al. 2007b, for more details).

We analyse the four cleaned data maps, applying the common, CL58 and CL37 masks. The admitted fraction of the sky is respectively $78\%, 58\%$ and $37\%$. The northern and southern ecliptic hemispheres outside the U73 mask are also considered independently in order to check for possible hemispherical asymmetries as those analysed in Sect. 5.3. The results are shown in Fig. 2 and Table 6. In the U73 case, the lower tail probabilities are low. Applying the two CL58 and CL37 masks that permit less sky coverage, gives data that are consistent with simulations. The low $\chi^2$ value appears to be localized in the northern ecliptic hemisphere. These results are directly comparable to the anomalous variance mentioned in Sect. 4.1. Note that the four maps show similar values, but the differences are larger when using the U73 mask. This could indicate the presence of some residual foreground contamination near the Galactic plane. Therefore, the frequency dependence of our estimator is analysed in order to identify any possible foreground contamination. The results are shown in Fig. 3 and Table 7. A moderate frequency dependence is found when using the U73 mask, which could indicate the presence of some foreground residuals near the Galactic plane. The frequency dependence of the results vanishes when using the CL58 and CL37 masks that exclude more of the sky from analysis.

**Table 6.** Lower tail probablity for the $N$-pdf, using different masks.

| Mask | Probability | | | |
|---|---|---|---|---|
| | C-R | NILC | SEVEM | SMICA |
| U73, $f_{\mathrm{sky}} = 78\%$ . . . . . . . . . . . | 0.027 | 0.028 | 0.019 | 0.030 |
| CL58, $f_{\mathrm{sky}} = 58\%$ . . . . . . . . . . | 0.137 | 0.137 | 0.147 | 0.146 |
| CL37, $f_{\mathrm{sky}} = 37\%$ . . . . . . . . . . | 0.409 | 0.415 | 0.420 | 0.436 |
| Ecliptic North, $f_{\mathrm{sky}} = 39\%$ . . . | 0.024 | 0.022 | 0.021 | 0.021 |
| Ecliptic South, $f_{\mathrm{sky}} = 39\%$ . . . | 0.170 | 0.196 | 0.183 | 0.193 |

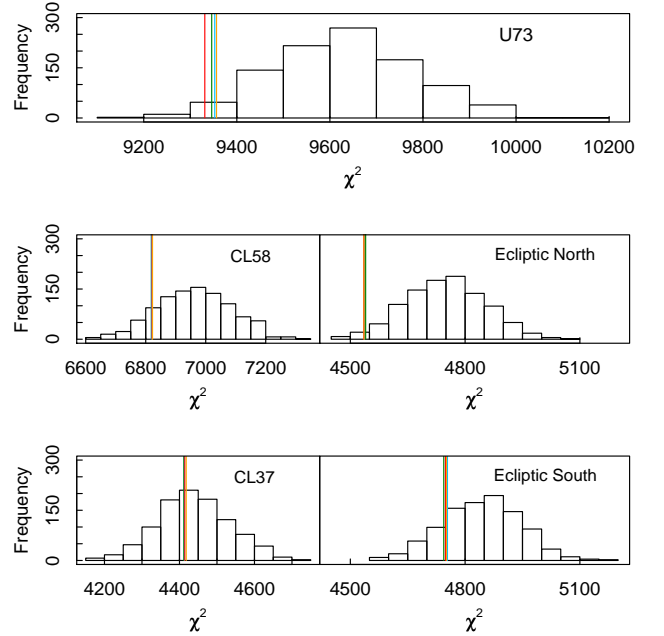

**Fig. 2.** $N$-pdf $\chi^2$ for the U73 mask, CL58, CL37, ecliptic North and ecliptic South. The different colours represent the four component separation methods, namely C-R (green), NILC (blue), SEVEM (red), and SMICA (orange).

**Table 7.** Frequency dependence of the lower tail probablity for the $N$-pdf, using different masks.

| Mask | Probability | | | |
|---|---|---|---|---|
| | 70 GHz | 100 GHz | 143 GHz | 217 GHz |
| U73, $f_{\mathrm{sky}} = 78\%$ . . . . . . . . . . . | 0.037 | 0.058 | 0.013 | 0.124 |
| CL58, $f_{\mathrm{sky}} = 58\%$ . . . . . . . . . . | 0.169 | 0.123 | 0.111 | 0.169 |
| CL37, $f_{\mathrm{sky}} = 37\%$ . . . . . . . . . . | 0.422 | 0.366 | 0.376 | 0.386 |
| Ecliptic North, $f_{\mathrm{sky}} = 39\%$ . . . | 0.028 | 0.050 | 0.015 | 0.083 |
| Ecliptic South, $f_{\mathrm{sky}} = 39\%$ . . . | 0.225 | 0.233 | 0.166 | 0.330 |

### 4.3. N-point correlation functions

In this section we present tests of the non-Gaussianity of the *Planck* CMB maps using real-space $N$-point correlation functions. While harmonic-space methods are often preferred over real-space methods for studying primordial fluctuations, real-space methods may have an advantage with respect to systematics and foregrounds, since such effects are usually localized in real space. It is therefore important to analyse the data in both spaces in order to highlight different features.

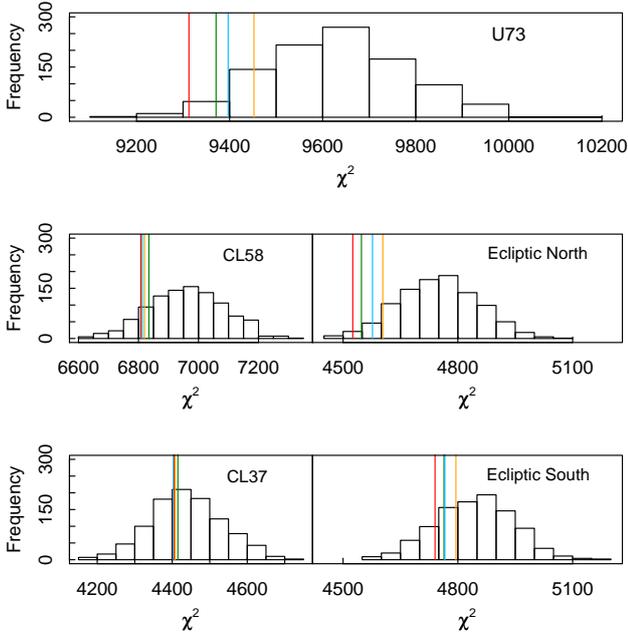

**Fig. 3.** Frequency dependence for 70 GHz (green), 100 GHz (blue), 143 GHz (red), and 217 GHz (orange), for different masks.

An $N$-point correlation function is by definition the average product of $N$ temperatures, measured in a fixed relative orientation on the sky,

$$C_N(\theta_1, \ldots, \theta_{2N-3}) = \langle \Delta T(\hat{\mathbf{n}}_1) \cdots \Delta T(\hat{\mathbf{n}}_N) \rangle, \tag{9}$$

where the unit vectors $\hat{\mathbf{n}}_1, \ldots, \hat{\mathbf{n}}_N$ span an $N$-point polygon on the sky. By assuming statistical isotropy, the $N$-point functions are only functions of the shape and size of the $N$-point polygon, and not on its particular position or orientation on the sky. Hence, the smallest number of parameters that uniquely determines the shape and size of the $N$-point polygon is $2N - 3$. In practice, the functions are estimated by simple product averages over all sets of $N$ pixels fulfilling the geometric requirements set by $\theta_1, \ldots, \theta_{2N-3}$ characterizing the shape and size of the polygon,

$$C_N(\theta_1, \ldots, \theta_{2N-3}) = \frac{\sum_i \left( w_1^i \cdots w_N^i \right) \left( \Delta T_1^i \cdots \Delta T_N^i \right)}{\sum_i w_1^i \cdots w_N^i}. \tag{10}$$

Pixel weights $w_1^i, \cdots, w_N^i$ can be introduced in order to reduce noise or mask boundary effects. Here they represent masking by being set to 1 for included pixels and 0 for excluded pixels.

The main difficulty with computing $N$-point functions is their computational scaling. The number of independent pixel combinations scales as $O(N_{\text{pix}}^N)$, and for each combination of $N$ pixels, $2N - 3$ angular distances must be computed to uniquely determine the properties of the corresponding polygon. Computing the full $N$-point function for $N > 2$ and $N_{\text{pix}} \gtrsim 10^5$ is therefore computationally challenging. However, it is not necessary to include all possible $N$-point configurations in order to produce interesting results. For instance, one may focus only on small angular scales, or on configurations with some special symmetry properties. By using the methods described in Eriksen et al. (2004b), the computational expense then becomes tractable, since no CPU time is spent on excluded configurations. In this paper several such subsets are computed, covering three distinct ranges of scales, namely small (up to 3°), intermediate (up to 10°) and large angular scales (the full range between 0° and 180°).

The shapes of the polygons selected for the analysis are the pseudo-collapsed and equilateral configurations for the 3-point function, and the rhombic configuration for the 4-point function, composed of two equilateral triangles that share a common side. We use the same definition of pseudo-collapsed as in Eriksen et al. (2005) i.e., as isosceles triangle where the length of the baseline falls within the second bin of the separation angles. The length of the longer edge of the triangle, $\theta$, parametrizes its size. Analogously, in the case of the equilateral triangle and rhombus, the size of the polygon is parametrized by the length of the edge, $\theta$. Note that these functions are chosen because of ease of implementation, not because they are better suited for testing Gaussianity than other configurations. In the following, all results refer to the reduced 4-point function, i.e., corrected for the Gaussian contribution due to Wick's theorem.

We analyse the CMB estimates downgraded to $N_{\text{side}} = 64$ and $N_{\text{side}} = 512$ as well as at the original resolution of $N_{\text{side}} = 2048$. In the case of the analysis at $N_{\text{side}} = 64$ the maps were additionally smoothed with FWHM of 165′ (three times the pixel size for the downgraded map). Due to computational limitations, an analysis is possible on the full sky only in the case of resolution $N_{\text{side}} = 64$. For the higher resolution maps, we perform the analysis on a set of non-overlapping discs. For $N_{\text{side}} = 512$ we uniformly retain, after masking, part of the sky with approximately 100 discs of radius 10°. Analogously to the analysis by Eriksen et al. (2005), we consider two disc sets, A and B, with a relative offset between their grids such that the centres of the discs of set B are located in parts of the sky not covered by disc set A (see Fig. 4). For studies at the original resolution $N_{\text{side}} = 2048$, we restrict the analysis to 20 discs with a radius of 3° located randomly on an unmasked part of the sky (Fig. 4).

As in Eriksen et al. (2005), we consider the $N$-point correlation functions averaged over the disc sets. In order to minimize correlations between the discs, we subtract from the maps at resolutions $N_{\text{side}} = 512$ and $N_{\text{side}} = 2048$ the best-fit multipoles computed for the ranges $\ell \leq 18$ and $\ell \leq 60$, respectively. This procedure corresponds in practice to a high-pass filtering of the maps. One consequence of this procedure is the reduced sensitivity of the method to diffuse Galactic foreground residuals, which contribute most significantly to the low order multipoles. However, such a contribution is tested for in the analysis at resolution $N_{\text{side}} = 64$, in which high-pass filtering is not employed beyond the removal of the monopole and dipole.

In the case of the analysis of the maps at resolution $N_{\text{side}} = 512$ the $N$-point correlation functions were averaged over all discs for a given set, as well as separately for discs as a function of Galactic latitude. Such analysis allows the study of possible contamination of the maps by diffuse Galactic foreground residuals distributed symmetrically around the Galactic plane.

The low resolution versions of the U73 mask described earlier were used as required. Residual monopole and dipole contributions were then removed from the maps.

A simple $\chi^2$ test is chosen to quantify the degree of agreement between the simulations and the observations, where $\chi^2$ as usual is defined by

$$\chi^2 = \sum_{i,j=1}^{N_{\text{bin}}} \left( C_N(\theta_i) - \left\langle C_N(\theta_i) \right\rangle \right) M_{ij}^{-1} \left( C_N(\theta_j) - \left\langle C_N(\theta_j) \right\rangle \right). \tag{11}$$

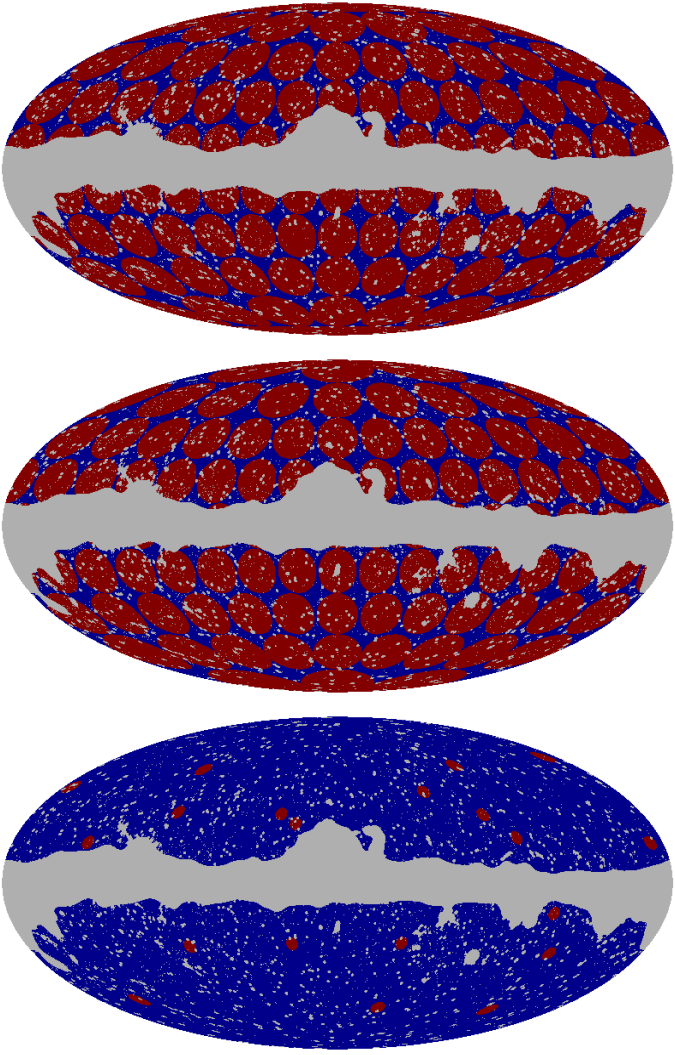

**Fig. 4.** Two sets of discs of radius 10° A (*top*) and B (*middle*), for the $N_{side} = 512$ CMB estimates; and a set of 20 randomly placed discs of radius 3° super-imposed on the U73 mask (blue region) for the CMB estimates at $N_{side} = 2048$ (*bottom*).

Here $C_N(\theta_i)$ is the $N$-point correlation function for the bin of separation angle, $\theta_i$ (corresponding to the length of the longest side for the pseudo-collapsed configuration and the length of the polygon's side for the equilateral triangle and rhombus), $\langle C_N(\theta_i) \rangle$ is the corresponding average from the Monte Carlo (MC) ensemble, and

$$M_{ij} = \frac{1}{N_{sim}} \sum_{k=1}^{N_{sim}} \left( C_N^{(k)}(\theta_i) - \left\langle C_N(\theta_i) \right\rangle \right) \left( C_N^{(k)}(\theta_j) - \left\langle C_N(\theta_j) \right\rangle \right) \quad (12)$$

is the covariance matrix. Although the inverse of the covariance matrix constructed from MC simulations can be biased, it is relatively small for 1000 simulations and has a negligible impact on the significance levels estimated from the simulations, as described below.

This statistic is optimized for studying Gaussian distributed data. However, usually it also works quite well for mildly non-Gaussian distributions, and in particular symmetric ones. Nevertheless, as for any statistic constructed from MC simulations, it can also be used for non-Gaussian and asymmetrically distributed data. Below, we quote the significance level in terms of the fraction of simulations with a larger $\chi^2$ value than the observed map.

We analyse the mask dependence of the non-Gaussianity of the maps using the pseudo-collapsed 3-point correlation function. The low order multipoles were estimated and subtracted from the maps for each mask separately. These estimates thus depend on the specific sky coverage employed, so that the comparison of the correlation functions for different masks is non-trivial. The function averaged over disc set A is shown in Fig. 5. The corresponding probabilities of obtaining values of the $\chi^2$ statistic for the concordance $\Lambda$CDM model at least as large as the observed values are given in Table 8.

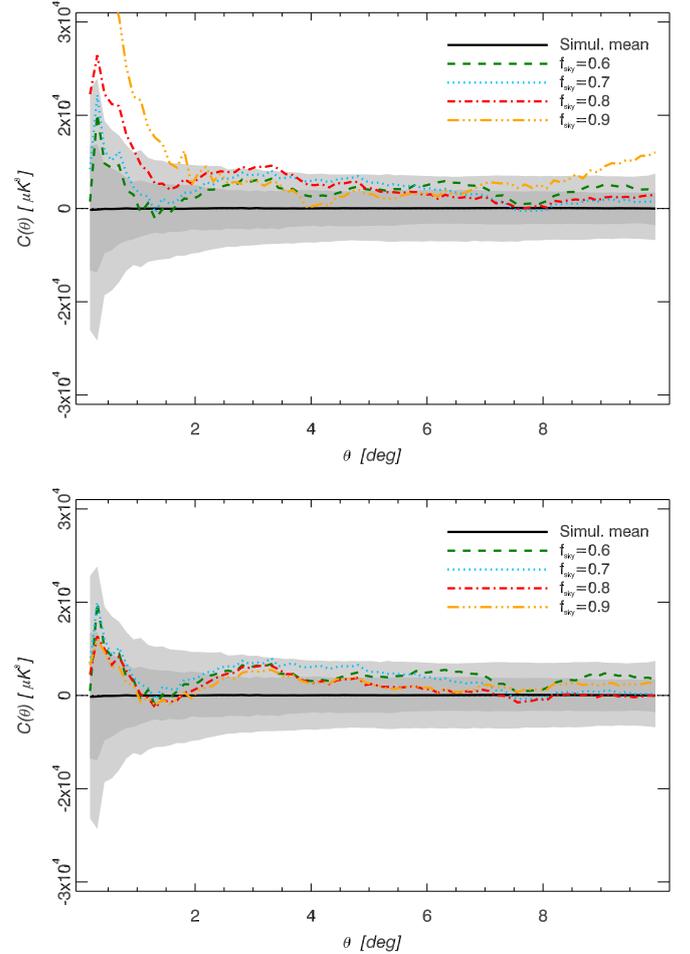

**Fig. 5.** Pseudo-collapsed 3-point function averaged over disc set A for the raw (upper figure) and `SEVEM` foreground corrected (lower figure) 143 GHz map at $N_{side} = 512$ for different values of sky coverage ($f_{sky}$). Estimates of the multipoles for $\ell \leq 18$ are removed from the sky maps. The black solid line indicates the mean for 1000 MC simulations and the shaded dark and light grey regions indicate the 68% and 95% confidence regions, respectively, for the CG90 ($f_{sky} = 0.9$) mask. See Sect. 4.3 for the definition of the separation angle $\theta$.

In summary, the pseudo-collapsed 3-point function does not show any significant deviation from Gaussianity for the raw 143 GHz map masked with the CG60 ($f_{sky} = 0.6$) and CG70 ($f_{sky} = 0.7$) masks. To a lesser extent, this is true also for the CG80 ($f_{sky} = 0.8$) mask. We do not see any significant deviation

**Table 8.** Probabilities of obtaining values of the $\chi^2$ statistic for the concordance $\Lambda$CDM model at least as large as the observed values of the statistic for the raw 143 GHz (first row) and foreground corrected 143 GHz SEVEM CMB maps (second row). In this table $f_{sky}$ is the fraction of unmasked pixels.

| Map | Probability | | | |
|---|---|---|---|---|
| | $f_{sky} = 0.6$ | $f_{sky} = 0.7$ | $f_{sky} = 0.8$ | $f_{sky} = 0.9$ |
| Raw . . . . . . . . . . . | 0.907 | 0.889 | 0.563 | <0.001 |
| SEVEM . . . . . . . . . | 0.959 | 0.959 | 0.905 | 0.940 |

**Table 9.** Probabilities of obtaining values for the $\chi^2$ statistic of the $N$-point functions shown in Fig. 6 for the *Planck* fiducial $\Lambda$CDM model at least as large as the observed values of the statistic for the *Planck* CMB maps with resolution parameter $N_{side} = 64$, estimated using the C-R, NILC, SEVEM and SMICA methods.

| Function | Probability | | | |
|---|---|---|---|---|
| | C-R | NILC | SEVEM | SMICA |
| Two-pt. . . . . . . . . . . . . . . . . . | 0.883 | 0.859 | 0.884 | 0.855 |
| Pseudo-coll. three-pt. . . . . . . . | 0.922 | 0.918 | 0.945 | 0.908 |
| Equil. three-pt. . . . . . . . . . . . | 0.962 | 0.966 | 0.978 | 0.968 |
| Four-pt. . . . . . . . . . . . . . . . . . | 0.975 | 0.977 | 0.979 | 0.977 |

for any of the analysed masks after cleaning the 143 GHz map using the SEVEM method.

The correlation functions for the four component separation methods and resolution parameters $N_{side} = 64$, $N_{side} = 512$ and $N_{side} = 2048$ are shown in Fig. 6, Fig. 7 (disc set A), Fig. 8 (disc set B). and Fig. 9, respectively. The probabilities of obtaining values of the $\chi^2$ statistic for the *Planck* fiducial $\Lambda$CDM model at least as large as the observed values are given in the Tables 9, 10, and 12, respectively. In addition, Table 11 reports the probabilities for the $N$-point functions averaged over sets of discs as a function of Galactic latitude for the SMICA map at $N_{side} = 512$.

**Table 10.** Probabilities of obtaining values of the $\chi^2$ statistic of the $N$-point functions shown in Figs. 7 and 8 for the *Planck* fiducial $\Lambda$CDM model at least as large as the observed values of the statistic for *Planck* CMB maps with resolution parameter $N_{side} = 512$. estimated using the C-R, NILC, SEVEM and SMICA methods.

| Set | Probability | | | |
|---|---|---|---|---|
| | C-R | NILC | SEVEM | SMICA |
| | Two-pt. | | | |
| A . . . . . . . . . . . | 0.858 | 0.902 | 0.886 | 0.904 |
| B . . . . . . . . . . . | 0.351 | 0.370 | 0.404 | 0.376 |
| | Pseudo-coll. three-pt. | | | |
| A . . . . . . . . . . . | 0.568 | 0.565 | 0.651 | 0.603 |
| B . . . . . . . . . . . | 0.483 | 0.526 | 0.550 | 0.540 |
| | Equilateral three-pt. | | | |
| A . . . . . . . . . . . | 0.004 | 0.032 | 0.045 | 0.043 |
| B . . . . . . . . . . . | 0.452 | 0.485 | 0.443 | 0.479 |
| | Four-pt. | | | |
| A . . . . . . . . . . . | 0.104 | 0.102 | 0.102 | 0.107 |
| B . . . . . . . . . . . | 0.521 | 0.569 | 0.537 | 0.579 |

**Table 11.** Probabilities of obtaining values of the $\chi^2$ statistic of the $N$-point functions shown in Figs. 7 and 8 for the *Planck* fiducial $\Lambda$CDM model at least as large as the observed values for SMICA at $N_{side} = 512$. The analysis was performed separately for every set of discs centred on the Galactic latitudes with the same absolute value. The Galactic latitudes of the disc centres are given in the first two rows of the table.

| Set | Probability | | | | |
|---|---|---|---|---|---|
| A . . . . . . . . . . . | ±90° | ±70° | ±50° | ±30° | ±10° |
| B . . . . . . . . . . . | ±80° | ±60° | ±40° | ±20° | 0° |
| | Two-pt. | | | | |
| A . . . . . . . . . . . | 0.057 | 0.672 | 0.780 | 0.594 | 0.556 |
| B . . . . . . . . . . . | 0.158 | 0.401 | 0.741 | 0.667 | <0.001 |
| | Pseudo-coll. three-pt. | | | | |
| A . . . . . . . . . . . | 0.006 | 0.669 | 0.153 | 0.414 | 0.783 |
| B . . . . . . . . . . . | 0.014 | 0.211 | 0.639 | 0.763 | 0.240 |
| | Equilateral three-pt. | | | | |
| A . . . . . . . . . . . | 0.083 | 0.922 | 0.161 | 0.086 | 0.082 |
| B . . . . . . . . . . . | 0.956 | 0.800 | 0.827 | 0.288 | <0.001 |
| | Four-pt. | | | | |
| A . . . . . . . . . . . | 0.014 | 0.390 | 0.324 | 0.492 | <0.001 |
| B . . . . . . . . . . . | 0.098 | 0.920 | 0.987 | 0.537 | <0.001 |

**Table 12.** Probabilities of obtaining values of the $\chi^2$ statistic of the $N$-point functions shown in Fig. 9 for the *Planck* fiducial $\Lambda$CDM model at least as large as the observed values of the statistic for *Planck* CMB maps with resolution parameter $N_{side} = 2048$ estimated using the C-R, NILC, SEVEM, and SMICA methods.

| Function | Probability | | | |
|---|---|---|---|---|
| | C-R | NILC | SEVEM | SMICA |
| Two-pt. . . . . . . . . . . . . . . . . | 0.335 | 0.474 | 0.573 | 0.497 |
| Pseudo-coll. three-pt. . . . . . | 0.522 | 0.463 | 0.469 | 0.448 |
| Equil. three-pt. . . . . . . . . . . | 0.853 | 0.789 | 0.819 | 0.796 |
| Four-pt. . . . . . . . . . . . . . . . | 0.532 | 0.534 | 0.579 | 0.526 |

The results show consistency between the CMB maps estimated using the different component separation methods. No statistically significant deviations of the CMB maps from Gaussianity are found on any scale, with the exception of some analyses performed on sets of discs centred at low Galactic latitudes. However, for disc set B the fraction of such discs available for analysis after masking is very small, reducing the robustness of the results. A more intriguing case with very small probability arises for the disc set A 4-point function averaged over discs centred at Galactic latitude ±10°. The fraction of discs available for the analysis is much larger in this case than for disc set B, and may indicate the presence of Galactic emission residuals.

It is also worth noticing that the CMB map smoothed and downgraded to $N_{side} = 64$ show the largest deviation, with a tendency towards super-Gaussianity, especially for the 4-point correlation function, in comparison to the intermediate and small angular scale analyses. Since the CMB maps at this resolution were not high-pass filtered, they may be more susceptible to Galactic foreground residual artefacts that survive the component separation processes.

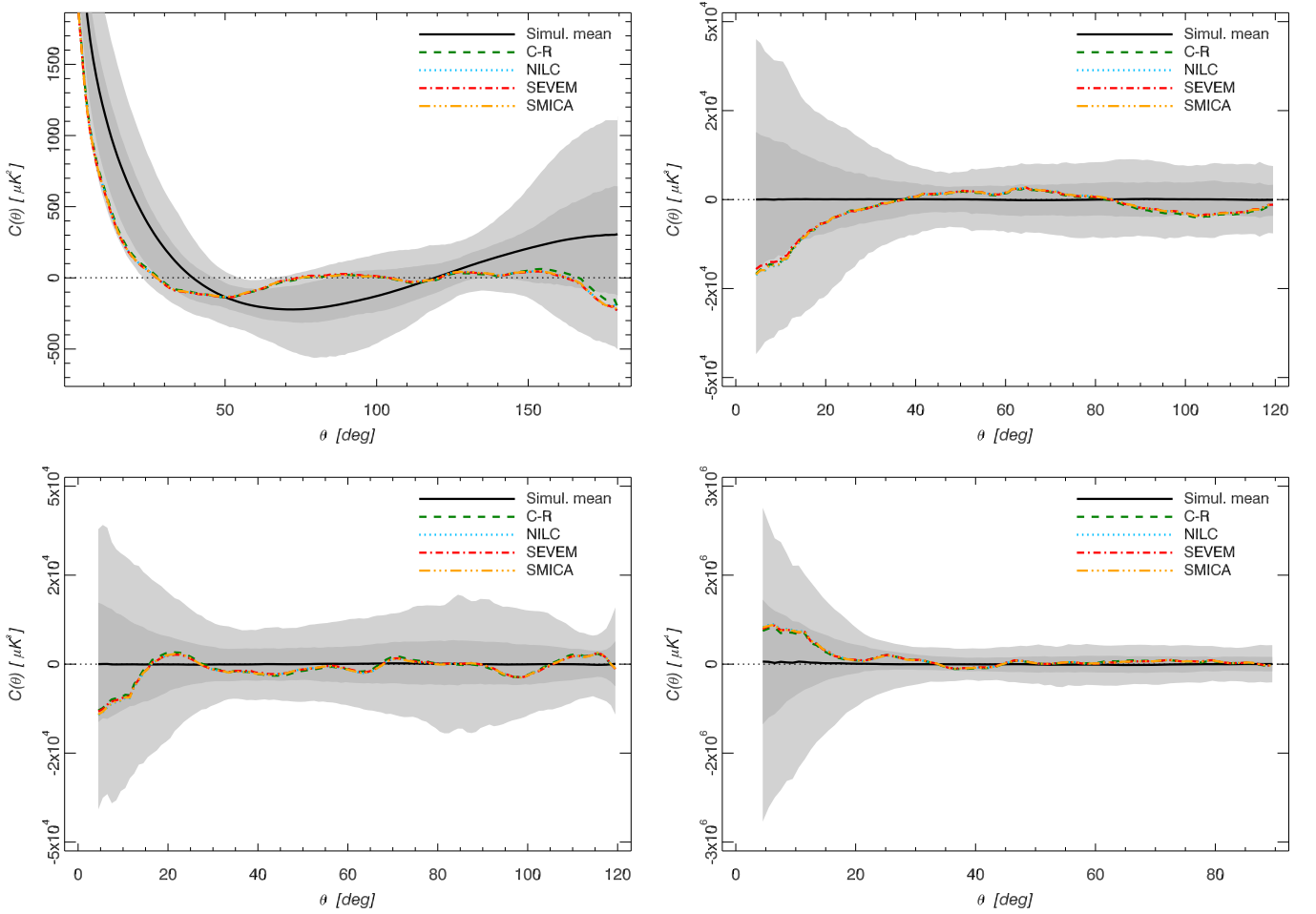

**Fig. 6.** The 2-point (upper left), pseudo-collapsed (upper right), equilateral (lower left) 3-point and reduced rhombic 4-point (lower right) functions for the $N_{side} = 64$ CMB estimates. The black solid line indicates the mean for 1000 Monte Carlo simulations and the shaded dark and light grey regions indicate the 68% and 95% confidence regions, respectively. See Sect. 4.3 for the definition of the separation angle $\theta$.

### 4.4. Minkowski functionals

Minkowski functionals (hereafter MFs) describe the morphology of fields in any dimension and have long been used as estimators of non-Gaussianity and anisotropy in the CMB (see e.g., Mecke et al. 1994; Schmalzing & Buchert 1997; Schmalzing & Gorski 1998; Komatsu et al. 2003; Eriksen et al. 2004c; Curto et al. 2007; De Troia et al. 2007; Spergel et al. 2007; Curto et al. 2008; Hikage et al. 2008; Komatsu et al. 2009). They are additive for disjoint regions of the sky and invariant under rotations and translations. Traditionally in the literature, the contours are defined by a threshold $\nu$, usually given in units of the sky standard deviation ($\sigma_0$). We compute MFs for the regions colder and hotter than a given threshold $\nu$. Thus, the three MFs, the area $V_0(\nu) = A(\nu)$, the perimeter $V_1(\nu) = C(\nu)$, and the genus $V_2(\nu) = G(\nu)$, are defined respectively as:

$$V_0(\nu) = A(\nu) = \frac{N_\nu}{N_{pix}}; \tag{13}$$

$$V_1(\nu) = C(\nu) = \frac{1}{4A_{tot}} \sum_i S_i; \tag{14}$$

$$V_2(\nu) = G(\nu) = \frac{1}{2\pi A_{tot}} \left( N_{hot} - N_{cold} \right). \tag{15}$$

Here $N_\nu$ is the number of pixels where $\Delta T/\sigma_0 > \nu$, $N_{pix}$ is the total number of available pixels, $A_{tot}$ is the total area of the available sky, $N_{hot}$ is the number of compact hot spots, $N_{cold}$ is the number of compact cold spots, and $S_i$ is the contour length of each hot spot. We construct a fourth functional $V_3(\nu) = N_{cluster}(\nu)$ which corresponds to $N_{cold}$ for negative $\nu$ and $N_{hot}$ for positive $\nu$ (Ducout et al. 2012). Analytical expressions for a Gaussian random field can be derived in terms of $\nu$ (see e.g. Vanmarcke 1983; Matsubara 2010) and give

$$V_k(\nu) = A_k v_k(\nu), \tag{16}$$

with

$$v_k(\nu) = \exp(-\nu^2/2) H_{k-1}(\nu), \quad k \leq 2 \tag{17}$$

$$v_3(\nu) = \frac{e^{-\nu^2}}{erfc\left(\nu/\sqrt{2}\right)}, \tag{18}$$

and

$$H_n(\nu) = e^{\nu^2/2} \left(-\frac{d}{d\nu}\right)^n e^{-\nu^2/2}. \tag{19}$$

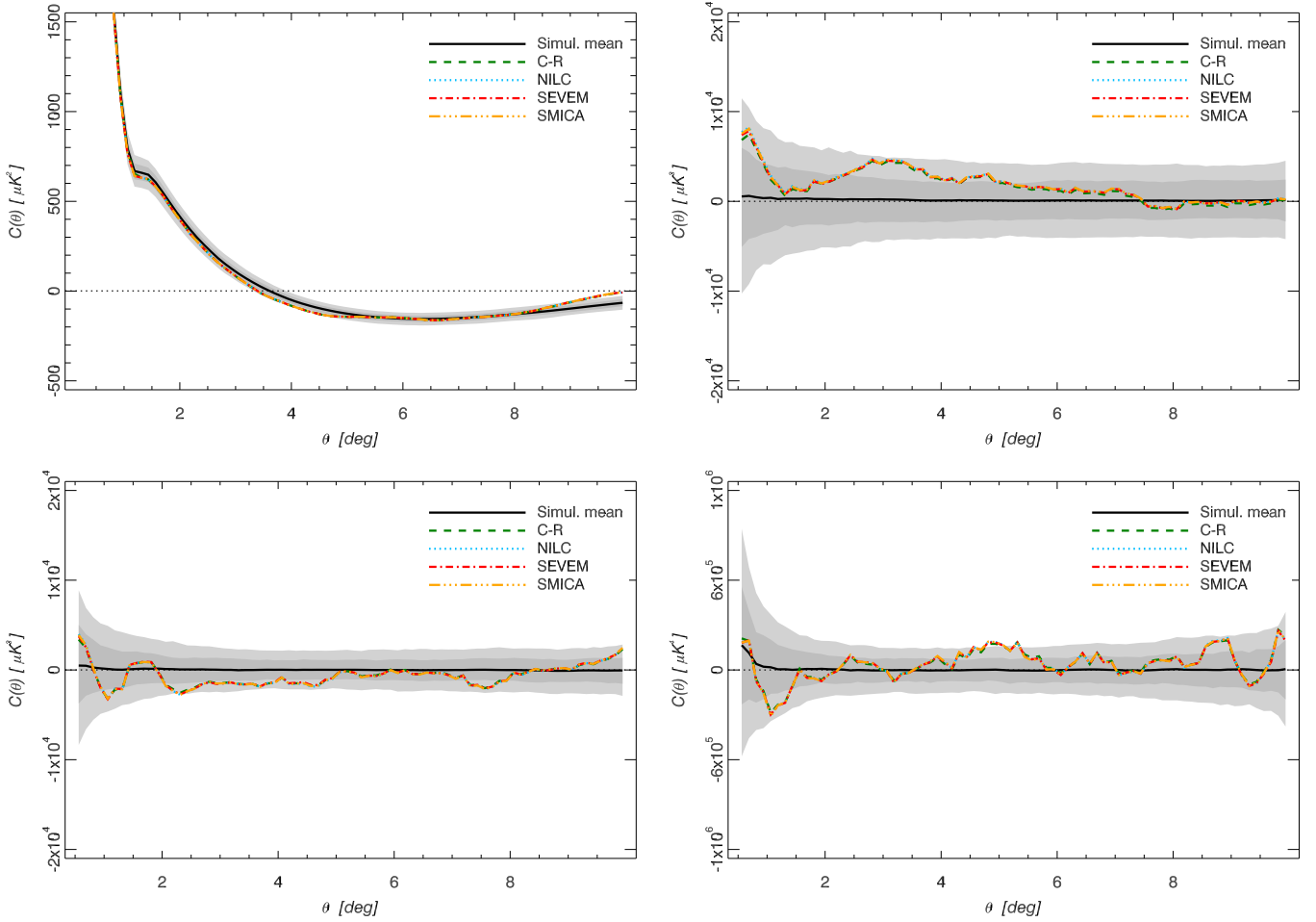

**Fig. 7.** The 2-point (upper left), pseudo-collapsed (upper right), equilateral (lower left) 3-point and reduced rhombic 4-point (lower right) functions averaged over disc set A for the $N_{side} = 512$ CMB estimates. Estimates of the multipoles for $\ell \leq 18$ are removed from the sky maps. The black solid line indicates the mean for 1000 Monte Carlo simulations and the shaded dark and light grey regions indicate the 68% and 95% confidence regions, respectively. See Sect. 4.3 for the definition of the separation angle $\theta$.

The amplitude $A_k$ only depends on the shape of the power spectrum $C_\ell$:

$$A_k = \frac{1}{(2\pi)^{(k+1)/2}} \frac{\omega_2}{\omega_{2-k}\omega_k} \left(\frac{\sigma_1}{\sqrt{2}\sigma_0}\right)^k, \quad k \leq 2 \tag{20}$$

$$A_3 = \frac{2}{\pi} \left(\frac{\sigma_1}{\sqrt{2}\sigma_0}\right)^2 \tag{21}$$

where $\omega_k \equiv \pi^{k/2}/\Gamma(k/2 + 1)$, which gives $\omega_0 = 1$, $\omega_1 = 2$, and $\omega_2 = \pi$ with $\sigma_0$ and $\sigma_1$ beign respectively, the rms of the field and its first derivatives. These analytical expressions represent useful descriptions of the MFs which, for the case of a Gaussian random field, can be factorized as a function of the threshold and another of the shape of the $C_\ell$. We will use both the unnormalized ($V_k$) and normalized ($v_k$) MFs in the Gaussianity tests performed in this section. The unnormalized functionals are computed with a code that was used for the analysis of Archeops data (Curto et al. 2007) and has been thoroughly validated with *Planck* simulations, while for the normalized ones a code adapted to the high resolution *Planck* data and described in Ducout et al. (2012) is used.

By combining the MFs curves into a vector $y$ of size $n = n_{thresholds} \times n_{functionals}$, a null hypothesis test can be performed using a $\chi^2$ statistic given by

$$\chi^2(y) = (y - \langle y_G \rangle)^T C^{-1} (y - \langle y_G \rangle) \tag{22}$$

where $y$ represents the MFs of the data, $y_G$ those of the simulations and $C$ is the covariance matrix. In order to assure convergence, in the case of the four normalized MFs $C$ is estimated from $10^4$ Gaussian simulations, drawn from the *Planck* fiducial power spectrum, having the same instrumental properties of effective beam and noise as the data, the same applied mask and which have been processed in the same way to reach the corresponding resolution. For the three unnormalized MFs, $C$ was estimated from only $10^3$ FFP6 simulations that proved to be sufficient for convergence. We compare the $\chi^2_{Planck}$ value obtained from the data to the $\chi^2$ obtained from those simulations, and report the probability of having a value of $\chi^2$ larger than the measured one, $P\left(\chi^2 > \chi^2_{Planck}\right)$. We explore different resolutions represented by the parameter $N_{side}$, different methods of component separation (**Commander-Ruler**, **NILC**, **SEVEM**, and **SMICA**) and different sky coverages.

First, the three unnormalized MFs ($V_k$ as a function of $\nu$, $k = 0, 1, 2$) are used to construct a test of the null hypothesis.

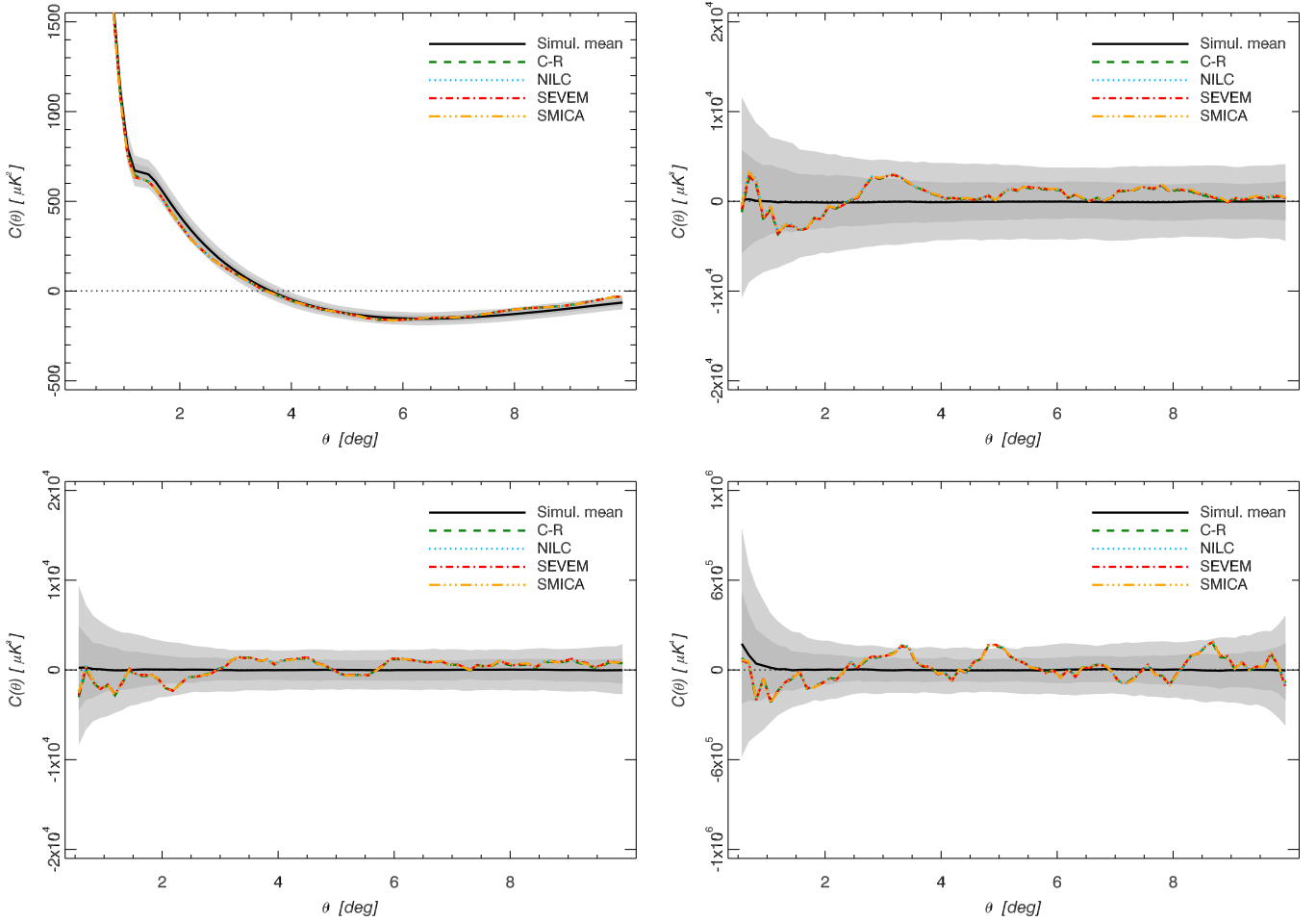

**Fig. 8.** As Fig. 7 for disc set B.

The test assesses not only the primordial Gaussian hypothesis, but also whether the data is correctly represented by the simulations in terms of power spectrum, systematics and the lensing effect. A set of 17 thresholds in units of $T/\sigma_0$ between $-4$ and $+4$ in steps of 0.5 are considered. The comparison between the MFs of the data provided by the four component separation methods and those corresponding to each of the four sets of $10^3$ FFP6 simulations representing each method, for the standard U73 mask, are shown in Fig. 10. From that figure, a deviation at a level of about $2\sigma$ can be seen for the contour and genus curves at a resolution $N_{\text{side}} = 512$. The situation is very similar for the analyses performed at other resolutions, $N_{\text{side}} = 1024$, 256 and 128. Although the deviation is not particularly compelling, because of the correlations among neighbouring thresholds, it is worth mentioning that a possible explanation is the background of unresolved sources that has been detected in *Planck* data with the bispectrum estimators (see Planck Collaboration XXIV (2014)). In order to understand the effect of unresolved sources on the MFs, we added the point source residuals derived from the FFP6 simulations as processed by the SEVEM algorithm to 100 realizations which were then analysed. We conclude that the background of unresolved sources may be responsible for at least part of the excess signal that is detected. The corresponding probabilities $P\left(\chi^2 > \chi^2_{\text{Planck}}\right)$ derived from the MF values for each of the four component separation methods and resolutions are given in Table 13. The full resolution maps have been degraded to the lower resolution ones following the procedure described in Sect. 2. All the cases considered are compatible with the null hypothesis.

**Table 13.** Non-directional Gaussianity tests using unnormalized MFs: $P\left(\chi^2 > \chi^2_{\text{Planck}}\right)$ as a function of sky resolution for different component separation methods.

| | Probability | | | |
|---|---|---|---|---|
| Method | $N_{\text{side}} = 1024$ | $N_{\text{side}} = 512$ | $N_{\text{side}} = 256$ | $N_{\text{side}} = 128$ |
| C-R . . . . . . . . | 0.812 | 0.299 | 0.482 | 0.357 |
| NILC . . . . . . . | 0.993 | 0.567 | 0.354 | 0.234 |
| SEVEM . . . . . . | 0.925 | 0.911 | 0.738 | 0.094 |
| SMICA . . . . . . | 0.874 | 0.675 | 0.426 | 0.213 |

In the second case, the four normalized MFs $v_k = V_k/A_k$ ($k = 0, 1, 2, 3$) are used for the null hypothesis test. A set of 26 thresholds equally spaced between $-3.5$ and $+3.5$, in units of $T/\sigma_0$, are considered. The normalization factor $A_k$ is estimated directly from the maps, having previously computed the moments $\sigma_0$ and $\sigma_1$. This normalization minimizes the dependence of the MFs on the power spectrum, thereby decreasing the cosmic variance and improving their sensitivity to deviations from Gaussianity. The resolutions considered in this case are $N_{side}$ = 2048, 1024, 512, 256, and 128. For the highest resolution $N_{side}$ = 2048, the map is smoothed with a Gaussian smoothing kernel with a width $\theta_{FWHM}$ = 5′, in order to decrease the noise level. We use the standard U73 mask, inpainting the smallest point sources. The maps at lower resolution are constructed by the standard simple degrading process applied to the original map at $N_{side}$ = 2048, and the corresponding masks are degraded following a conservative procedure such that any degraded pixel with a value < 0.8 is set to zero (as explained in Sect. 2). The results of the analysis performed on the different resolutions are presented in Table 14. The results of the analysis performed on the different component separation methods at the highest resolution ($N_{side}$ = 2048) are presented in Table 15. The difference of the normalized MFs with respect to the expected values of the null hypothesis as a function of the threshold $\nu$ are shown in Fig. 11. A slight deviation in $N_{cluster}(\nu)$ is noticeable at thresholds $\nu \approx 0$, however it is not very compelling, since the values at neighboring thresholds are very correlated and this correlation is taken into account in the $\chi^2$ statistics. Finally, we analyse the dependence of the normalized MFs on the sky coverage. We use the standard U73 mask and then decrease the sky coverage by using CL65, CL48 and CL25 masks in combina-different sky coverages are presented in Fig. 12, for the `SMICA` method. Results of the data as a function of sky coverage are compiled in Table 16. All the cases considered are compatible with the null hypothesis.

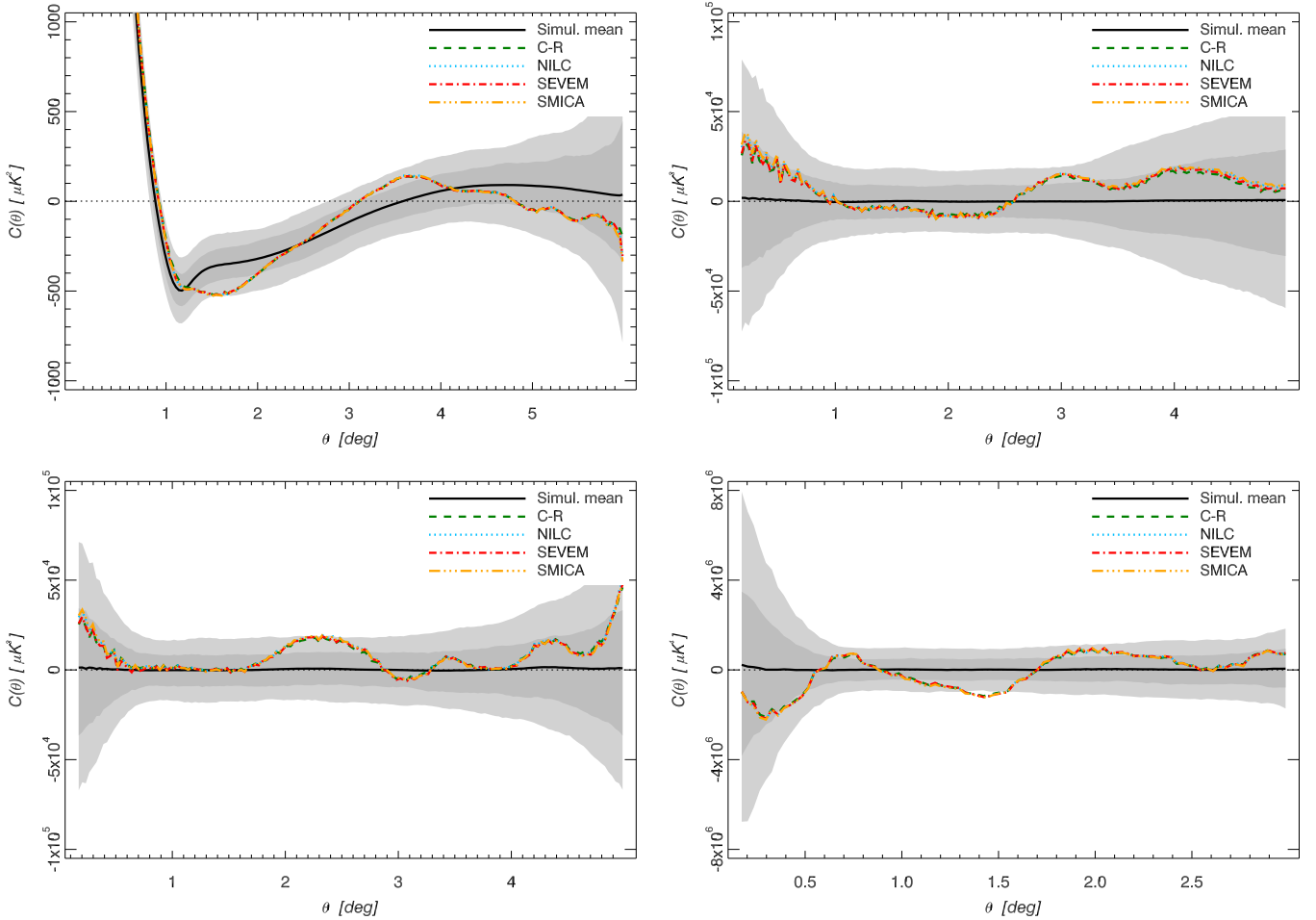

**Fig. 9.** The 2-point (upper left), pseudo-collapsed (upper right), equilateral (lower left) 3-point and reduced rhombic 4-point (lower right) functions averaged over the disc set for the $N_{side}$ = 2048 CMB estimates. Estimates of the multipoles for $\ell \leq 60$ are removed from the sky maps. The black solid line indicates the mean for 1000 Monte Carlo simulations and the shaded dark and light grey regions indicate the 68% and 95% confidence regions, respectively. See Sect. 4.3 for the definition of the separation angle $\theta$.

**Table 14.** Non-directional Gaussianity tests using normalized Minkowski Functionals: dependence of $P\left(\chi^2 > \chi^2_{Planck}\right)$ on resolution.

| $N_{side}$ | Probability |
|---|---|
| 2048 . . . . . . . | 0.358 |
| 1024 . . . . . . . | 0.356 |
| 512 . . . . . . . . | 0.245 |
| 256 . . . . . . . . | 0.225 |
| 128 . . . . . . . . | 0.223 |

**Table 15.** Non-directional Gaussianity tests using normalized Minkowski Functionals: dependence of $P\left(\chi^2 > \chi^2_{Planck}\right)$ on component separation method.

| Method | Probability |
|---|---|
| C-R . . . . . . . | 0.288 |
| NILC . . . . . . | 0.303 |
| SEVEM . . . . . | 0.415 |
| SMICA . . . . . | 0.358 |

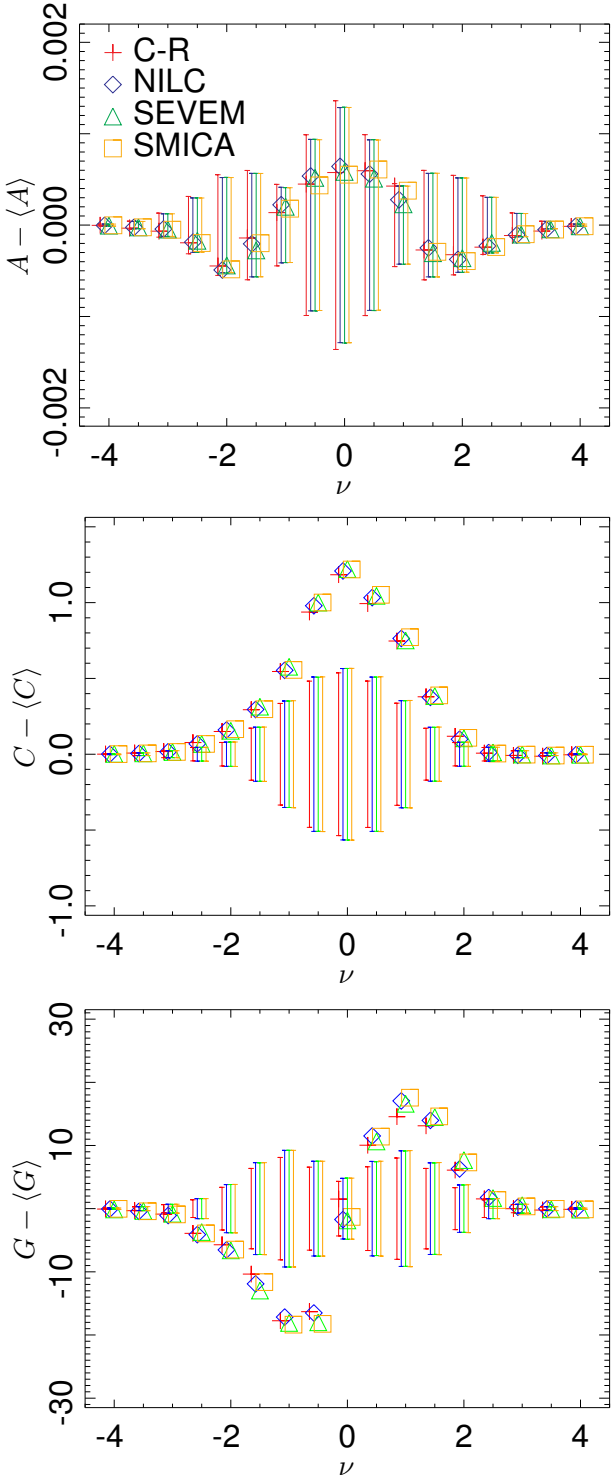

**Fig. 10.** Difference of the data MFs (unnormalized) with respect to the average of the curves obtained with realistic *Planck* simulations for several cleaned maps. From top to bottom: Area, Contour, Genus. The error-bars represent the $\pm 1\sigma$ (68% CL) dispersions around the mean obtained with simulations.

In summary, we find that the data are globally consistent with the primordial Gaussian hypothesis, and no strong deviation is found between the data and realistic simulations for both the unnormalized and normalized MFs. We would like to remark that a certain level of non-Gaussianity is expected from lensing and, in particular, from the ISW-lensing signal, thus it is important to compare the data to realistic *lensed* simulations.

**Table 16.** Non-directional Gaussianity tests using normalized Minkowski Functionals: dependence of $P\left(\chi^2 > \chi^2_{\mathrm{Planck}}\right)$ on sky coverage.

| $f_{\mathrm{sky}}$ | Probability |
|---|---|
| 0.73 . . . . . . . . | 0.358 |
| 0.62 . . . . . . . . | 0.042 |
| 0.46 . . . . . . . . | 0.670 |
| 0.23 . . . . . . . . | 0.780 |

### 4.5. Wavelet statistics

A broad range of wavelets have been used in the analysis of CMB data, but in this paper we specifically consider the Spherical Mexican Hat wavelet (SMHW, Martínez-González et al. 2002).

The SMHW is an example of a continuous, non-orthogonal wavelet. Given a signal on the sky, $T(p)$, where $p$ represents a given position/pixel wich is a function of the co-latitude $\theta$ and longitude $\phi$ (also defined by the unit direction vector $\mathbf{x}$), the SMHW coefficients at a given scale $R$, $\omega_T(R, p)$, are obtained by convolution:

$$\omega_T(R, p) = \sum_{\ell=0}^{\ell_{\max}} \sum_{m=-\ell}^{m=\ell} t_{\ell m} W_{\ell}^{\mathrm{SMHW}}(R) Y_{\ell m}(p), \tag{23}$$

where $W_{\ell}^{\mathrm{SMHW}}(R)$ is the window function associated with the SMHW, $\ell_{\max}$ is the maximum multipole allowed by the corresponding `HEALPix` pixelization, $Y_{\ell m}(p)$ is the spherical harmonic basis, and $t_{\ell m}$ are the spherical harmonic coefficients of the analysed map:

$$t_{\ell m} = \int \mathrm{d}\Omega Y_{\ell m}^*(p) T(p), \tag{24}$$

where $\mathrm{d}\Omega = \mathrm{d}\theta \sin\theta \mathrm{d}\phi$ and $^*$ denotes complex conjugation.

Several statistics can be computed from the wavelet coefficients map, in particular, the first moments: the dispersion $\sigma_R$, the skewness $S_R$, and the kurtosis $K_R$ (as a function of scale $R$). It is interesting to notice that in the case of Gaussian temperature fluctuations the linear transformation involved in the determination of the wavelet coefficients (eqs. 23 and 24) guarantees that Gaussianity is preserved.

The study of the moments of the distribution of the CMB temperature fluctuations, as a function of the scale, is a standard approach to test the null hypothesis. We have performed a full resolution multi-scale analysis of the four CMB clean maps and computed the quantities $\sigma_R$, $S_R$, and $K_R$ from the SMHW wavelet coefficients at 18 scales, $R = \{2, 4, 7, 14, 25, 50, 75, 100,$ tion with a special point source mask that is based on the U73 mask. The fraction of sky left unmasked in the combined masks is 62%, 46% and 23%, respectively. The point source mask is inpainted previously to the analysis. The curves obtained for the

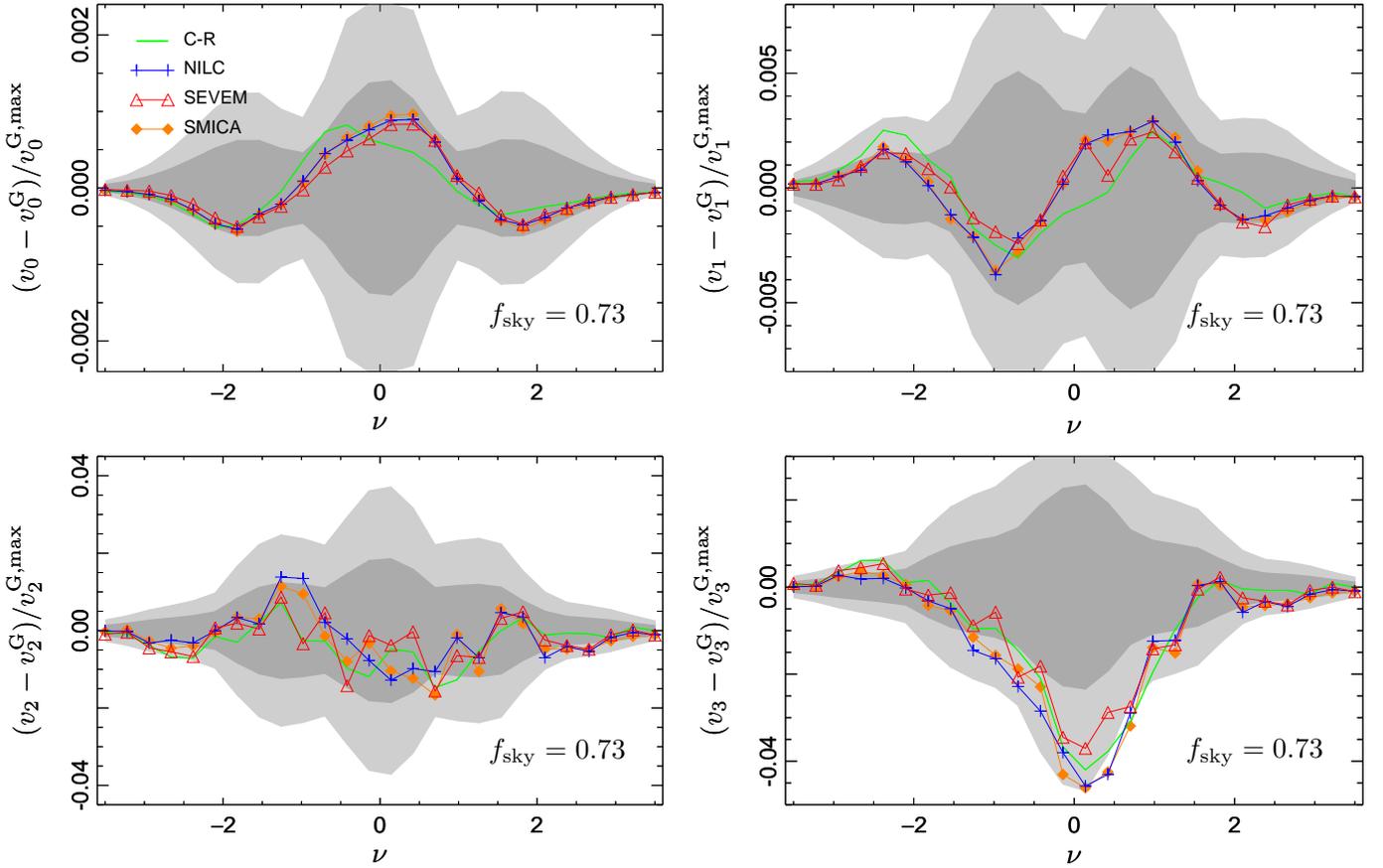

**Fig. 11.** Difference of the normalized MFs obtained from the data with respect to the expected values of the null hypothesis, for the different component separation methods. From left to right and top to bottom: Area, Contour, Genus and $N_{\mathrm{cluster}}$. The grey bands represent the $\pm 1$ and $\pm 2\sigma$ dispersions around zero, based on realistic *Planck* simulations including lensing, for the C-R method.

150, 200, 250, 300, 400, 500, 600, 750, 900, 1050}, in arcminutes. These are compared to the standard *Planck* simulations.

As explained in Vielva et al. (2004), when computing the SMHW coefficients of a masked data set, artefacts are introduced close to the mask that degrade the performance of any null hypothesis tests. We therefore define a set of *exclusion masks* such that, at each scale, an extra region of the sky is excluded when performing any statistical test. The exclusion mask for a given scale $R$ is defined as follows: we build an auxiliary mask by removing from the U73 mask all the features associated with compact objects, and degrade this auxiliary mask to $N_{\mathrm{side}} = 1024$ (imposing a restrictive cut); a first temporary mask is obtained by extending the borders of this auxiliary mask by a distance of twice $R$; a second temporary mask is obtained, first, by convolving the auxiliary mask with the SMHW at that particular scale $R$ and, second, by imposing that any pixel of that second temporary mask with an absolute value lower than 0.1 is masked, whereas the remaining ones are set to 1; the two temporary masks are multiplied to yield a single mask that is upgraded to $N_{\mathrm{side}} = 2048$; and finally, the final exclusion mask is obtained by multiplying this mask by the parent U73 mask.

The comparison of the four CMB maps with the corresponding simulations is summarized in Fig. 13. The three panels show (from left to right) the statistical significance of the standard deviation, the skewness and the kurtosis (as a function of the SMHW scale). The points represent the upper tail probability associated to a given statistic, i.e., the fraction of the simulations that present a value of a given statistic equal to or greater than the one obtained for the data. In fact, we define a modified upper tail probability: if an upper tail probability is larger than 0.5, then a new quantity is defined as 1 minus that upper tail probability. Hence, this modified definition of the upper tail probability is constrained between $10^{-3}$ (the minimum value that can be imposed with 1 000 simulations) and 0.5. Overall, the agreement between the four CMB maps is quite good, showing that all of them provide a consistent estimation of the true CMB. However, several aspects need to be discussed. Let us clarify that the differences among the CMB methods for small modified upper tail probabilities are expected to be larger than for large modified upper tail probabilities. This is because a small modified upper tail probability is determined by a small number of simulations and, therefore, has a relatively large error bar. In other words, the tails of the distributions of the different statistics are quite sparsely sampled.

We will distinguish between the small ($R \lesssim 10'$), intermediate ($10' \lesssim R \lesssim 500'$), and large ($R \gtrsim 500'$) scale regimes. Let us focus on the three statistics independently. We will highlight the most important features and, afterwards, we will try to find an explanation for them.

- On the smallest scales, the four CMB maps show a dispersion in SMHW coefficients significantly larger than seen in the simulations. However on larger scales, the dispersion is systematically below the median of the simulations and, on scales of $R \approx 5°$, the modified upper tail probability is approximately 0.015.

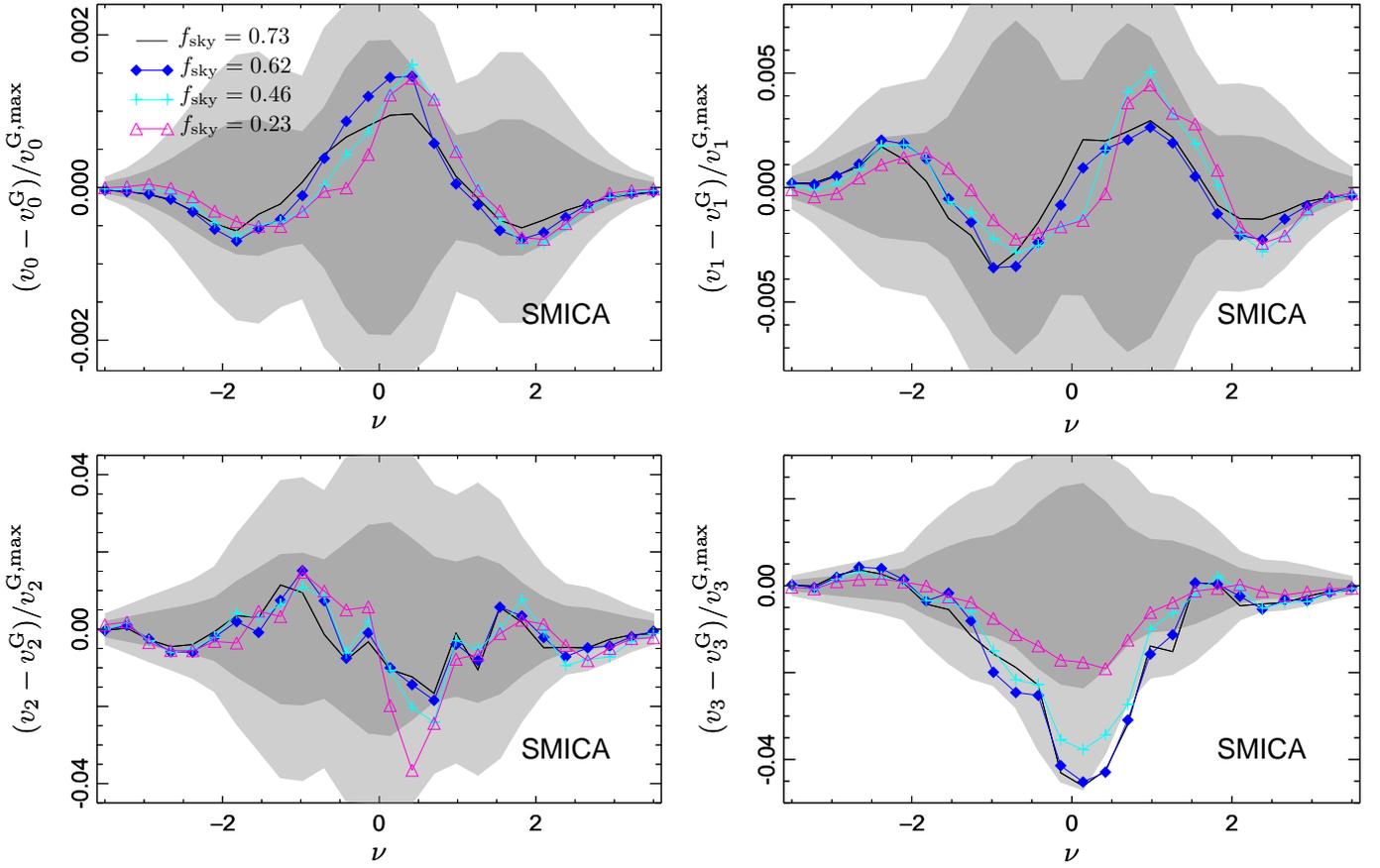

**Fig. 12.** Difference of the normalized MFs obtained from the data with respect to the expected value of the null hypothesis for several sky coverages. The `SMICA` map is considered. From left to right and top to bottom: Area, Contour, Genus and $N_{cluster}$. The grey bands represent the 1 and $2\sigma$ dispersions around zero, based on realistic *Planck* simulations including lensing, for $f_{sky} = 0.23$.

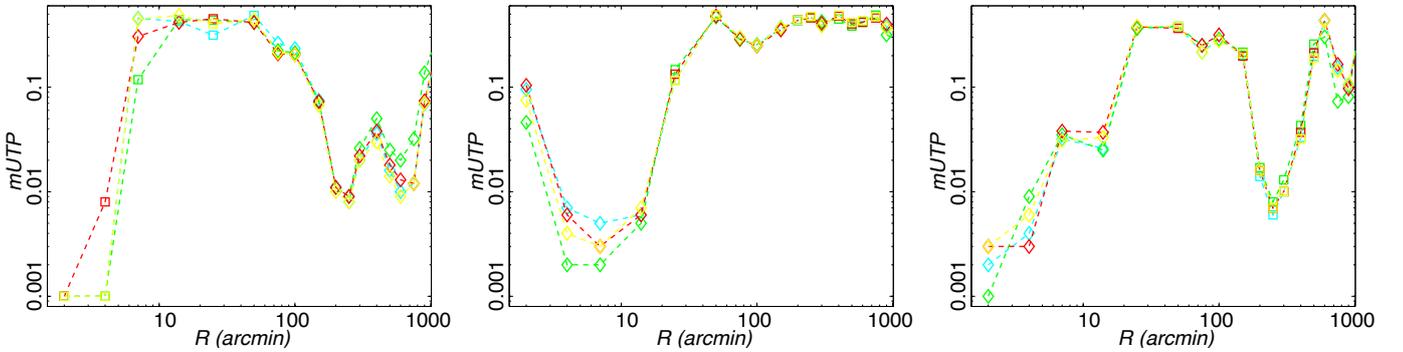

**Fig. 13.** Standard deviation (left), skewness (centre) and kurtosis (right) of the SMHW coefficients as a function of the wavelet scale $R$. Results are given for the four *Planck* CMB maps (green, `Commander-Ruler`; light-blue, NILC; red, `SEVEM`; and yellow, `SMICA`). Modified upper tail probabilities (mUTP, see text for details) are obtained by comparing with 1000 simulations processed through the component separation pipelines. Squares represent modified upper tail probabilities that correspond to an actual upper tail probability above 0.5; diamonds represent upper tail probabilities below 0.5.

– Regarding the skewness, all four maps yield a value that is significantly lower (with a modified upper tail probability of around 0.004) than expected from the simulations in the small scale regime (except for the smallest one, where the deviation is around 0.07). The rest of the scales are fairly compatible with the null hypothesis.
– The kurtosis is also smaller than expected in the small scale regime. Overall, the modified upper tail probability is about

0.03. At scales of around 300′, an anomalously large value (modified upper tail probability of approximately 0.01) is found.

These results are compatible with the values reported for *WMAP* data (Vielva et al. 2004; Cruz et al. 2005), over the scales common to both experiments (i.e., $R > 10'$). In particular, the large value of the kurtosis has been associated with the Cold Spot (Vielva et al. 2004). We will return to this topic specifically in Sect. 5.9. The low variance of the wavelet coefficients was previously seen in Vielva et al. (2004) and Wiaux et al. (2008). In addition, the low dispersion at scales above a few degrees is likely to be related to the low variance anomaly detected in *WMAP* (Monteserín et al. 2008; Cruz et al. 2011), that is also seen in the *Planck* data (see Sect. 4.1).

We have also studied the robustness of the results for different masking scenarios. In particular, we have investigated variations in the results when we adopt, as auxiliary masks to define the exclusion masks, the two CG70 and CG60 masks removing 30% and 40% of the sky, respectively. Note that the auxiliary masks obtained from the U73 mask already cut around 20% of the sky. The corresponding results for the SMICA map are presented in Fig. 14. The conclusions are similar for the other CMB maps. For the dispersion of the wavelet coefficients, we do not notice any significant change in the anomalously high value obtained for the SMICA map at the smallest scales However, some changes are observed at larger scales. In this regime, it seems that the most significant deviation occurs for the CG70 mask (modified upper tail probability of around 0.005), whereas similar (and slightly less significant) modified upper tail probabilities are obtained for both the U73 (modified upper tail probability of approximately 0.015) and the CG60 (modified upper tail probability of about 0.01) masks. A possible explanation for this behaviour would be that a less restrictive mask admits some residual contamination from Galactic foregrounds, thus increasing the dispersion of the wavelet coefficients, and artificially increasing their inconsistency with the null hypothesis. In principle, the larger the Galactic cut, the lower would be the dispersion of the wavelet coefficients (assuming that some residual contamination of the Galactic foregrounds is left) and, therefore, the smaller the upper tail probability. However, as we already said, the modified upper tail probability for the dispersion is higher for the CG60 mask than for the CG70 mask. This apparent contradiction could be resolved by accounting for the larger sampling variance for smaller areas, which would result in a lower significance for the anomaly.

The anomalous kurtosis at scales of $R \approx 300'$ shows an overall stable modified upper tail probability of around $0.01 - 0.03$. In the small scale regime, the differences are better defined: the smaller the mask, the more significant the deviation (characterized by the low value of the kurtosis). In particular, the modified upper tail probability associated with the CG60 mask is 0.001, around 0.009 for CG70, and approximately 0.03 for the U73 mask. A similar pattern is also observed for the skewness on these scales, although the three masks results in more similar upper tail probabilities, between around 0.001 and 0.007 (except for the smallest scales).

It is therefore clear that there is some inconsistency between the CMB data and the corresponding simulations. On intermediate scales, both the low dispersion and the high kurtosis could be related to previously known anomalies: the low variance and the Cold Spot. On the smallest scales, the three statistics report a low upper tail probability independently of the mask coverage — it is important to determine what this inconsistency is due to. Besides the possibility that it is an intrinsic cosmological signal, the non-Gaussianity could be caused either by instrumental systematics or residual foreground contamination.

In the former case, we have considered whether the origin of the signal could be related to properties of the noise that are inadequately modelled by the simulations. In particular, we have studied the statistical properties of the half-ring half-difference maps generated by the four component separation algorithms as proxies for the noise present in the CMB maps. Although in detail there are some discrepancies between these noise estimates and the simulated ones, they are not compatible with the inconsistencies observed between the CMB map and simulations. Therefore, a systematic effect associated with the instrumental noise does not provide a satisfactory explanation for the small-scale deviations.

In the latter case, an obvious candidate is due to the contribution from residual unresolved point sources in the clean CMB maps. Although the brightest point sources are masked, and the component separation process itself can suppress the amplitude of the unresolved background of point sources, some signal will remain. Indeed, in Planck Collaboration XXIV (2014) it has been determined that the bispectrum of this contribution is clearly detected in the four CMB *Planck* maps, at a significance in excess of $4\sigma$. In addition, the dispersion of the wavelet coefficients is higher than expected, which is also compatible with the presence of an additional signal. We therefore consider this as the most likely non-CMB explanation for the observed signal.

## 4.6. Bispectrum

The CMB bispectrum is the three point correlator of the $a_{\ell m}$ coefficients: $B^{m_1 m_2 m_3}_{\ell_1 \ell_2 \ell_3} \equiv \langle a_{\ell_1 m_1} a_{\ell_2 m_2} a_{\ell_3 m_3} \rangle$. In this paper, we focus on the bispectrum reconstruction as a blind test of non-Gaussianity. Therefore, we assume we are seeking a non-trivial bispectrum that has arisen through a physical process which is statistically isotropic, that is, we can employ the angle-averaged bispectrum $B_{\ell_1 \ell_2 \ell_3}$ and the reduced bispectrum $b_{\ell_1 \ell_2 \ell_3}$. Note that $b_{\ell_1 \ell_2 \ell_3}$ is defined on a tetrahedral domain of multipole triples $\{\ell_1 \ell_2 \ell_3\}$ satisfying both a triangle condition and a limit given by the maximum resolution $\ell_{max}$ of the experiment. A much more extensive introduction to the bispectrum can be found in Planck Collaboration XXIV (2014).

Modal, wavelet and binned bispectrum estimators filter the CMB map with separable basis functions

$$Q_{ijk}(\ell_1, \ell_2, \ell_3) = q_i(\ell_1) \, q_j(\ell_2) \, q_k(\ell_3) + \text{perms}, \qquad (25)$$

to find the corresponding modal coefficients. For appropriately orthonormalized basis functions $Q_{ijk}(\ell_1, \ell_2, \ell_3)$, these coefficients can be used to reconstruct the CMB bispectrum through a signal-to-noise weighted expansion (Planck Collaboration XXIV 2014). This reconstruction method has been extensively validated, showing the accurate recovery of CMB bispectra from non-Gaussian simulated maps, and it has been applied to the *WMAP* seven year data to reconstruct the full 3D CMB bispectrum (Fergusson et al. 2010). To quantify whether or not there is a model-independent deviation from Gaussianity, we can consider the total integrated bispectrum. By summing over all multipoles, we use the integrated nonlinearity parameter $\bar{F}_{NL}$ defined in Planck Collaboration XXIV (2014). For ideal Gaussian CMB maps, the quantity $\bar{F}^2_{NL}$ should obey a $\chi^2$-distribution with a mean given by the number of degrees of freedom (the modes) $\mu = n_{max}$ and with a variance $\sigma^2 = 2n_{max}$. Assuming that the three-point correlator is the leading non-Gaussian contribution, then $\bar{F}_{NL}$ provides a blind test for the presence of any integrated CMB bispectrum (once the expected two-point term is subtracted). We note that this is less sensitive than targeted searches for particular bispectrum shapes.

First, we discuss reconstructions from the modal estimator which has passed successfully through the full suite of non-Gaussian bispectrum validation tests (for further details about bispectrum estimators, see Planck Collaboration XXIV 2014). We have applied this to the *Planck* temperature maps for the foreground-separation techniques `NILC SEVEM` and `SMICA`, using two alternative sets of hybrid basis functions in order to cross-check results and identify particular signal (Planck Collaboration XXIV 2014). As with all the other bispectrum analyses based on spherical harmonic coefficients, we used the U73 mask to which we applied inpainting. Together with the foreground separated maps, noise simulations were provided which were used to calibrate the estimator's linear correction term and to determine the variance.

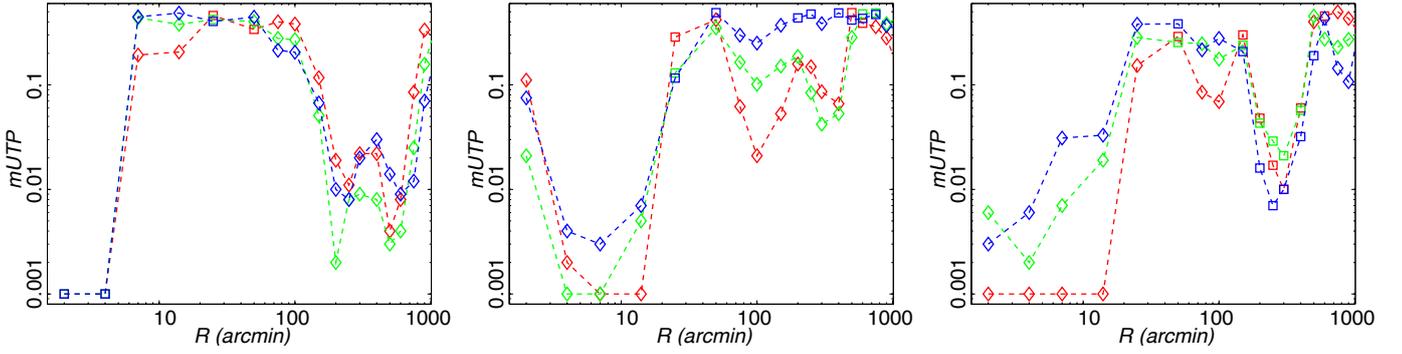

**Fig. 14.** Standard deviation (left), skewness (centre) and kurtosis (right) of the SMHW coefficients as a function of the wavelet scale $R$. Results are given for the `SMICA` CMB map. Several masking scenarios are compared: red, CG60 mask (cutting out 40% of the sky); green, CG70 mask (cutting out 30% of the sky); and blue, U73 mask. The modified upper tail probabilities (*mUTP*) are defined in the text.

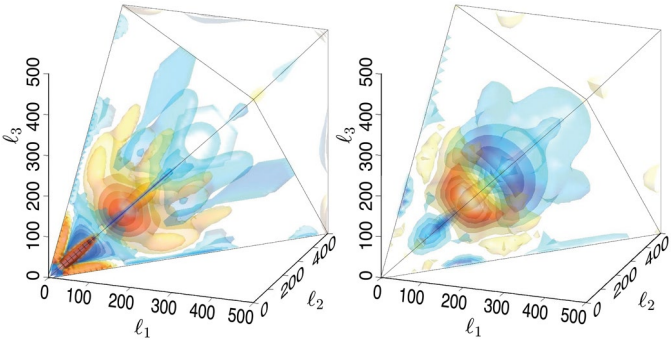

**Fig. 15.** Comparison between the *WMAP* seven-year spectrum signal (left) ((Fergusson et al. 2010)) and the low-$\ell$ signal of *Planck* (right) reconstructed from the `SMICA` foreground-separated map (in both cases using polynomial modes). The same basic patterns are observed in both bispectra, including an apparent central "oscillatory" feature.

The modal coefficients extracted from the *Planck* `NILC`, `SEVEM`, and `SMICA` maps show remarkable consistency between the different maps, with shape cross-correlations better than 96% and the overall amplitudes to within 7% agreement (see corresponding figure in Planck Collaboration XXIV 2014). This demonstrates that the indebent foreground separation techniques do not appear to be introducing spurious non-Gaussianity. The corresponding quantity $\bar{F}_{\mathrm{NL}}^2$ shows consistency with the null hypothesis (Planck Collaboration XXIV 2014). Using the modal expansion, we have reconstructed the full 3D *Planck* bispectrum (Planck Collaboration XXIV 2014) for `SMICA` but also `NILC` and `SEVEM`; the reconstructions are visually indistinguishable.

In Fig. 15, we show a comparison of the $\ell < 500$ *Planck* bispectrum signal and that reconstructed from the *WMAP* seven-year data (Fergusson et al. 2010). Here for consistency we show the *Planck* signal using polynomial modes. The *Planck* signal pattern correlates well with the *WMAP* bispectrum obtained pre-viously, despite the different domains used for the modal analysis of the two different experiments.

Similarly to the modal bispectrum, a wavelet decomposition can be used to reconstruct the bispectrum. Here we use the continuous, non-orthogonal Spherical Mexican Hat Wavelet (SMHW, Martínez-González et al. 2002). Cubic moments $q_{ijk}$ are defined in terms of the SMHW coefficients for three different angular scales $R_i$, $R_j$, $R_k$ (Curto et al. 2009b,a, 2010, 2011a,c,b). See Planck Collaboration XXIV (2014) for more details on the implementation of this wavelet estimator for *Planck* data analysis. Considering the covariance matrix of the $q_{ijk}$ moments, $\mathsf{C} \equiv \langle \boldsymbol{q}\boldsymbol{q}^{\mathsf{T}} \rangle$, and its eigenvector decomposition, $\mathsf{C} = \mathsf{R}\mathsf{D}\mathsf{R}^{\mathsf{T}}$, with $\mathsf{R}$ the eigenvector matrix and $\mathsf{D}$ the eigenvalue matrix, a new set of quantities $\boldsymbol{y} \equiv \mathsf{D}^{-1/2}\mathsf{R}^{\mathsf{T}}\boldsymbol{q}$ is defined. Considering the decorrelation produced by the convolution of the SMHW on the temperature anisotropies and applying the central limit theorem to the averages defined in $q_{ijk}$, then these quantities are expected to have a nearly Gaussian distribution. Therefore, the $\boldsymbol{y}$ quantities are nearly multinormal and satisfy $\langle \boldsymbol{y}\boldsymbol{y}^{\mathsf{T}} \rangle = \mathsf{I}$ and $\langle \boldsymbol{y} \rangle = \boldsymbol{0}$ (Curto et al. 2011a).

We have computed this reconstruction using the *Planck* data and compared with the null hypothesis (Gaussian *Planck* simulations). The considered data map is the resulting map after foreground cleaning based on different cleaning procedures: `Commander-Ruler`, `NILC`, `SEVEM`, and `SMICA`. The mask used is the U73 one (contrary to the modal reconstruction, no inpainting of the point sources is made in this case). In Fig. 16 the $\boldsymbol{y}$ statistics corresponding to the *Planck* data are plotted and compared with the $3\sigma$ error-bars obtained with *Planck* Gaussian simulations. From the list of different $q_{ijk}$ statistics corresponding to the 16 angular scales described in Planck Collaboration XXIV (2014), there are 11, 4, 3, 3 statistics with $|y_i| \geq 3$ (corresponding to `Commander-Ruler`, `NILC`, `SEVEM`, and `SMICA`, respectively). The error-bars are obtained with *Planck* simulations for each type of component separation cleaned map. The error-bars of the $y_i$ statistics for low indices $i$ are associated to large scales where the `q` statistics have a less Gaussian-like shape. The $\boldsymbol{y}$ statistics are combined into a $\chi^2$ test after a principal component analysis with a threshold of $10^{12}$ (Curto et al. 2011a) and compared with the $\chi^2$ statistics obtained from *Planck* Gaussian simulations for each type of component separation method (see Table 17). The $\chi^2$ statistic corresponding to the data is compatible with the values obtained from Gaussian simulations according to the cumulative probability $P(\chi^2 \geq \chi^2_{\mathrm{data}})$, as can be seen in Table 17. Therefore the wavelet bispectrum reconstruction does not de-tect a significant amplitude of bispectrum in the considered data maps. Details on the constraints on the amplitude of different bispectrum shapes are presented in Planck Collaboration XXIV (2014).

**Table 17.** $\chi^2$ statistics based on the wavelet bispectrum reconstruction $y_i$ statistics for the foreground cleaned data map. Considered data map: combined map cleaned with `C-R`, `NILC`, `SEVEM`, and `SMICA`.

| Method | $\chi^2_{data}$ | DOF | $\langle \chi^2 \rangle$ | $\sigma$ | $P(\chi^2 \geq \chi^2_{data})$ |
|---|---|---|---|---|---|
| `C-R`............ | 874 | 690 | 740 | 87 | 0.074 |
| `NILC`............ | 883 | 682 | 731 | 83 | 0.045 |
| `SEVEM`............ | 858 | 682 | 731 | 83 | 0.070 |
| `SMICA`............ | 878 | 682 | 732 | 83 | 0.058 |

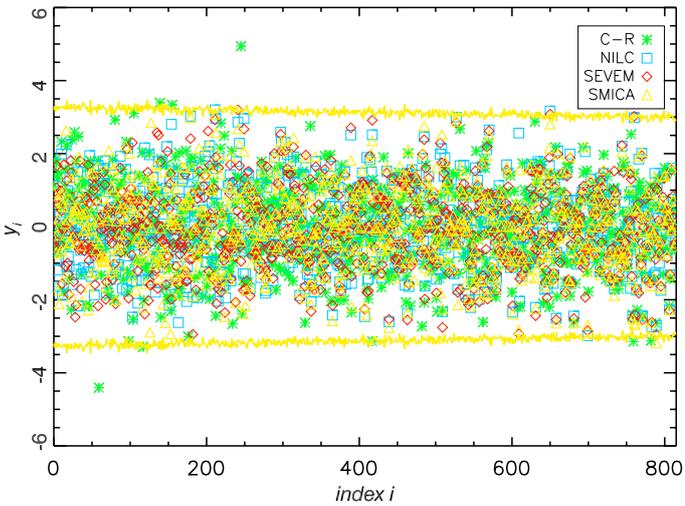

**Fig. 16.** Wavelet bispectrum reconstruction $y_i$ statistics for the foreground cleaned *Planck* data map. Considered data map: combined map cleaned with `C-R`, `NILC`, `SEVEM` and `SMICA`. The solid yellow lines represent the $3\sigma$ error-bars for `SMICA` (similar error-bars are obtained for `C-R`, `NILC`, and `SEVEM` maps).

# 5. Intriguing inconsistencies – *WMAP* anomalies revisited

In the previous section, we established that the *Planck* data show little evidence for non-Gaussianity beyond that expected due to the ISW-lensing effect (which is accounted for directly by simulations) and contributions from residual unresolved point sources. The exceptions are on large-angular scales, where features consistent with various anomalies previously seen in the *WMAP* data have been observed. In this section, we explicitly consider several of the most important anomalies detected in the *WMAP* data, namely the quadrupole-octopole alignment (Sect. 5.1), the low variance (Sect. 5.2), the hemispherical asymmetry (Sect. 5.3), phase correlations (Sect. 5.4), power asymmetry (Sect. 5.5), dipolar modulation (Sect. 5.6), generalized power modulation (Sect. 5.7), parity asymmetry (Sect. 5.8), and the Cold Spot (Sect. 5.9). Each of these anomalies may represent different violations of the fundamental properties of isotropy and/or Gaussianity in the CMB data that are assumed for the estimation of the CMB power spectrum.

There is an ongoing debate about the significance of these anomalies in the literature. A critical issue relates to the role of a posteriori choices — whether interesting features in the data bias the choice of statistical tests, or if arbitrary choices in the subsequent data analysis enhance the significance of the features. Indeed, the *WMAP* team (Bennett et al. 2011) contends that the anomalies are significantly over-interpreted due to such selections, whilst other authors claim highly significant and robust detections. Therefore, care must be taken to address the issue of significance, since our analyses necessarily follow up tests of the previous *WMAP* investigations. However, a careful and fair statistical treatment can enable us to study possible links among the anomalies and to search for a physical interpretation.

## 5.1. Mode alignment

Tegmark et al. (2003) first reported a significant alignment between the orientation of the quadrupole and the octopole in the *WMAP* first year temperature data. We study this quadrupole-octopole alignment in the *Planck* data using the maximization of the angular momentum dispersion as described in de Oliveira-Costa et al. (2004). Specifically, we determine the orientation of the multipoles by finding the axis $\boldsymbol{n}$ around which the angular momentum dispersion

$$\sum_m m^2 |a_{\ell m}(\boldsymbol{n})|^2 \qquad (26)$$

is maximized. Here, $a_{\ell m}(\boldsymbol{n})$ denotes the spherical harmonic coefficients of the CMB map in a rotated coordinate system with its $z$-axis in the $\boldsymbol{n}$-direction. This definition of the multipole-orientation has been devised for planar multipoles and is simply the direction perpendicular to the plane in which most of the power of the multipole lies. It is thus intuitive and easy to use. Note that the value of the statistic in Eq. (26) is the same for $-\boldsymbol{n}$ as for $\boldsymbol{n}$, i.e., the multipole orientation is only defined up to a sign.

An alternative method, based on the multipole vector decomposition of the data (Copi et al. 2004; Schwarz et al. 2004; Bielewicz et al. 2005; Bielewicz & Riazuelo 2009), has also been used to verify the robustness of the results presented here, and excellent consistency is found.

Residual foregrounds (mostly on the Galactic plane) present in the four *Planck* CMB estimates could influence the reconstruction of the low-order multipoles. However, when a mask is applied, the resulting mode-coupling can also affect the reconstruction of the low-$\ell$ multipoles. Since our understanding of the relationship between the size of the mask and the extent to which residual foregrounds and mode-coupling can affect the result is incomplete, we therefore utilize Wiener-filtered maps computed from the data to which the U73 mask has been applied. Specifically, we utilize the same implementation of the Wiener filter as used in Planck Collaboration XXIV (2014), i.e., a "messenger" method, as described by Elsner & Wandelt (2013). A direct inversion method for masked data (Efstathiou 2004; Bielewicz et al. 2004, 2013) is a possible alternative, but the Wiener-filtered maps result in a significantly smaller uncertainty in the reconstructed orientation of the multipoles.

We search for the preferred orientation by explicitly rotating the CMB map such that the $z$-axis pierces the centres of each of the low resolution pixels defined at $N_{side} = 16$, and then subsequently refine the search by testing the $N_{side} = 2048$ pixels that lie within the preferred lower resolution pixel. The angular resolution for the orientation of the multipoles is thus given by the distance between the pixel centres of the $N_{side} = 2048$ map, which is of order $1.94'$. Figure 17 shows the Wiener-filtered `SMICA` CMB sky, with the corresponding reconstruction of the quadrupole and octopole moments. The reconstructed orientations are quite robust with respect to the component separation method used for reconstructing the CMB. The significance of the alignment between the quadrupole and the octopole is assessed from the scalar product of their orientations, compared to values derived from the standard set of 1000 MC simulations. The orientation, the angular distance, the scalar-product between quadrupole and octopole, and the probability of at least as strong an alignment to occur in an isotropic universe, are summarized in Table 18 for each CMB map.

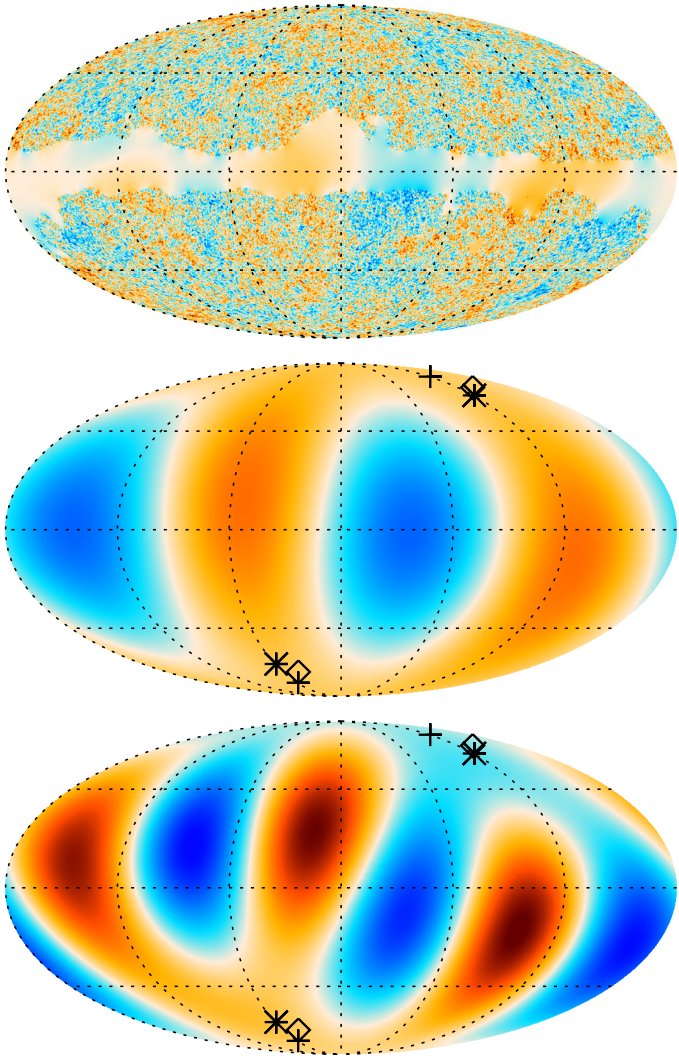

**Fig. 17.** *Upper*: Wiener-filtered `SMICA` CMB sky (temperature range ± 400 μK). *Middle*: derived quadrupole (temperature range ± 35 μK). *Lower*: derived octopole (temperature range ± 35 μK). The plus and star symbols indicate the axes of the quadrupole and octopole, respectively, around which the angular momentum dispersion is maximized. The diamond symbols correspond to the quadrupole axes after correction for the kinematic quadrupole.

We find that, depending on the component separation method, the quadrupole and octopole orientations are misaligned by an amount between 9° and 13°. Such a range of results is reasonable given the observed differences, typically of order 5 μK in amplitude at high Galactic latitudes, between the CMB estimates, as shown in figures 5 and 6 of Planck Collaboration XII (2014). This then implies that it may not be possible to determine the alignment with a precision of better than a few degrees, due to uncertainties related to component separation.

Note that the misalignment determined from the *Planck* data is larger than the 3° reported[3] recently by Bennett et al. (2012) for the 9-year *WMAP* ILC map. However, when considering the plausible range of errors introduced by the covariance between the CMB and foregrounds and their impact on the derived ILC map, Bennett et al. (2012) cite a median misalignment angle of 6°.

Finally, as noted by Tegmark et al. (2003), the observed quadrupole includes a contribution due to our motion relative to the CMB rest frame. This kinematic quadrupole (hereafter KQ) has a well-defined morphology and frequency dependence, as described in Kamionkowski & Knox (2003). The latter is of particular importance for the HFI channels. Schwarz et al. (2004) have demonstrated that correcting for the KQ contribution can be relevant for the evaluation of the significance of the alignment between the cosmological quadrupole and octopole. We therefore apply map-based corrections to the `NILC`, `SEVEM`, and `SMICA` maps, taking into account both the weighted contributions of each *Planck* frequency to the final CMB maps and the corresponding KQ scaling factors. Such corrections are possible for these maps, since, unlike for `Commander-Ruler`, the frequency weights used to generate them do not vary locally on the scales of interest. As seen in Table 18, after KQ-correction the three estimates of the quadrupole-octopole alignment are more consistent, with the misalignment angle decreasing to approximately 8°, and the significance increasing to 99%.

### 5.2. Variance, skewness, and kurtosis anomalies

A low value for the variance on the CMB sky was previously observed in the *WMAP* data by Monteserín et al. (2008) and Cruz et al. (2011), and confirmed for *Planck* in Sect. 4.1. Furthermore, the effect has also been seen in the wavelet analysis of Sect. 4.5, where the variance of the SMHW coefficients is low at scales between 400 and 600 arcmin (Fig. 13). In addition, anomalous behaviour was also observed for the skewness and kurtosis in low resolution maps at $N_{side} = 16$. Here, we reassess these results and determine their robustness to masking and data selection. The former will allow us to determine whether a particular region is causing the anomalous behaviour, whilst the latter can establish whether foreground residuals could be responsible.

Table 19 and Fig. 18 present the results for the variance, skewness and kurtosis determined from the four CMB maps with the U73, CL58 and CL37 masks applied. Results are also computed for data within the ecliptic hemispheres surviving the U73 mask. The variance is low in all cases, with only small differences in significance observed for the different maps. Interestingly, the low variance seems to be localized in the northern ecliptic hemisphere. Conversely, anomalous values for the skewness and kurtosis only are apparent for the southern ecliptic hemisphere.

---

[3] Although Bennett et al. (2012) used a variant on the estimator adopted in this paper, we also find a 3° misalignment of the quadrupole and octopole determined from the ILC map.

**Table 18.** Orientations of the quadrupole and octopole components extracted from the four *Planck* CMB maps, as determined by a maximization of the angular momentum dispersion (see Eqn. 26). The absolute value of the scalar-product between the orientation vectors of the quadrupole and the octopole are provided in the fifth column. For an isotropic universe, this is uniformly distributed on the interval [0, 1]. The last column gives the probability for such an alignment (or stronger) to occur, as determined from the FFP6 simulations. Results computed after correction for the kinematic quadrupole (KQ) are denoted by "KQ corrected". In practice, we multiply the usual expression for the angular dependence of the KQ (see Kamionkowski & Knox 2003) in the Rayleigh-Jeans regime by factors that account for both its frequency dependence and the relative contributions of the different *Planck* frequencies to the component separated maps. For `NILC`, `SEVEM`, and `SMICA` these factors are 1.7, 1.5, and 1.7, respectively.

| Method | (l,b) quadrupole [°] | (l,b) octopole [°] | Ang. distance [°] | Scalar product | Probability |
|---|---|---|---|---|---|
| `C-R`. . . . . . . . . . . . . . . . . | (228.2,60.3) | (246.1,66.0) | 9.80 | 0.985 | 0.019 |
| `NILC`. . . . . . . . . . . . . . . . | (241.3,77.3) | (241.7,64.2) | 13.1 | 0.974 | 0.033 |
| `SEVEM`. . . . . . . . . . . . . . | (242.4,73.8) | (245.6,64.8) | 9.08 | 0.988 | 0.016 |
| `SMICA`. . . . . . . . . . . . . . . | (238.5,76.6) | (239.0,64.3) | 12.3 | 0.977 | 0.032 |
| `NILC, KQ corrected` ... | (225.6,69.7) | (241.7,64.2) | 8.35 | 0.989 | 0.011 |
| `SEVEM, KQ corrected`... | (228.3,68.3) | (245.6,64.8) | 7.69 | 0.991 | 0.009 |
| `SMICA, KQ corrected`... | (224.2,69.2) | (239.0,64.3) | 7.63 | 0.991 | 0.009 |

Since these results might be indicative of the presence of Galactic foreground residuals near the Galactic plane, we analyse the frequency dependence of the statistics as summarized in Table 20 and Fig. 19. The variance shows little frequency dependence for the considered masks and regions, whereas the skewness and kurtosis show a moderate frequency dependence when the U73 mask is applied, as also seen for the *N*-pdf analysis in Sect. 4.2. Cruz et al. (2011) found that a small region of the sky localized to both the ecliptic and Galactic south and near to the Galactic plane (their so-called "gp10" region) exhibited particularly high variance. Thus, since the skewness is negative, we consider a prominent cold spot at $b = -8°$, $l = 32°$, partially masked by the Galactic plane. However, when masking the seven coldest pixels of the spot, the significance of the skewness and kurtosis only drops slightly, with lower tail probabilities of approximately 0.03 and 0.93 respectively. If the whole "gp10" region ($f_{sky} = 7\%$) is masked, the skewness and kurtosis drop drastically and have lower tail probabilities of approximately 0.30 and 0.50, respectively, whereas the variance is highly significant since none of the 1000 simulations has a value below that of the data. In order to check the possible leakage of Galactic contamination due to the Gaussian smoothing applied to the low resolution data, we repeated our calculations for the Wiener filtered maps used in Sect. 5.1, but found little variation to the existing results. Therefore, it is unlikely that any leakage impacts the estimators.

The incompatibility of the observed variance with simulations based on a cosmological model that has been determined from the same data set might appear puzzling at first, but can be understood as follows. The map-based variance is dominated by contributions from large angular scales on the sky, whilst the cosmological parameter fits are relatively insensitive to these low-order ℓ-modes, and are instead largely dominated by scales corresponding to $\ell > 50$. Thus, the best-fit spectrum in the context of a 6-parameter ΛCDM model can have a mismatch with the data on these scales, so that the corresponding simulations will not adequately capture the dearth of power at low-ℓ. The results presented here do indeed imply that the large-angular scale power is low relative to the fiducial sky model. In fact, when subtracting the quadrupole and octopole from both the data and simulations outside the U73 mask, the results are more consistent. In this case, the lower tail probabilities for the variance, skewness and kurtosis are 0.192, 0.637 and 0.792, respectively. This result was already found in Cruz et al. (2011). It is then plausible that the low multipole alignment could have the same cause as the anomalies considered here. Interestingly, Sarkar et al. (2011) have demonstrated that the low-variance and mode alignment is very unlikely to have a common origin in Gaussian random and statistically isotropic CMB anisotropies. However, when subtracting the quadrupole and octopole outside the CL58 mask, the lower tail probability for the low variance is 0.036, which remains rather low. The connection with the very low power in the ecliptic northern hemisphere also remains to be explored.

### 5.3. Hemispherical Asymmetry

In Eriksen et al. (2004a) and Hansen et al. (2004) it was discovered that the angular power spectrum of the first year *WMAP* data, when estimated locally at different positions on the sphere, appears not to be isotropic. In particular, the power spectrum calculated for a hemisphere centred at $(l, b) = (237°, -20°)$ (in Galactic longitude and latitude) was larger than when calculated in the opposite hemisphere over the multipole range $\ell = 2$–40. Simultaneously, Park (2004) also presented evidence for the ex-istence of such hemispherical asymmetry — in which a particular statistical measure is considered to change discontinuously between two hemispheres on the sky — with the application of Minkowski functionals to the *WMAP* data. Since the preferred direction of Eriksen et al. (2004a) lies close to the ecliptic plane, it was also demonstrated that the large-angular scale $N$-point correlation functions showed a difference in behaviour when com-puted on ecliptic hemispheres. Many studies have subsequently been undertaken focusing on hemispheres in the ecliptic coordinate system, with Schwarz et al. (2004) particularly emphasizing the connection. Hemispherical asymmetry has also been seen with other measures of non-Gaussianity (Eriksen et al. 2004c, 2005; Räth et al. 2007a).

**Table 19.** Lower tail probablty for the variance, skewness, and kurtosis at $N_{side} = 16$, using different masks.

| | Probability | | | |
|---|---|---|---|---|
| Mask | C-R | NILC | SEVEM | SMICA |
| *Variance* | | | | |
| U73, $f_{sky} = 78\%$ . . . . . . . . . . . | 0.019 | 0.017 | 0.014 | 0.019 |
| CL58, $f_{sky} = 58\%$ . . . . . . . . . . | 0.004 | 0.003 | 0.003 | 0.003 |
| CL37, $f_{sky} = 37\%$ . . . . . . . . . . | 0.028 | 0.017 | 0.018 | 0.016 |
| Ecliptic North, $f_{sky} = 39\%$ . . . . | 0.001 | 0.001 | 0.001 | 0.002 |
| Ecliptic South, $f_{sky} = 39\%$ . . . . | 0.464 | 0.479 | 0.454 | 0.490 |
| *Skewness* | | | | |
| U73, $f_{sky} = 78\%$ . . . . . . . . . . . | 0.016 | 0.015 | 0.023 | 0.012 |
| CL58, $f_{sky} = 58\%$ . . . . . . . . . . | 0.208 | 0.139 | 0.162 | 0.147 |
| CL37, $f_{sky} = 37\%$ . . . . . . . . . . | 0.517 | 0.467 | 0.503 | 0.469 |
| Ecliptic North, $f_{sky} = 39\%$ . . . . | 0.502 | 0.526 | 0.526 | 0.521 |
| Ecliptic South, $f_{sky} = 39\%$ . . . . | 0.004 | 0.006 | 0.008 | 0.004 |
| *Kurtosis* | | | | |
| U73, $f_{sky} = 78\%$ . . . . . . . . . . . | 0.972 | 0.973 | 0.966 | 0.982 |
| CL58, $f_{sky} = 58\%$ . . . . . . . . . . | 0.630 | 0.726 | 0.711 | 0.711 |
| CL37, $f_{sky} = 37\%$ . . . . . . . . . . | 0.069 | 0.135 | 0.130 | 0.124 |
| Ecliptic North, $f_{sky} = 39\%$ . . . . | 0.094 | 0.229 | 0.196 | 0.245 |
| Ecliptic South, $f_{sky} = 39\%$ . . . . | 0.933 | 0.916 | 0.886 | 0.948 |

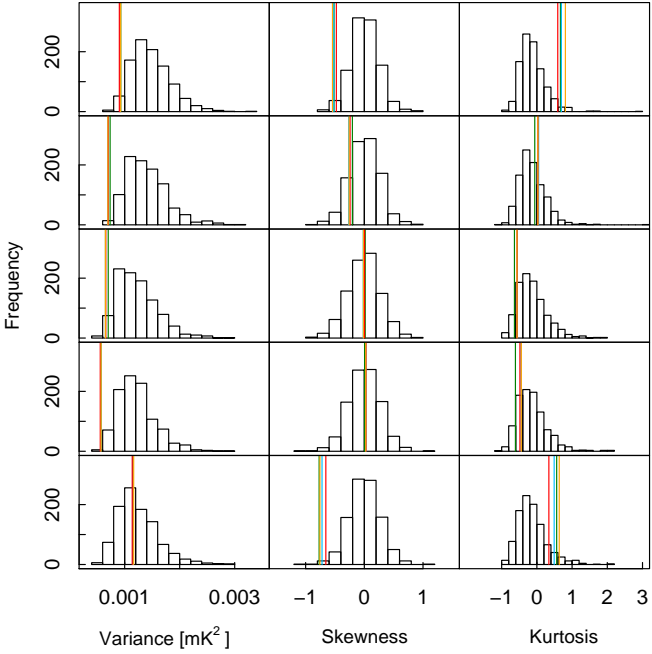

**Fig. 18.** Variance, skewness and kurtosis at $N_{side} = 16$, for the U73 mask, CL58, CL37, ecliptic North, and ecliptic South (from top to bottom). The different lines represent the four component separation methods C-R (green), NILC (blue), SEVEM (red), and SMICA (orange).

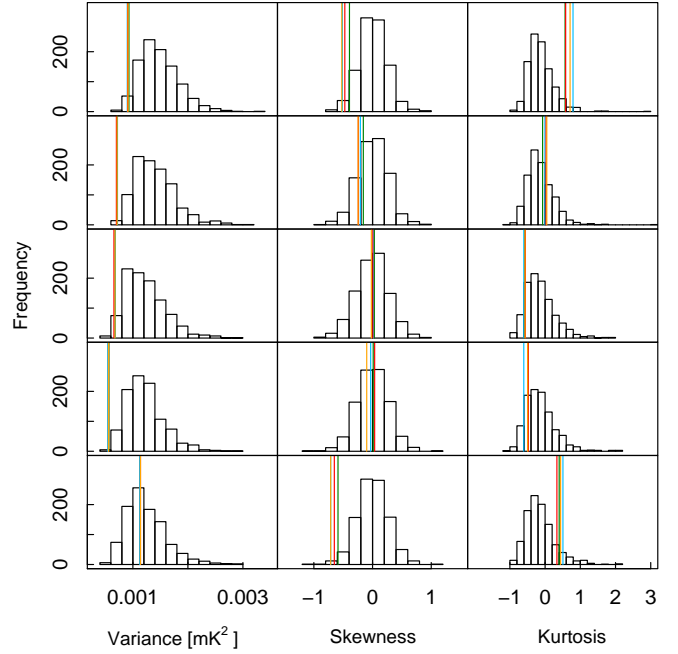

**Fig. 19.** Variance, skewness and kurtosis at $N_{side} = 16$, for the U73 mask, CL58, CL37, ecliptic North, and ecliptic South (from top to bottom). The different lines represent the four considered frequencies, namely 70 GHz (green), 100 GHz (blue), 143 GHz (red), and 217 GHz (orange).

**Table 20.** Frequency dependence of the lower tail probablity for the variance skewness and kurtosis at $N_{side} = 16$, using different masks.

| Mask | Probability | | | |
|---|---|---|---|---|
| | 70 GHz | 100 GHz | 143 GHz | 217 GHz |
| **Variance** | | | | |
| U73, $f_{sky} = 78\%$ . . . . . . . . . . . | 0.019 | 0.013 | 0.014 | 0.016 |
| CL58, $f_{sky} = 58\%$ . . . . . . . . . . | 0.003 | 0.003 | 0.003 | 0.003 |
| CL37, $f_{sky} = 37\%$ . . . . . . . . . . | 0.024 | 0.020 | 0.018 | 0.020 |
| Ecliptic North, $f_{sky} = 39\%$ . . . | 0.001 | 0.002 | 0.001 | 0.001 |
| Ecliptic South, $f_{sky} = 39\%$ . . . | 0.446 | 0.436 | 0.455 | 0.455 |
| **Skewness** | | | | |
| U73, $f_{sky} = 78\%$ . . . . . . . . . . . | 0.045 | 0.016 | 0.024 | 0.015 |
| CL58, $f_{sky} = 58\%$ . . . . . . . . . . | 0.254 | 0.205 | 0.162 | 0.157 |
| CL37, $f_{sky} = 37\%$ . . . . . . . . . . | 0.503 | 0.471 | 0.468 | 0.515 |
| Ecliptic North, $f_{sky} = 39\%$ . . . | 0.505 | 0.447 | 0.541 | 0.352 |
| Ecliptic South, $f_{sky} = 39\%$ . . . | 0.015 | 0.006 | 0.009 | 0.006 |
| **Kurtosis** | | | | |
| U73, $f_{sky} = 78\%$ . . . . . . . . . . . | 0.962 | 0.981 | 0.965 | 0.974 |
| CL58, $f_{sky} = 58\%$ . . . . . . . . . . | 0.619 | 0.684 | 0.710 | 0.725 |
| CL37, $f_{sky} = 37\%$ . . . . . . . . . . | 0.114 | 0.091 | 0.130 | 0.121 |
| Ecliptic North, $f_{sky} = 39\%$ . . . | 0.180 | 0.096 | 0.203 | 0.180 |
| Ecliptic South, $f_{sky} = 39\%$ . . . | 0.902 | 0.920 | 0.882 | 0.909 |

Here, we repeat the analysis of Eriksen et al. (2005) on the *Planck* component separated data, smoothed and then downgraded to $N_{side} = 64$, as described in Sect. 2. However, in this section the $N$-point correlation functions are not averaged over the full sky and depend on a choice of specific direction, thus, they constitute tools to study statistical isotropy rather than non-Gaussianity (Ferreira & Magueijo 1997). The latter study was already presented in Sect. 4.3, where it was found that the results for the low resolution maps are the most deviant relative to the MC simulations based on the *Planck* fiducial model.

The $N$-point correlation functions computed on the northern and southern hemispheres determined in the ecliptic coordinate frame and using the U73 mask are shown in Fig. 20. The correlation functions for the four *Planck* maps are very consistent, and the observed behaviour is in agreement with that seen in the *WMAP* data (Eriksen et al. 2004a). Specifically, the northern hemisphere correlation functions are relatively featureless (both the 3- and 4-point functions lie very close to zero), whereas the southern hemisphere functions exhibit a level of structure that is in good agreement with the confidence regions computed from the Gaussian simulations.

The probabilities of obtaining a value for the $\chi^2$ statistic for the *Planck* fiducial $\Lambda$CDM model at least as large as the observed values for different CMB component-separated maps are presented in Table 21. The probabilities for the 3-point and 4-point functions in the northern ecliptic hemisphere are especially large, and in the case of the pseudo-collapsed configuration all simulations yielded a larger value of the $\chi^2$ than observed. This suggests that the tendency to super-Gaussianity in the maps downgraded to $N_{side} = 64$, as noted previously in Sect. 4.3, is associated with the northern hemisphere. Nominally, this value is even more remarkable than that found with the *WMAP* data (Eriksen et al. 2004a), although to interpret it correctly one has to keep in mind that the analysis presented here is an example of an a posteriori statistic. Specific choices have been made about the smoothing scale used for downgrading the data, and, in particular, for the reference direction that defines the hemispheres. This results in a tendency to overestimate the significance of the results. Nevertheless, the observed properties of the *Planck* data are consistent with a remarkable lack of power in a direction towards the north ecliptic pole, consistent with the simpler one-point statistics presented in Sect. 5.2.

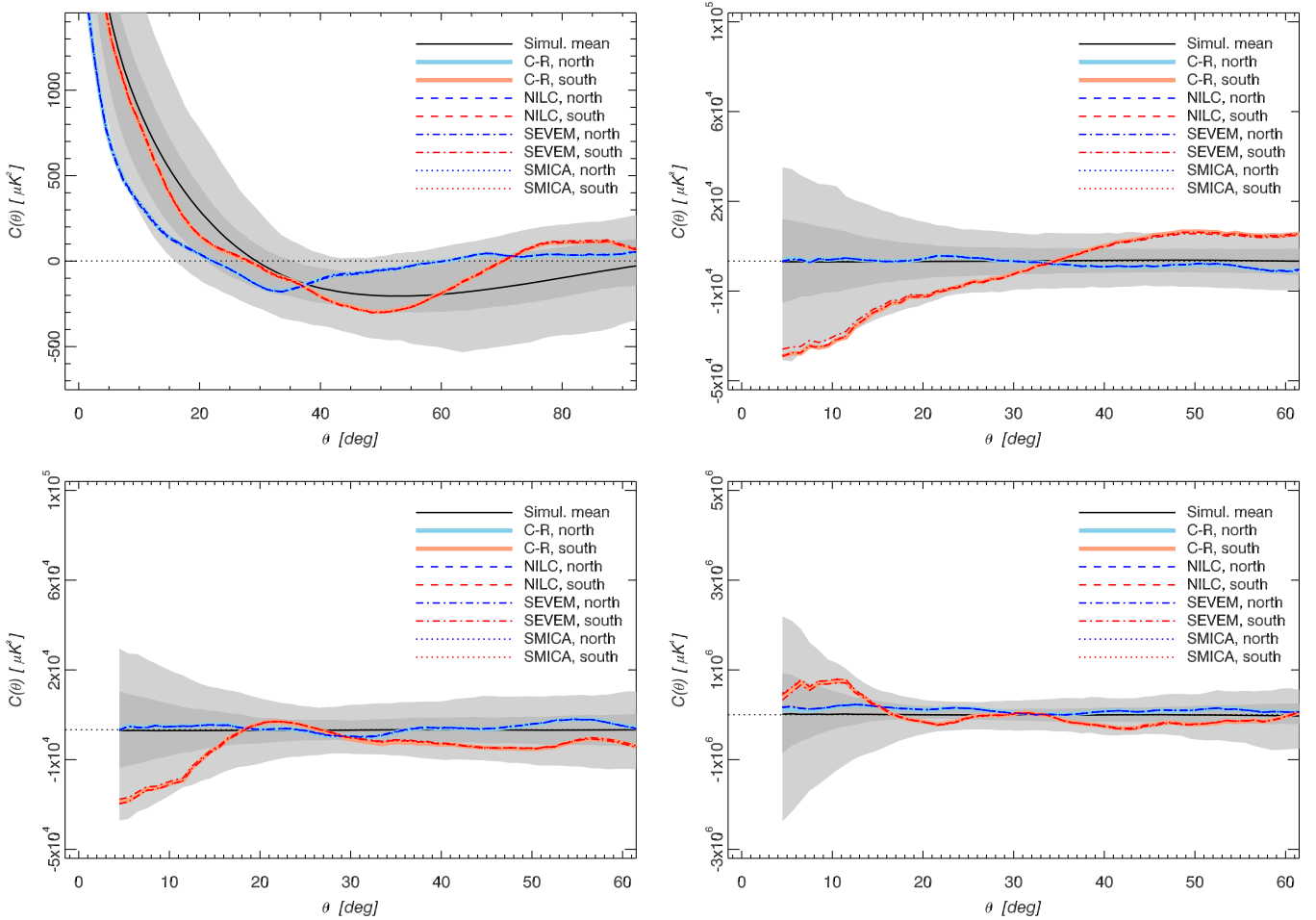

**Fig. 20.** The 2-point (upper left), pseudo-collapsed (upper right), equilateral 3-point (lower left), and rhombic 4-point (lower right) correlation functions ($N_{side} = 64$). Correlation functions are shown for the analysis performed on northern (blue) and southern (red) hemispheres determined in the Ecliptic coordinate frame. The shaded dark and light grey bands indicate the 68% and 95% confidence regions, respectively. See Sect. 4.3 for the definition of the separation angle $\theta$.

We have also carried out the corresponding calculations in Galactic coordinates, as well as in reference frames preferred by other analyses, such as the power asymmetry (PA, reference frame for $\ell_{max} = 1500$, see Sect. 5.5), the dipole modulation (DM, see Sect. 5.6), and the Doppler boost (DB, see Planck Collaboration XXVII 2014). The corresponding $N$-point correlation functions are shown in Fig. 21. Note that the positive hemispheres for the ecliptic and Galactic reference frames correspond to the southern hemispheres in other tables/figures. Interestingly, the largest asymmetry is seen for the hemispheres determined in ecliptic coordinates.

In Table 22 we show the corresponding probabilities for the $N$-point functions of the SMICA map in these reference frames. Whilst the largest asymmetry is seen for ecliptic coordinates, a substantial asymmetry is present also for Galactic coordinate hemispheres and, to a lesser extent, for the DM direction. The asymmetry for the latter can be explained by the fact that the DM direction is relatively close to the ecliptic pole. For the PA and DB directions we do not find any significant asymmetry. In the latter case, this is not surprising, since the Doppler boost effect is significant only for small angular scales, while the map used in our analysis was smoothed and downgraded to much lower resolution.

Since the largest asymmetry is seen in ecliptic coordinates, we might conjecture an association with the peculiar configuration of the quadrupole and octopole (although the *Planck* results from Sect. 5.1 do not favour as significant an alignment as with the *WMAP* data). In order to study the influence of the lowest order multipoles on the hemispherical asymmetry, we therefore re-analysed the SMICA map corrected for multipoles of order $\ell \leq 3$ and $\ell \leq 5$ and present the results in Table 23. It should be ap-parent that, in general, the asymmetry is less significant after removing the lowest order multipoles. However, in a few cases, such as for the 4-point function computed in the ecliptic and DM frames, we do not see such behaviour. Furthermore, in all cases, with the exception of the DB frame, we observe a tendency to super-Gaussianity for the negative hemisphere. This presumably indicates that the hemispherical asymmetry is not only related to the properties of the very lowest order multipoles, but also extends beyond this.

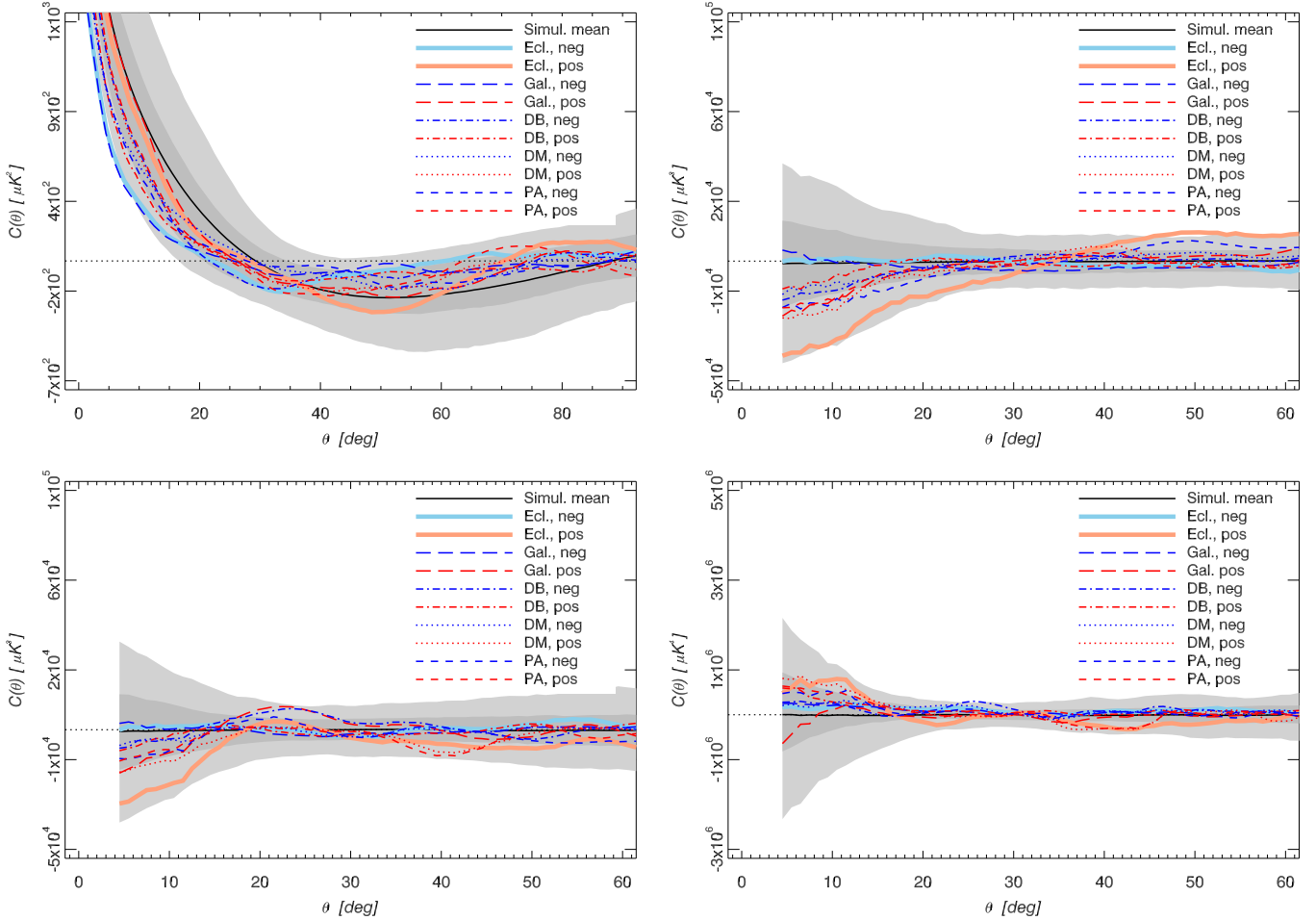

**Fig. 21.** The 2-point (upper left), pseudo-collapsed (upper right), equilateral 3-point (lower left), and rhombic 4-point (lower right) correlation functions for the `SMICA` map ($N_{side} = 64$). Correlation functions are shown for the analysis performed on negative (blue) and positive (red) hemispheres determined in the Ecliptic, Galactic, Doppler boost (DB), dipole modulation (DM), and power asymmetry (PA) coordinate frames. The shaded dark and light grey bands indicate the 68% and 95% confidence regions, respectively. See Sect. 4.3 for the definition of the separation angle $\theta$.

### 5.4. Phase correlations

The statistical properties of a Gaussian random field on the sphere can be fully described by its two-point correlation function (or power spectrum) and should exhibit Fourier phases that are independent and identically distributed (i.i.d.) and follow a uniform distribution in the interval $[-\pi, \pi]$. Thus, demonstrating the existence of Fourier phase correlations in CMB maps could indicate the presence of non-Gaussianities in the primordial density fluctuations.

One could search for phase correlations using different approaches. However, without there being a specific model for the origin of any such correlations, one would like to test for scale-dependent non-Gaussianity and deviations from isotropy in a completely model-independent ("blind") way. One such approach is offered through the method of "scaling indices" and "surrogates" (which we explain below).

Previous studies using these methods, based on the *WMAP* three-, five- and seven-year data (Räth et al. 2009, 2011; Rossmanith et al. 2012; Modest et al. 2013), showed significant evidence for intrinsic phase correlations at low $\ell$ values in the CMB. The signal was demonstrated to be robust with respect to the *WMAP* data release, to the component separation methods and to the selected test statistics. In this section we apply these methods to the *Planck* component separated data sets.

The scaling index method represents one way to estimate the local scaling properties of a point set in an arbitrary $d$-dimensional embedding space. The technique provides the possibility of revealing local structural characteristics of a given point distribution. A number of analyses have used scaling indices to test the Gaussian nature and statistical isotropy of the CMB as represented by the *WMAP* data (Räth et al. 2007a, 2009; Rossmanith et al. 2009a; Räth et al. 2011).

**Table 21.** Probabilities of obtaining values of the $\chi^2$ statistic for the *Planck* fiducial model at least as large as the observed values of the statistic for the *Planck* maps with resolution parameter $N_{\text{side}} = 64$, estimated using the C-R, NILC, SEVEM and SMICA methods.

| Hemisphere | Probability | | | |
|---|---|---|---|---|
| | C-R | NILC | SEVEM | SMICA |
| *Two-point function* | | | | |
| Northern Ecliptic . . . . . | 0.935 | 0.924 | 0.927 | 0.932 |
| Southern Ecliptic . . . . . | 0.633 | 0.599 | 0.639 | 0.592 |
| *Pseudo-collapsed 3-point function* | | | | |
| Northern Ecliptic . . . . . | >0.999 | >0.999 | >0.999 | >0.999 |
| Southern Ecliptic . . . . . | 0.349 | 0.310 | 0.381 | 0.301 |
| *Equilateral 3-point function* | | | | |
| Northern Ecliptic . . . . . | 0.996 | 0.999 | 0.999 | 0.999 |
| Southern Ecliptic . . . . . | 0.627 | 0.644 | 0.678 | 0.656 |
| *Rhombic 4-point function* | | | | |
| Northern Ecliptic . . . . . | 0.999 | 0.999 | 0.999 | 0.999 |
| Southern Ecliptic . . . . . | 0.559 | 0.548 | 0.574 | 0.553 |

**Table 22.** Probabilities of obtaining values of the $\chi^2$ statistic for the *Planck* fiducial model at least as large as the observed values for the SMICA map at $N_{\text{side}} = 64$. The probabilities were computed for hemispheres determined in the ecliptic, Galactic, Doppler boost (DB), dipole modulation (DM), and power asymmetry (PA) frames.

| Hemisphere | Probability | | | | |
|---|---|---|---|---|---|
| | Ecl. | Gal. | DB | DM | PA |
| *Two-point function* | | | | | |
| Negative . . . . . . . | 0.932 | 0.974 | 0.573 | 0.735 | 0.747 |
| Positive . . . . . . . . | 0.592 | 0.275 | 0.918 | 0.913 | 0.882 |
| *Pseudo-collapsed 3-point function* | | | | | |
| Negative . . . . . . . | >0.999 | 0.998 | 0.978 | 0.912 | 0.826 |
| Positive . . . . . . . . | 0.301 | 0.768 | 0.698 | 0.300 | 0.452 |
| *Equilateral 3-point function* | | | | | |
| Negative . . . . . . . | 0.999 | 0.994 | 0.844 | 0.893 | 0.492 |
| Positive . . . . . . . . | 0.656 | 0.379 | 0.917 | 0.583 | 0.425 |
| *Rhombic 4-point function* | | | | | |
| Negative . . . . . . . | 0.999 | >0.999 | 0.771 | 0.970 | 0.784 |
| Positive . . . . . . . . | 0.553 | 0.568 | 0.993 | 0.525 | 0.548 |

In general, the method is a mapping that calculates, for each member $\boldsymbol{p}_i$, $i = 1, \ldots, N_{\text{pix}}$ of a point set $P$, a single value that depends on the spatial position of $\boldsymbol{p}_i$ relative to the group of other points in its neighbourhood, in which the point under consideration is embedded. A three-dimensional point set $P$ is generated for two-dimensional spherical CMB-data by transforming the temperature values $T(\theta_i, \phi_i)$ of each pixel to a radial jitter around a sphere of radius $R$ at the position of the pixel centre $(\theta_i, \phi_i)$. To obtain scaling indices the local weighted cumulative point distribution is calculated:

$$\rho(\boldsymbol{p}_i, r) = \sum_{j=1}^{N_{\text{pix}}} s_r(d(\boldsymbol{p}_i, \boldsymbol{p}_j)). \tag{27}$$

Here $r$ describes the scaling range, while $s_r$ and $d$ denote a shaping function and a distance measure, respectively. The scaling index $\alpha(\boldsymbol{p}_i, r)$ is then defined as the logarithmic derivative of $\rho(\boldsymbol{p}_i, r)$ with respect to $r$:

$$\alpha(\boldsymbol{p}_i, r) = \frac{\partial \log \rho(\boldsymbol{p}_i, r)}{\partial \log r}. \tag{28}$$

Using a quadratic Gaussian shaping function $s_r(x) = \exp(-(x/r)^2)$ and an isotropic Euclidian norm $d(\boldsymbol{p}_i, \boldsymbol{p}_j) = \|\boldsymbol{p}_i - \boldsymbol{p}_j\|$ as distance measure, one obtains the following analytic formula for the scaling indices:

$$\alpha(\boldsymbol{p}_i, r) = \frac{\sum_{j=1}^{N_{\text{pix}}} 2\left(\frac{d_{ij}}{r}\right) e^{-\left(\frac{d_{ij}}{r}\right)^2}}{\sum_{j=1}^{N_{\text{pix}}} e^{-\left(\frac{d_{ij}}{r}\right)^2}}, \tag{29}$$

where we use the abbreviation $d_{ij} \equiv d(\boldsymbol{p}_i, \boldsymbol{p}_j)$.

As seen from the derivation of the expression for $\alpha$, the scaling index analysis can be considered as a structural decomposition of the point set under study. Points in well-localized, cluster-like structures lead to scaling indices of $\alpha \approx 0$, filamentary structures have $\alpha \approx 1$, and sheet-like structures $\alpha \approx 2$. A uniform, three-dimensional distribution of points would result in $\alpha \approx 3$. One can schematically think of the scaling indices applied to CMB data as estimating the roughness of the last scattering surface on different scales.

As should be clear from equation (29), the calculation of scaling indices depends on the scale parameter $r$. Ten scaling range parameters $r_k = 0.05, 0.1, \ldots, 0.5, k = 1, 2, \ldots, 10$, in the notation of Räth et al. (2007a), are used in this analysis. In order to calculate scaling indices on large scales, as in previous studies, we couple the $r$-jitter $a$ to $r_k$ via $a = 0.5r_k$. The mean $\langle \alpha(r_k) \rangle$ and standard deviation $\sigma_{\alpha(r_k)}$, derived from the full sky and from a set of 768 rotated hemispheres, are used to test for non-Gaussianity and deviations from statistical isotropy.

In order to quantify the significance of the scaling index results, and focus the study on the phase properties of the observed CMB sky, we utilize the method of "surrogate maps" (Räth et al. 2009). This surrogate approach suppresses the sensitivity of the null tests to the assumed fiducial power spectrum, which is particularly pertinent, given the potential mismatch of the *Planck* data to the fiducial spectrum on large angular scales (Planck Collaboration XV 2014).

The basic idea of the method of surrogates is to generate for a given data set an ensemble of so-called surrogate data, which mimic some properties of the original data being defined by a null hypothesis, while all other properties are subject to randomization. Analysing both the original and surrogate data sets with test statistics, which are sensitive to the complement of the null hypothesis, then reveals whether the original data were consistent with the null hypothesis or not. In our case, the null hypothesis is that the CMB map is Gaussian and thus fully described by the power spectrum and free of any phase correlations. The test statistics being employed are those that are sensitive to higher-order correlations (i.e., correlations beyond the two-point correlation function) in real space. Alternatively, one can think of introducing Fourier phase statistics, but those methodologies are currently less developed.

However, the Gaussianity of the temperature distribution and the randomness of the set of Fourier phases of the map to be studied are a necessary prerequisite for the application of the surrogate-generating algorithm. Therefore the following preprocessing steps are applied to generate a zero order surrogate map. First, the maps are remapped onto a Gaussian distribution in a rank-ordered way. Then we ensure the randomness of the set of Fourier phases by making a rank-ordered remapping of the phases onto a set of uniformly distributed ones. These two preprocessing steps only have marginal influence on the maps. Now the set of surrogates to be used for assessing the statistical properties of the data sets can be generated by shuffling the phases in the space of the spherical harmonics, while exactly preserving the modulus of the $a_{\ell m}$. Moreover, by introducing a two-step shuffling scheme for previously specified $\ell$-ranges, a scale-dependent analysis is made possible. It is worth noting that while in all surrogate maps the modulus of each $a_{\ell m}$ is exactly preserved, null tests involving a comparison to an assumed fiducial power spectrum only preserve the $C_\ell$ values, which are average values of the $|a_{\ell m}|$ when summed over $m$. Thus, the linear properties of the surrogate maps are more tightly constrained, and specifically kept constant, than they would be in tests involving simulated maps generated on the basis of the $C_\ell$s.

**Table 23.** Probabilities of obtaining values of the $\chi^2$ statistic for the *Planck* fiducial model at least as large as the observed values of the statistic for the *Planck* SMICA map with resolution parameter $N_{\text{side}} = 64$ and removed multipoles of order $\ell \leq 3$ (columns denoted as "$\ell > 3$") or order $\ell \leq 5$ (columns denoted as "$\ell > 5$") . The probabilities were computed for hemispheres determined in the ecliptic, Galactic, Doppler boost (DB), dipole modulation (DM), and power asymmetry (PA) frames.

| | Probability | | | | | | | | | |
|---|---|---|---|---|---|---|---|---|---|---|
| | Ecl. | | Gal. | | DB | | DM | | PA | |
| Hemisphere | $\ell > 3$ | $\ell > 5$ | $\ell > 3$ | $\ell > 5$ | $\ell > 3$ | $\ell > 5$ | $\ell > 3$ | $\ell > 5$ | $\ell > 3$ | $\ell > 5$ |
| Two-point function | | | | | | | | | | |
| Negative . . . . . . . . . . . | 0.886 | 0.951 | 0.875 | 0.957 | 0.344 | 0.509 | 0.746 | 0.869 | 0.745 | 0.819 |
| Positive . . . . . . . . . . . | 0.609 | 0.906 | 0.123 | 0.330 | 0.824 | 0.973 | 0.811 | 0.979 | 0.756 | 0.947 |
| Pseudo-collapsed 3-point function | | | | | | | | | | |
| Negative . . . . . . . . . . . | 0.985 | 0.999 | 0.981 | 0.978 | 0.965 | 0.879 | 0.966 | 0.948 | 0.986 | 0.872 |
| Positive . . . . . . . . . . . | 0.262 | 0.643 | 0.468 | 0.789 | 0.807 | 0.989 | 0.132 | 0.390 | 0.224 | 0.516 |
| Equilateral 3-point function | | | | | | | | | | |
| Negative . . . . . . . . . . . | 0.984 | 0.997 | 0.963 | 0.952 | 0.531 | 0.790 | 0.945 | 0.998 | 0.536 | 0.943 |
| Positive . . . . . . . . . . . | 0.776 | 0.790 | 0.347 | 0.727 | 0.966 | 0.991 | 0.358 | 0.694 | 0.485 | 0.649 |
| Rhombic 4-point function | | | | | | | | | | |
| Negative . . . . . . . . . . . | 0.951 | >0.999 | 0.998 | 0.992 | 0.482 | 0.842 | 0.991 | 0.997 | 0.929 | 0.989 |
| Positive . . . . . . . . . . . | 0.677 | 0.584 | 0.746 | 0.870 | 0.994 | 0.966 | 0.334 | 0.368 | 0.314 | 0.359 |

So-called first- and second-order surrogate maps are then obtained as follows. We initially generate a first-order surrogate map, in which any phase correlations for the scales that are not of interest are randomized by shuffling the phases $\phi_{\ell m}$ for $\ell \notin \Delta l = [\ell_{\min}, \ell_{\max}], 0 < m \leq \ell$. In a second step, $N$ (=1000) realizations of second-order surrogate maps are generated from the first-order surrogate map, in which the remaining phases $\phi_{\ell m}$ with $\ell \in \Delta l, 0 < m \leq \ell$ are shuffled, while the previously randomized phases for the other scales are preserved. The generation of surrogates is always performed using the maps with the highest resolution, i.e., $N_{\text{side}} = 2048$. Given the evidence for anomalies on the largest angular scales, and to ensure consistency with the previous *WMAP* analyses, we perform dedicated scale-dependent tests for the scales defined by $\Delta l = [2, 20]$.

The exclusion of the phases belonging to the zonal modes ($m = 0$) from the shuffling raises the question as to whether the results depend on the choice of coordinate system, since these modes are only defined with respect to the selected coordinate system. This issue is particularly important in the specific case of the CMB when analysing data in the Galactic coordinate system, because possible residuals in and around the Galactic plane are likely to be reflected mostly in these modes. To test the dependence of the results on the coordinate system, we repeated the surrogate analysis with $N = 100$ second-order surrogates for each member of a set of 48 rotated CMB maps, where the rota-tions, defined by a HEALPix $N_{\text{side}} = 2$ map, such that the pixel centres indicate the orientation of the $z$-axis of the rotated frame, uniformly cover the unit sphere.

Since the methodology in its simplest form requires orthonormality of the set of basis functions $Y_{\ell m}$, we apply the method to the full sky foreground-cleaned maps as obtained after component separation with Commander-Ruler, NILC, SEVEM and SMICA. For the selected $\ell$-interval $\Delta l = [2, 20]$, the generation of the first-order surrogate map removes the phase signature of the small scale residuals in the data and this can be interpreted as a procedure for inpainting procedure for small masked patches in the Galactic plane. The differences between the first- and second-order surrogates are quantified by the $\sigma$-normalized deviation

$$S(Y) = \frac{Y_{\text{surr1}} - \langle Y_{\text{surr2}} \rangle}{\sigma_{Y_{\text{surr2}}}} \tag{30}$$

with $Y = \langle \alpha(r_k) \rangle, \sigma_{\alpha(r_k)}, \chi^2$. Here, $\chi^2$ represents either a diagonal combination of the mean $\langle \alpha(r_k) \rangle$ and standard deviation $\sigma_{\alpha(r_k)}$ at a certain scale $r_k$, or for the full scale-independent $\chi^2$ statistics:

$$\chi^2 = (\boldsymbol{M} - \langle \boldsymbol{M} \rangle)^{\mathsf{T}} \mathsf{C}^{-1} (\boldsymbol{M} - \langle \boldsymbol{M} \rangle), \tag{31}$$

where the test statistics to be combined are contained in the vector $\boldsymbol{M}$ and C is obtained by cross-correlating the elements of $\boldsymbol{M}$. With the mean and the standard deviation as input for $\boldsymbol{M}$ we obtain $\chi^2_{\langle \alpha \rangle}$ and $\chi^2_{\sigma_\alpha}$ statistics with $\boldsymbol{M}^{\mathsf{T}} = (\langle \alpha(r_1) \rangle, \dots, \langle \alpha(r_{10}) \rangle)$ and $\boldsymbol{M}^{\mathsf{T}} = (\sigma_{\alpha(r_1)}, \dots, \sigma_{\alpha(r_{10})})$, respectively.

Figure 22 shows the deviation $S(\langle \alpha(r_5) \rangle)$ values for the set of rotated hemispheres for the SMICA map. Each pixel of the full sky map with a HEALPix resolution of $N_{\text{side}} = 8$ specifies one of the 768 $S$-values for a rotated hemisphere, where the pixel position indicates the orientation of the $z$-axis of the rotated coordinate system. We find a large number of hemispheres for which $|S|$ is larger than 3, indicating a significant difference between the first- and second-order surrogate maps. Therefore, one has to conclude that the low-$\ell$ phases are correlated. In addition, one can clearly see that the deviation $S$ goes in opposite directions for a number of pairs of opposite hemispheres. This indicates the presence of pronounced anisotropy in the CMB. We further find that the $S$-map derived for the Galactic coordinate system and the $S$-map from the average of rotated coordinate systems are nearly indistinguishable. This shows that the zonal modes are not special and that the choice of Galactic coordinates as reference frame does not induce any bias. Therefore, all of the results that follow are only shown for Galactic coordinates. We also find very similar yet less significant patterns of asymmetry when considering the standard deviation of the scaling indices $\sigma_\alpha$, which leads to a significant detection of anomalies when combining both quantities into a $\chi^2$-statistic (see Fig. 23). The results are consistent for `Commander-Ruler`, `NILC`, `SEVEM` and `SMICA`.

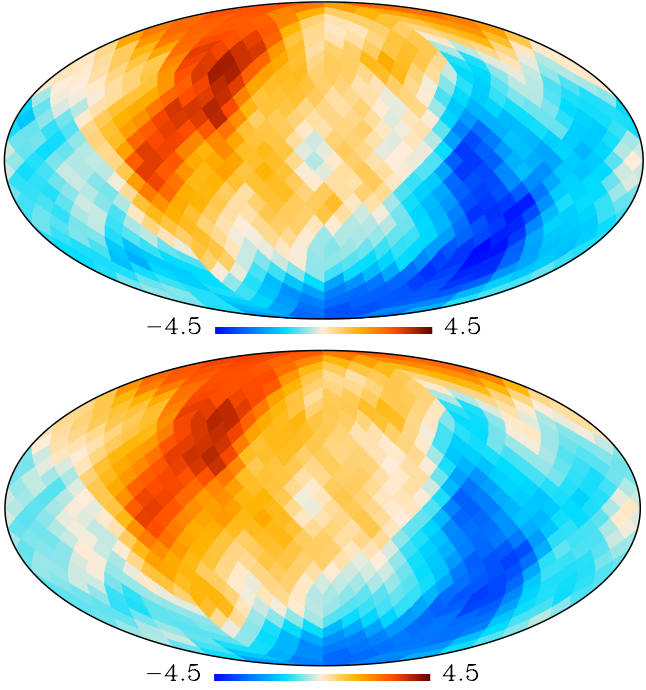

**Fig. 22.** Deviations $S(\langle\alpha\rangle)$ of the rotated hemispheres derived from the mean of the scaling indices for the scale $r_5$ determined from the `SMICA` map. *Upper panel*: Result for the calculations with respect to the Galactic coordinate system. *Lower panel*: Result obtained after averaging over 48 rotated coordinate systems.

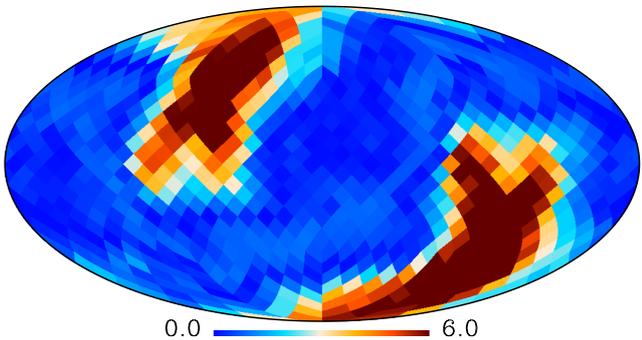

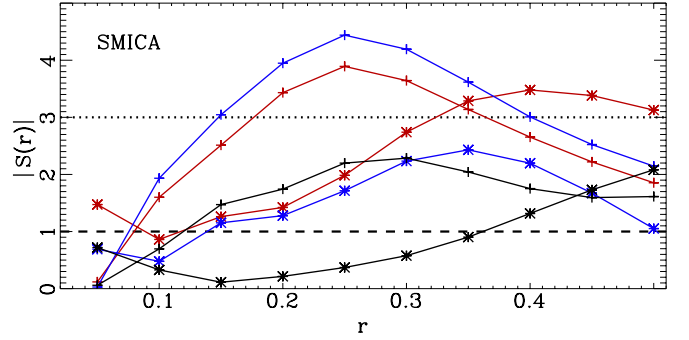

**Fig. 24.** Deviations $|S(r)|$ for the `SMICA` map as a function of the scale parameter $r$ for the full sky (black) and positive (red) and negative (blue) rotated hemisphere. The reference frame for defining the positive and negative hemisphere is chosen such that the difference $\Delta S = S_{pos} - S_{neg}$ becomes maximal for $\langle\alpha(r_5)\rangle$. The plus signs denote the results for the mean $\langle\alpha(r_k)\rangle$, the star-signs represent the standard deviation $\sigma_{\alpha(r_k)}$. The dashed (dotted) line indicates the 1 (3) $\sigma$ significance level.

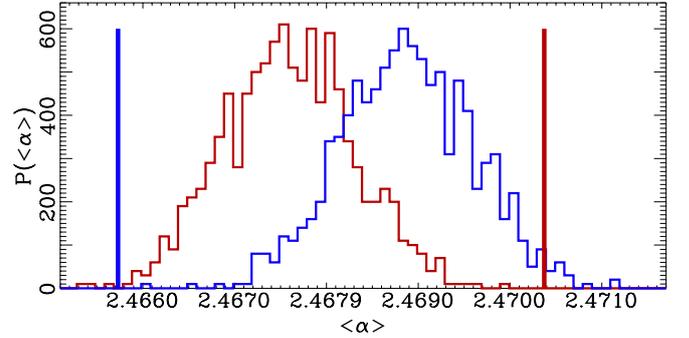

**Fig. 25.** Probability density $P(\langle\alpha(r_5)\rangle)$ for the second-order surrogates using the `SMICA` map for the positive (red) and negative (blue) rotated hemispheres. The vertical lines show the respective values for $\langle\alpha(r_5)\rangle$ for the first-order surrogates.

In Fig. 24, the deviations $S(Y)$ are displayed for the mean and standard deviation. We only show the results for the `SMICA` map. The other three maps yield very similar results. For all four maps the values for $S(\langle\alpha\rangle)$ extend beyond 3 for $r = 0.2$–0.25 when rotated hemispheres are considered separately. Since the effect in the separate hemispheres goes in opposing directions, no signal is observed for the full sky. The results for the scale-independent $\chi^2$ statistics are summarized in Table 24. The results suggest a significant detection of both non-Gaussianity and anisotropy in the *Planck* data, consistent with those obtained previously with *WMAP* data (for comparison see Modest et al. 2013).

Figure 25 presents results that allow us to attempt to elucidate the origin of the detected asymmetry and its relationship to other anomalies. Considering the first-order surrogate maps, in which the low-$\ell$ phases are unchanged by the shuffling procedure, there is a clear shift between the mean of the scaling indices for the positive (red) and negative (blue) hemispheres. If we now examine the probability density of the mean of the scaling indices for the positive and negative rotated hemispheres, as determined from the set of second-order surrogates, again a sys-tematic shift is observed. Since the phases are uncorrelated in the second-order surrogate maps, the effect can only be attributed to properties of the distribution of $|a_{lm}|$, i.e., to power asymmetries. However, it should be apparent that the shift in mean values is in the opposite sense to the first-order surrogates, implying that an additional pattern of asymmetry is induced solely by correlations in the phases.

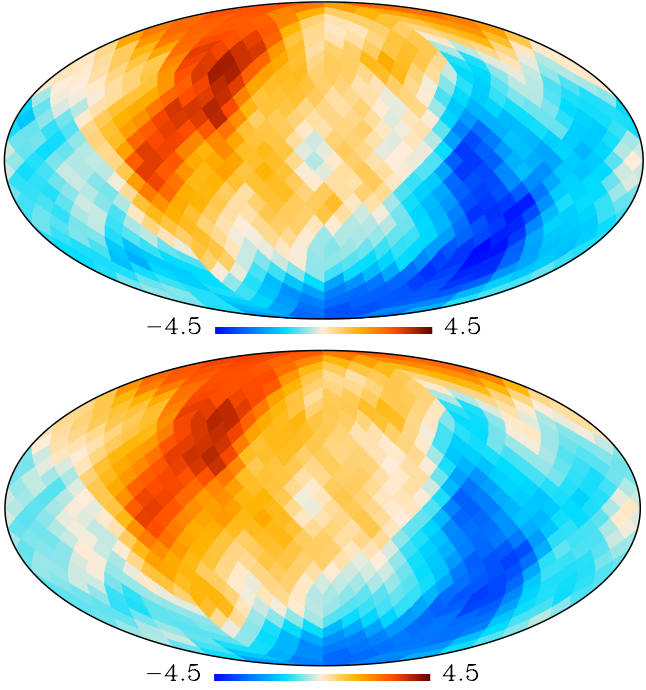

**Fig. 23.** Deviations $S(\chi^2)$ of the rotated hemispheres derived from a combination of the mean and the standard deviation of the scaling indices for the scale $r_5$ determined from the `SMICA` map.

**Table 24.** Deviations $S$ and empirical probabilities $p$ for the scale-independent $\chi^2$-statistics derived from the C-R, NILC, SEVEM, and SMICA maps.

| | $S / p$ [%] | | |
| Model | Full Sky | Positive Hemisphere | Negative Hemisphere |
| --- | --- | --- | --- |
| C-R, $\chi^2_{(\alpha)}$ . . . . . . . . | 0.86 / 82.6 | 4.21 / 99.7 | 3.18 / 99.0 |
| C-R, $\chi^2_{\sigma_\alpha}$ . . . . . . . . | 0.88 / 85.2 | 3.94 / 99.5 | 3.10 / 99.2 |
| NILC, $\chi^2_{(\alpha)}$ . . . . . . | 0.86 / 81.8 | 3.74 / 99.6 | 4.41 / >99.9 |
| NILC, $\chi^2_{\sigma_\alpha}$ . . . . . . | 0.79 / 78.8 | 3.69 / 99.6 | 4.49 / >99.9 |
| SEVEM, $\chi^2_{(\alpha)}$ . . . . . | 0.00 / 58.0 | 3.22 / 99.3 | 5.02 / >99.9 |
| SEVEM, $\chi^2_{\sigma_\alpha}$ . . . . . | 0.05 / 60.8 | 3.20 / 99.0 | 5.11 / 99.9 |
| SMICA, $\chi^2_{(\alpha)}$ . . . . . | 0.75 / 80.1 | 3.80 / 99.8 | 4.70 / 99.8 |
| SMICA, $\chi^2_{\sigma_\alpha}$ . . . . | 0.01 / 54.4 | 3.64 / 99.3 | 4.81 / >99.9 |

Finally, it should be noted that certain choices made in the analysis, e.g., the selection of specific values for $r_k$ and $a$ parameters, together with the decision to make an analysis on hemispheres using the method of surrogates, constitute a posteriori selections and can lead to an overestimate of the significance of the results. Nevertheless, the claimed asymmetry is certainly consistent with results presented in the following sections. The analysis involving surrogate maps reveals that there *are* phase correlations at low $\ell$.

### 5.5. Power asymmetry

In their analysis of the 5-year *WMAP* data, Hansen et al. (2009) specifically searched for dipolar power asymmetry on the sky. In particular, a simple test was performed in which the power spectrum on discs was computed and binned into independent blocks of 100 multipoles from $\ell = 2$ to $\ell = 600$, then each block fitted for a dipolar asymmetry in the power distribution. The six $\ell$ ranges considered showed evidence of a consistent dipole direction. The significance of this was shown by the fact that not a single realization in a set of 10,000 simulations showed a similarly strong asymmetry. A further extension of the analysis introduced a model selection procedure taking into account the statistical penalty for introducing an asymmetric model with additional parameters (direction of asymmetry, amplitude of asymmetry and asymmetric multipole ranges). Even in this case, the asymmetry was found to be highly significant for the whole range $\ell = 2$ to 600.

Unfortunately, this procedure is highly CPU-intensive. Given the increased sensitivity and angular resolution of the *Planck* data, we have therefore chosen to focus on the simpler test, following the modified approach of Axelsson et al. (2013) as applied to the 9-year *WMAP* data. Specifically, we estimate the power spectrum amplitude on 12 non-overlapping patches of the sky in $\ell$-bins of 16 multipoles each, and fit a dipole to the spatial distribution of the amplitude for each $\ell$-bin. This allows us to probe further into a previously unexplored high-$\ell$ range and to answer at least in part any a posteriori criticisms of the study.

Since the analysis is power-spectrum based, the half-ring data sets for the different CMB estimators are used.[4] The approach is as follows.

1. We apodize the U73 Galactic, and point source mask by smoothing it twice with a $30'$ Gaussian beam followed by a $10'$ smoothing that is repeated three times. The apodized mask is multiplied with the original mask between each of the smoothing operations, thus all pixels which are zero in the U73 mask remain so in order to continue to fully eliminate foreground residuals. In addition, a smooth transition between masked and unmasked regions is created. The power spectrum estimation method has been shown to take into account the mode couplings introduced by sharp edges in the mask. Nevertheless, we have adopted an apodized mask, since the error bars at high $\ell$ decrease slightly due to the reduced couplings. The effective sky fraction of the U73 mask falls from 73% to 69% due to this procedure.

2. The MASTER technique (Hivon et al. 2002) is used to estimate local cross power spectra between the two half-ring maps on 12 non-overlapping patches on the sky. A given patch corresponds to one of the twelve base pixels of an $N_{\rm side} = 1$ HEALPix map, masked as appropriate by the apodized U73 Galactic and point source mask. This results in sky coverages of approximately 7% for each of the eight patches surrounding the poles, and of order 3% for each of the four patches aligned with the Galactic plane. This approach differs from that of the earlier analysis of Hansen et al. (2009) in which 3072 highly overlapping discs were used, but correlations were ignored. The revised method results in weaker correlations between the non-overlapping patches.

3. Following Hansen et al. (2009) the 12 local spectra are estimated in bins of 16 multipoles. The highest multipole included in the analysis is $\ell_{\rm max} = 1500$, since both noise and high-$\ell$ foregrounds become important at higher multipoles (Planck Collaboration XII 2014; Planck Collaboration XV 2014; Planck Collaboration XVI 2014). For some comparisons, it is useful to follow Hansen et al. (2009) and further combine these 16-multipole bins into blocks containing approximately 100 multipoles. Although no single block contains exactly 100 multipoles, they will be referred to as such. However, our main results are based on the 16-multipole blocks, which offer improved statistics over the more coarsely binned data for our statistic of choice.

4. Each 16-multipole block now has an associated local power map at $N_{\rm side} = 1$. The HEALPix routine remove_dipole is then used with inverse variance weighting (with the weight for each region determined from the variance of the local power spectra derived from simulations) to obtain the dipole direction of the power distribution for each map.

5. The alignment of the dipole directions between the different multipole blocks is then used to construct a measure of the power spectrum asymmetry. Despite the mask-induced correlations between adjacent multipoles, the power spectra estimated in 16-multipole blocks should be highly independent and the dipole directions determined for a Gaussian, isotropic CMB realization should be random, as confirmed by simulation.

6. The same procedure is then applied to a set of 500 simulated maps of the CMB and noise to derive significance.

---

[4] In fact, the simulated half-ring noise maps were generated by adjusting the properties of the existing 1000 noise simulations appropriately, thus explaining why only 500 simulations are used in this analysis.

In order to assess the significance of the asymmetry, it is necessary to compare the clustering of the dipole directions evaluated for the different scales to that observed in simulated maps. For this purpose, we compute the mean separation angle, $\theta_{\mathrm{mean}}$, between all possible pair-wise combinations of the 16-multipole dipole directions up to a given $\ell_{\mathrm{max}}$. The expectation value for this statistic is $90°$ when determined from the dipole directions computed on the full sky for an isotropic, Gaussian realization of the CMB. We have shown that this remains true for an analysis when applying the apodized U73 mask. We determine $\theta_{\mathrm{mean}}$ for the data as well as for all 500 simulated maps. The fraction of simulations with a higher value for $\theta_{\mathrm{mean}}$ than the data defines a $p$-value for the measurement. One advantage of this directional analysis is that it focuses on a central issue for tests of deviation from isotropy – whether there is a preferred direction. Moreover, as noted by Bunn & Scott (2000), the CMB may exhibit a pattern that cannot be identified from the power spectrum, but which would indicate some non-trivial large-scale structure. In the context of our analysis this means that the amplitudes of the power dipoles may not be discernible, in a frequentist sense, from the distribution of values expected for an ensemble of Gaussian, isotropic CMB realizations. However, evidence for the close correlation and alignment of directions on different angular scales presents a clear signature of power asymmetry, since in the standard model, these directions should all be independent random variables. Our claims about power asymmetry in this section are based *solely* on this statistic.

In Fig. 26 we show the dipole directions of the 15 lowest 100-multipole bins for the SMICA map. The preferred direction for WMAP-9 over the range $\ell = 2$ to 600 is also indicated, as is the preferred low-$\ell$ modulation direction determined in Sect. 5.6. Similar behaviour is seen for all of the Planck foreground-cleaned maps.

In Table 25 we present a summary of the power asymmetry results from the Planck data processed by all four Planck foreground cleaning methods — Commander-Ruler, NILC, SEVEM, and SMICA — computed on the apodized U73 mask. For comparison, we also include the latest WMAP 9-year result[5] computed with their KQ85 mask (which has a usable sky fraction of 74.8%). It should be apparent that the clustering seen visually in Fig. 26 is both significant and consistent for all foreground subtraction methods.

However, Doppler boosting, due to our motion with respect to the CMB rest frame, is expected to induce both a dipolar modulation of the temperature anisotropies, and an aberration that corresponds to a change in the apparent arrival directions of the CMB photons (Challinor & van Leeuwen 2002). Both of these effects are aligned with the CMB dipole. In a companion paper (Planck Collaboration XXVII 2014), a statistically significant detection of such an effect at small angular scales is reported. From an inspection of Fig. 26 it should be apparent that whilst the asymmetry direction on large and intermediate angular scales are in general tightly clustered around the same direction as found for WMAP, weak indications of this second preferred power asymmetry direction might be seen. However, we have determined that our statistic of choice is unable to cleanly

---

**Table 25.** Summary of the best-fit power dipole directions on the sky up to $\ell_{\mathrm{max}} = 600/1500$, for each of the four component separation methods, together with the 143 GHz map cleaned using the SEVEM method and denoted SEVEM-143. These are determined from maps of the power spectrum estimated on 12 non-overlapping patches in bins of 16 multipoles. The preferred direction is the mean of the separate 16-multipole dipole directions in the range $\ell = 2$ to $\ell_{\mathrm{max}}$. The angle $\theta_{\mathrm{mean}}$ is the mean separation angle between all possible pair-wise combinations of these 16-multipole dipoles over the same $\ell$ range. The $p$-value for the power asymmetry is given by the fraction of simulations with a larger mean separation angle, $\theta_{\mathrm{mean}}^{\mathrm{sim}}$, than the observed data, $\theta_{\mathrm{mean}}^{\mathrm{data}}$. SEVEM-143DB corresponds to data corrected for Doppler boosting, while all other results are derived without this correction. In fact, due to the frequency dependence of this effect, deboosting is not important for the WMAP data. For the Planck analysis, 500 FFP6 simulations were analysed, while 10 000 Gaussian simulations were used to study the WMAP data.

| Method | Mask | $(l, b)$ [°] | $\theta_{\mathrm{mean}}^{\mathrm{data}}$ [°] | Frac. $\theta_{\mathrm{mean}}^{\mathrm{sim}} > \theta_{\mathrm{mean}}^{\mathrm{data}}$ | $\ell_{\mathrm{max}}$ |
|---|---|---|---|---|---|
| C-R . . . . . . . . . | U73 | $(231, -2)$ | 86.7 | 493/500 | 1500 |
| NILC . . . . . . . . | U73 | $(214, 4)$ | 85.7 | 500/500 | 1500 |
| SEVEM . . . . . . . | U73 | $(215, 9)$ | 85.8 | 500/500 | 1500 |
| SMICA . . . . . . . | U73 | $(220, 4)$ | 86.2 | 499/500 | 1500 |
| SEVEM-143 . . . . | U73 | $(228, -1)$ | 86.7 | 496/500 | 1500 |
| SEVEM-143DB . . | U73 | $(218, -21)$ | 87.6 | 487/500 | 1500 |
| WMAP-9 . . . . . | KQ85 | $(209, -24)$ | 82.3 | 9966/10000 | 600 |
| C-R . . . . . . . . . | U73 | $(192, -5)$ | 81.7 | 500/500 | 600 |
| NILC . . . . . . . . | U73 | $(194, -3)$ | 80.8 | 500/500 | 600 |
| SEVEM . . . . . . . | U73 | $(195, -2)$ | 81.2 | 500/500 | 600 |
| SMICA . . . . . . . | U73 | $(196, -2)$ | 81.9 | 500/500 | 600 |
| SEVEM-143 . . . . | U73 | $(195, 3)$ | 81.6 | 500/500 | 600 |
| SEVEM-143DB . . | U73 | $(189, -10)$ | 81.9 | 500/500 | 600 |

separate a Doppler dipolar modulation from a power asymmetry imprinted on the data.

In order to investigate the power asymmetry further, it is necessary to analyse data that have been corrected for the Doppler modulation effect. As shown in Planck Collaboration XXVII (2014), the modulation amplitude is frequency dependent. It is therefore difficult to correct for the effect in the component separated maps used in this paper, since they are constructed from frequency maps to which both scale-dependent and spatially varying weighting schemes are applied. The SEVEM approach, however, adopts a single weight per frequency. In what follows, we utilize a version of the 143 GHz data cleaned using this method (which we will refer to as SEVEM-143), with the weights summarized in Table C.1. of Planck Collaboration XII (2014). The Doppler modulation can then be removed using the prescription and weights derived in Planck Collaboration XXVII (2014). We will refer to this process as deboosting, even though a correction for aberration is not performed. Such a de-aberration is not needed since the FFP6 simulations used in this analysis contain the aberration effect, but not the Doppler modulation of the CMB anisotropies. However, in order to match the noise properties of the demodulated data, the noise component of each simulation is separately demodulated.

In Fig. 27 we show the dipole directions of the 100-multipole bins after the deboosting correction. As compared to Fig. 26, the dipole directions have now moved away from the Doppler dipole direction and cluster more around the *WMAP*-9 and low multipole asymmetry directions. The mean dipole direction for $\ell = 2$ to 1500 is now $(l, b) = (218°, -21°)$, displaced by about 25° compared to the non-deboosted directions in Table 25. Note that the mean asymmetry directions for the $\ell = 2$ to 600 range for the deboosted SEVEM-143 and *WMAP*-9 data differ by about 20°, most likely due to the differences in the masks used and noise properties.

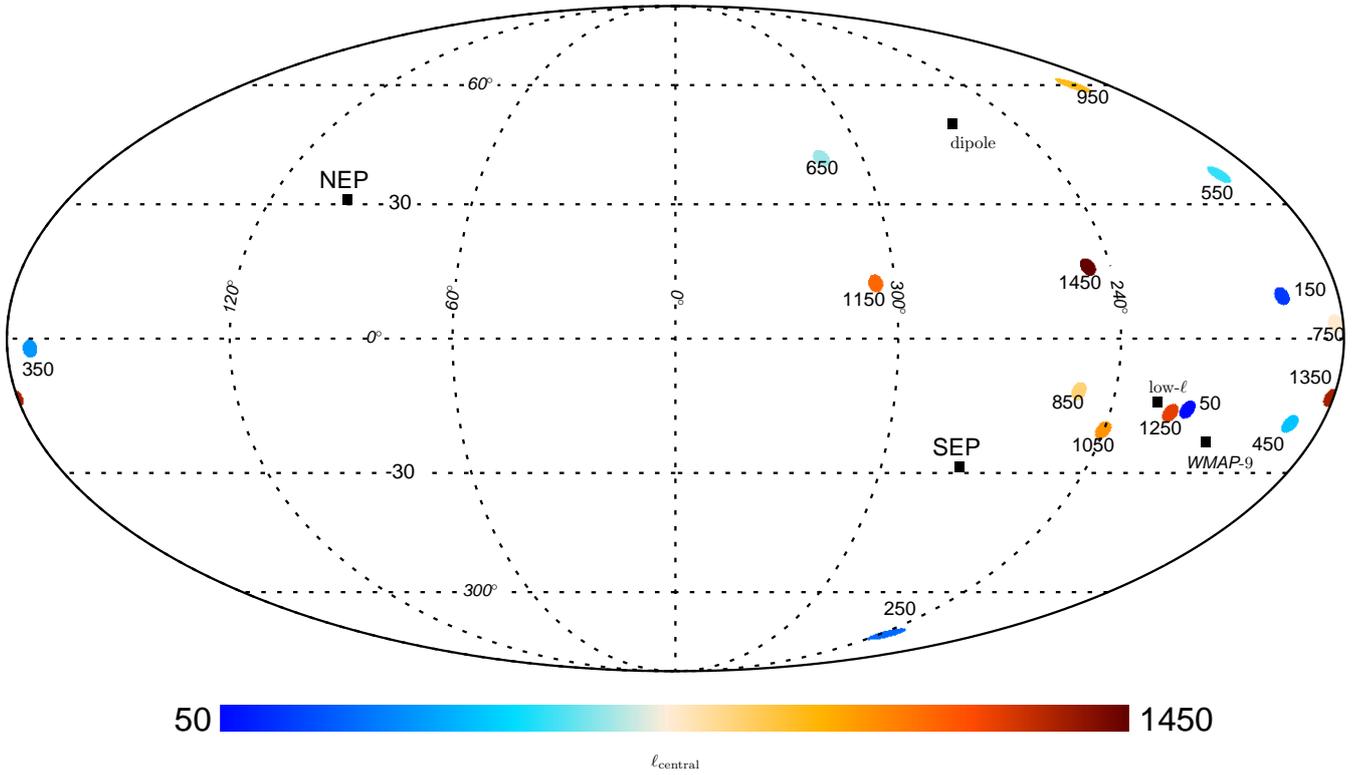

**Fig. 26.** Dipole directions for 100-multipole bins of the local power spectrum distribution from $\ell = 2$ to 1500 in the `SMICA` map with the apodized U73 mask applied. We also show the total direction for $\ell_{max} = 600$ determined from *WMAP*-9, as well as the preferred dipolar modulation axis (labelled as low-$\ell$) derived in Sect. 5.6.

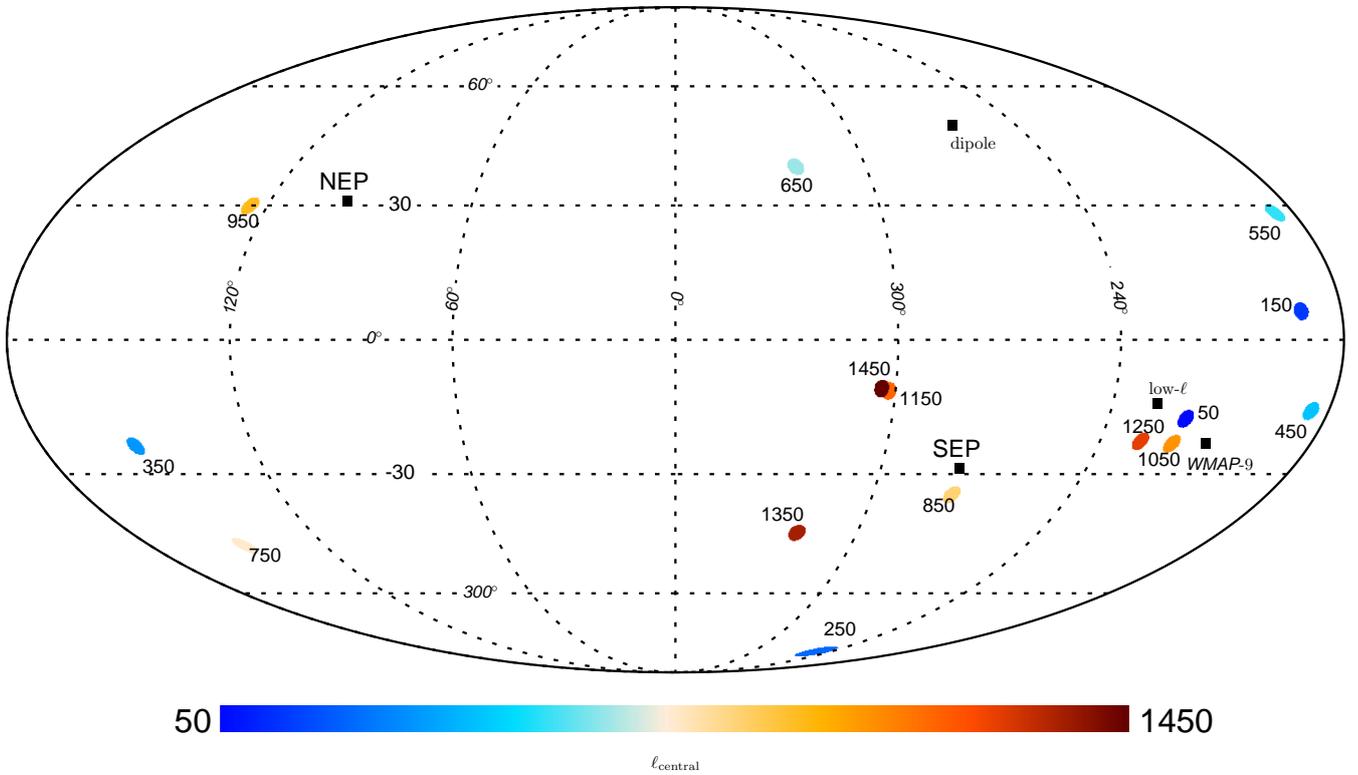

**Fig. 27.** Dipole directions for 100-multipole bins of the local power spectrum distribution from $\ell = 2$ to 1500 in the deboosted `SEVEM` 143 GHz map (`SEVEM-143DB`) with the apodized U73 mask applied. We also show the direction for $\ell_{max} = 600$ determined from *WMAP*-9, as well as the preferred dipolar modulation axis (labelled as low-$\ell$) derived in Sect. 5.6.

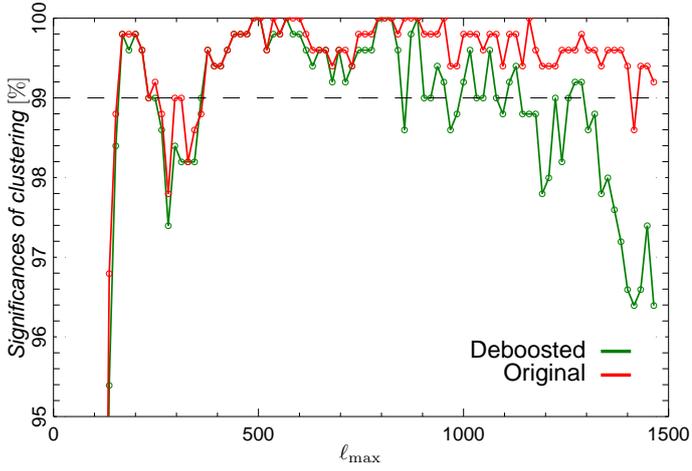

**Fig. 28.** Derived $p$-values as a function of $\ell_{max}$, for the SEVEM 143 GHz foreground-cleaned map, determined both before and after deboosting. The $p$-values are computed using 500 FFP6 simulated maps.

**Table 26.** Significance of the asymmetry using several global significance measures. The numbers in the table reflect the number of simulations that have clustering $p$-values of 100% and > 99% for less $\ell_{max}$ values than the data, as well as the number of simulations with lower mean $p$-values (as determined over all $\ell_{max}$) than the data. These values are calculated for the $\ell_{max}$ range 2–600 and 2–1500. The last two columns show the corresponding values when only multipoles $\ell > 100$ are considered in the analysis.

| Significance measure | [2, 600] | [2, 1500] | [100, 600] | [100, 1500] |
|---|---|---|---|---|
| No. of $p$=100% . . . . | 496/500 | 496/500 | 494/500 | 496/500 |
| No. of $p$ > 99% . . . . | 499/500 | 498/500 | 496/500 | 498/500 |
| Mean $p$-value . . . . . | 490/500 | 500/500 | 498/500 | 500/500 |

Fig. 28 presents the $p$-values (the fraction of simulations with a larger mean separation angle than that determined from the real sky) as a function of $\ell_{max}$. It is certainly the case that power asymmetry is observed to $\ell_{max} \sim 600$, even for deboosted data. Cursory inspection may lead to the conclusion that asymmetry at a significance level of 98–99% persists to at least $\ell_{max} = 1300$. However, this is a consequence of the cumulative nature of the statistic. Investigation of $\theta_{mean}$ for individual multipole ranges (see below) does not indicate the presence of asymmetry above $\ell_{max} \sim 600$. Similar conclusions were reached by Flender & Hotchkiss (2013) and Notari et al. (2013), although they only looked at the power ratio which is not directly related to the test performed here.

We have also attempted to assess the degree to which the significance of the asymmetry depends on a specific choice for

**Table 27.** Significance of the correlations of dipole directions between high and low multipoles. Significance levels are defined similarly to Table 26, but here $\theta_{mean}$ is defined as the mean of the angular distances between all pairs of dipole directions where one is computed for $\ell_{min} < \ell < \ell_{lim}$ and the second direction for $\ell_{lim} < \ell < \ell_{max}$. Note that the global significance levels are only based on $\ell_{max}$ values selected over the full range 2–1500, since high values of $\ell_{lim}$ yield poor statistics for the smaller range 2–600. The two last rows give significance levels for the case where we consider $\ell_{min} = 100$ in the analysis. In some cases the data do not have any $p = 100\%$ or $p > 99\%$ occurences in the given range. This is denoted by NA, as the given statistic is not applicable for these cases.

| $\ell_{lim}$ | No. of $p = 100\%$ | No. of $p > 99\%$ | Mean $p$-value |
|---|---|---|---|
| 100 . . . . . . . . . . . | 494/500 | 500/500 | 500/500 |
| 200 . . . . . . . . . . . | 499/500 | 500/500 | 500/500 |
| 300 . . . . . . . . . . . | 497/500 | 500/500 | 500/500 |
| 400 . . . . . . . . . . . | 496/500 | 500/500 | 500/500 |
| 500 . . . . . . . . . . . | NA | 473/500 | 489/500 |
| 600 . . . . . . . . . . . | NA | NA | 317/500 |
| 200 ($\ell_{min} = 100$) . . . | 500/500 | 500/500 | 500/500 |
| 300 ($\ell_{min} = 100$) . . . | 492/500 | 500/500 | 500/500 |
| 400 ($\ell_{min} = 100$) . . . | NA | 490/500 | 492/500 |
| 500 ($\ell_{min} = 100$) . . . | NA | 470/500 | 487/500 |
| 600 ($\ell_{min} = 100$) . . . | NA | NA | 267/500 |

$\ell_{max}$ by implementing several global statistics. Values for these are determined as follows.

1. The angle $\theta_{mean}$ is calculated as a function of $\ell_{max}$ both for the data and the simulations.
2. For a given $\ell_{max}$, the fraction of simulations (out of 500) with a larger mean separation angle than for the real data is defined as the $p$-value for that $\ell_{max}$.
3. This procedure is then repeated, treating each of the 500 simulations in turn as if it were the real data, and computing a corresponding $p$-value from the fraction of the other 499 simulations that gives a larger $\theta_{mean}$ than the selected simulation. Examples of such $p$-values as a function of $\ell_{max}$ for the data are shown in Figs. 28 and 29.
4. We define three global statistics in the following manner.
   - "number of $p = 100\%$": for certain values of $\ell_{max}$, each of the 500 simulations is found to have a larger mean separation angle than is determined for the real data, implying a corresponding $p$-value of 100%. We then evaluate the number of $\ell_{max}$ values for which this is the case. Each of the simulations is then treated in turn as if it were the real data, and effectively the same quantity is determined by comparison to the other 499 simulations. The fraction of all simulations that yield a smaller "number of $p = 100\%$" than the real data is then used as a global statistic, shown in the first row in Table 26.
   - "number of $p > 99\%$": for certain values of $\ell_{max}$, more than 99% of the simulations exhibit larger mean separation angles than the data, implying a $p$-value of > 99%. We count the number of $\ell_{max}$ values for which this applies. The same quantity is then evaluated for each simulation, treated in turn as if it were the real data (as above). The fraction of all simulations that yield a smaller "number of $p > 99\%$" than the real data is used as a second global statistic.

- "mean $p$-value": the mean $p$-value for the data is computed as an average over the set of $p$-values determined for each $\ell_{\max}$ (as in 2 above). A corresponding quantity is evaluated for each simulation, treated in turn as if it were the real data. The fraction of simulations with a lower "mean $p$-value" than for the data then forms the third global statistic.

We provide results for all three global statistics, since they reflect different aspects of the significance of the asymmetry. The fact that the global statistics indicate similar significance levels is evidence for the robustness of the results. We determine their values both for all $\ell_{\max}$ available ($\ell_{\max} = 2$–1500) and after restricting $\ell_{\max}$ to 2–600, since we do not claim any asymmetry beyond this value. The results are shown in Table 26. In addition, significance levels are also provided for the case when only multipoles $\ell > 100$ are included in the analysis. In this case, strong evidence (> 99%) is still found for asymmetry from all global measures.

We can also study the range of scales over which the asymmetry persists using these global statistics and a modified definition of $\theta_{\mathrm{mean}}$. Here, this is computed as the mean of the angular separations determined between all pairs of dipole directions where one direction corresponds to the range $\ell_{\min} < \ell < \ell_{\mathrm{lim}}$ and the second direction to $\ell_{\mathrm{lim}} < \ell < \ell_{\max}$. Figure 29 shows an example of the $p$-values corresponding to this definition of $\theta_{\mathrm{mean}}$ for $\ell_{\mathrm{lim}} = 500$. The high $p$-values indicated in the figure for multipoles in the range $\ell = 500$–600 show that the corresponding dipole directions are strongly correlated with the dipoles in the range $\ell = 2$–500. For most of the calculations, $\ell_{\min} = 2$ and $\ell_{\max} = 1500$. Table 27 presents the global significance levels for $\ell_{\mathrm{lim}} = 100, 200, 300, 400, 500$, and 600. We find that the dipole directions at high and low multipoles are strongly correlated up to $\ell = 600$, although the significance is weakening for $\ell > 500$. Indeed, when we consider multipoles $\ell > 600$ and $\ell_{\mathrm{lim}} > 600$, we find no evidence for asymmetry. The table also provides similar significance levels for the case $\ell_{\min} = 100$.

Although the global significance values in Table 27 show that the asymmetry is almost insignificant for $\ell_{\mathrm{lim}} = 500$, we must take into account the fact that the global statistics look at the entire range 2–1500. If the asymmetry disappears at, for example, $\ell = 600$, then the statistics would be too poor in the interval $\ell = 500$–600 for this to show up in the global significance measure for $\ell_{\mathrm{lim}} = 500$. For this reason we show in Fig. 29 the plot of $p$-values as a function of $\ell_{\max}$ for $\ell_{\mathrm{lim}} = 500$, showing that there is a strong signal all the way to $\ell_{\max} = 600$, after which it disappears. The continued evidence for asymmetry excludes the possibility that the results in Fig. 28 are dominated by the known low-$\ell$ signal, and supports the claim for asymmetry persisting over the range $\ell = 100$ to 600.

An alternative approach to studying power asymmetry is to determine the ratio of the local power spectra computed in two opposing directions on the sky (e.g., Eriksen et al. 2004a). Here, we consider such a ratio defined for the two hemispheres centred on the positive and negative poles of the power dipole fitted over a given $\ell$ range. A statistic can then be defined through the fractional power ratio as follows:

$$\frac{\Delta C_\ell}{C_\ell} = 2 \frac{C_\ell^{\mathrm{positive}} - C_\ell^{\mathrm{negative}}}{C_\ell^{\mathrm{positive}} + C_\ell^{\mathrm{negative}}}, \tag{32}$$

where $C_\ell^{\mathrm{positive}}$ corresponds to the power spectrum computed for the hemisphere centred on the positive pole, and $C_\ell^{\mathrm{negative}}$ to the spectrum in the antepodal direction. This can be compared to an ensemble of isotropic and Gaussian simulations to determine whether significant excess power is observed. Figure 30 presents this quantity, binned into blocks of 100 multipoles, for the hemispheres centred on the preferred dipole direction determined for the SEVEM 143 GHz map over the $\ell$ range 2 to 600, both before and after deboosting. It should be apparent that, although the ratio lies systematically above zero for $\ell < 600$, only a few bins lie significantly outside the range of values generated from simulations. The most significant bins are those centred on $\ell = 50$ and 150. Note that the observed values are not directly comparable to the explicit dipolar modulation fits in Sects. 5.6 and 5.7. The ratio on other scales is of smaller amplitude and lower significance. We re-emphasize that the claims of significant asymmetry presented in this section are based on the alignment of the power distribution as a function of angular scale, not on the corresponding amplitudes, nor on the ratio of power in the antipodal hemispheres. The ratio of power in the antipodal hemispheres is shown here for only illustrative purposes.

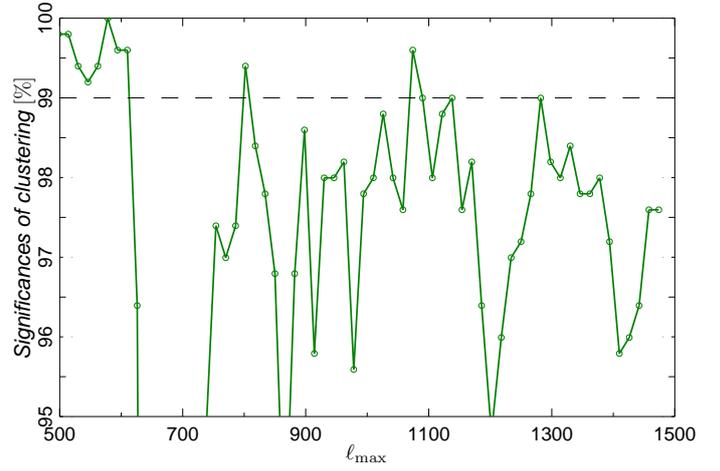

**Fig. 29.** Derived $p$-values as a function of $\ell_{\max}$ for the deboosted SEVEM 143 GHz foreground-cleaned map. The $p$-values in this plot are based on the the mean of the angular separations determined between all pairs of dipole directions where one direction falls in the range $[\ell_{\mathrm{lim}} = 500, \ell_{\max}]$ and the second direction in the range $[2, \ell_{\mathrm{lim}} = 500]$. The significance is computed using 500 FFP6 simulated maps.

In summary, we have presented evidence for power asymmetry in the *Planck* data. At high $\ell$, this is expected, since a dipolar modulation of the temperature anisotropy due to Doppler effects has been predicted, and subsequently detected, as detailed in a companion paper (Planck Collaboration XXVII 2014). When this is taken into account, significant power asymmetry can be claimed up to $\ell \sim 600$.

### 5.6. Dipole modulation

In Sect. 5.5 it was shown that power asymmetry is observed over a large range of angular scales in the *Planck* data, with a fairly consistent preferred axis. No explicit parametric model was assumed in the analysis. In this section, however, we only consider large angular scales and revisit the phenomenological model due to Gordon et al. (2005), who proposed that the power asymmetry could be described in terms of a multiplicative dipole modulation model of the form $\mathbf{d} = (1 + A\,\boldsymbol{p} \cdot \boldsymbol{n})\mathbf{s}_{\mathrm{iso}} + \mathbf{n} \equiv \mathbf{M}\mathbf{s}_{\mathrm{iso}} + \mathbf{n}$, where $A$ is the dipole amplitude, $\boldsymbol{p}$ is the dipole direction, $\mathbf{n}$ denotes in-strumental noise, and $\mathbf{s}_{\mathrm{iso}}$ is an underlying isotropic CMB field. Both $\mathbf{s}_{\mathrm{iso}}$ and $\mathbf{n}$ are assumed to be Gaussian random fields with covariance matrices $\mathsf{S}$ and $\mathsf{N}$, respectively. Since $\mathbf{s}_{\mathrm{iso}}$ is assumed to be isotropic, its covariance may be fully specified by some angular power spectrum $C_{\ell,\mathrm{iso}}$.

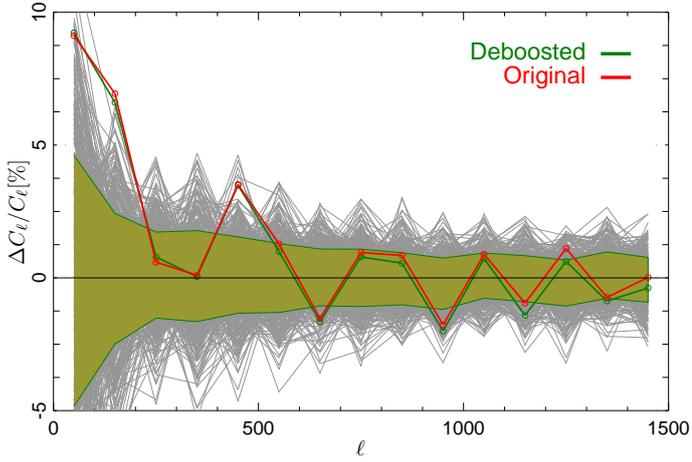

**Fig. 30.** Fractional power ratio, $\Delta C_\ell / C_\ell$, from antipodal sky regions, computed from the SEVEM 143 GHz map before and after deboosting along the mean dipole direction for $\ell = 2$ to 600. All spectra are evaluated on hemispheres using an apodized mask. The grey lines show the same quantity evaluated for each of the 500 FFP6 simulations along their respective asymmetry axes. The green band shows the 68% confidence region from these simulations.

In the following we present the results from a direct likelihood analysis of this model, similar to those described by Eriksen et al. (2007a) and Hoftuft et al. (2009) for the 3- and 5-year *WMAP* data, respectively. Since this method requires matrix inversions and determinant evaluations, the computational expense scales as $O(N_{\mathrm{pix}})$, and it is therefore only feasible at low resolution. Specifically, we consider maps downgraded to a `HEALPix` pixel resolution of $N_{\mathrm{pix}} = 32$, smoothed to angular resolutions ranging from 5° to 10°, ensuring sufficient bandwidth limitation at this pixelization. All four *Planck* CMB map solutions are included in the analysis; however, note that the Galactic plane is handled differently in each of the four approaches. Specifically: for the `Commander` map the region inside the corresponding analysis mask has been replaced with a Gaussian constrained realization, eliminating the possibility that bright Galactic foreground residuals might leak outside the mask during degradation (Planck Collaboration XV 2014); for `SMICA` and `NILC` a smaller region is replaced with Wiener filtered data; and for `SEVEM` no special precautions are taken.

After degrading each map to the appropriate resolution, we add random uniform Gaussian noise of 1 $\mu$K rms to each pixel to regularize the covariance matrix. All pixels inside the U73 mask are excluded, and we adopt the difference maps between the raw *Planck* LFI 30 GHz and HFI 353 GHz maps and the `SMICA` CMB solution as two foreground templates, tracing low- and high-frequency foregrounds, respecively. We marginalize over these Galactic foreground templates, $\boldsymbol{f}$, as well as four monopole and dipole templates, by adding corresponding terms of the form $\alpha \boldsymbol{f} \boldsymbol{f}^{\mathsf{T}}$ to the total data covariance matrix, where $\alpha$ is set to a numerically large value.

Before writing down the likelihood for $A$ and $\boldsymbol{p}$, a choice has to be made for the power spectrum, $C_{\ell,\mathrm{iso}}$. We follow Eriksen et al. (2007a), and adopt a simple two-parameter model of the form $C_{\ell,\mathrm{iso}} = q \left(\ell/\ell_{\mathrm{pivot}}\right)^n C_{\ell,\mathrm{fid}}$, where the fiducial spectrum, $C_{\ell,\mathrm{fid}}$, is the best-fit *Planck* spectrum, and $q$ and $n$ describe an amplitude scaling and spectral tilt with respect to this. The full model therefore includes five free parameters, namely three dipole parameters and two power spectrum parameters. Introducing the two parameters $q$ and $n$ addresses the known issue that the low-$\ell$ power spectrum is low by about 2–2.5 $\sigma$ compared with the overall best-fit $\Lambda$CDM spectrum (Planck Collaboration XV 2014; Planck Collaboration XVI 2014). Ignoring this creates a tension with the underlying isotropic model that results in the analysis measuring a combination of both asymmetry and power mismatch.

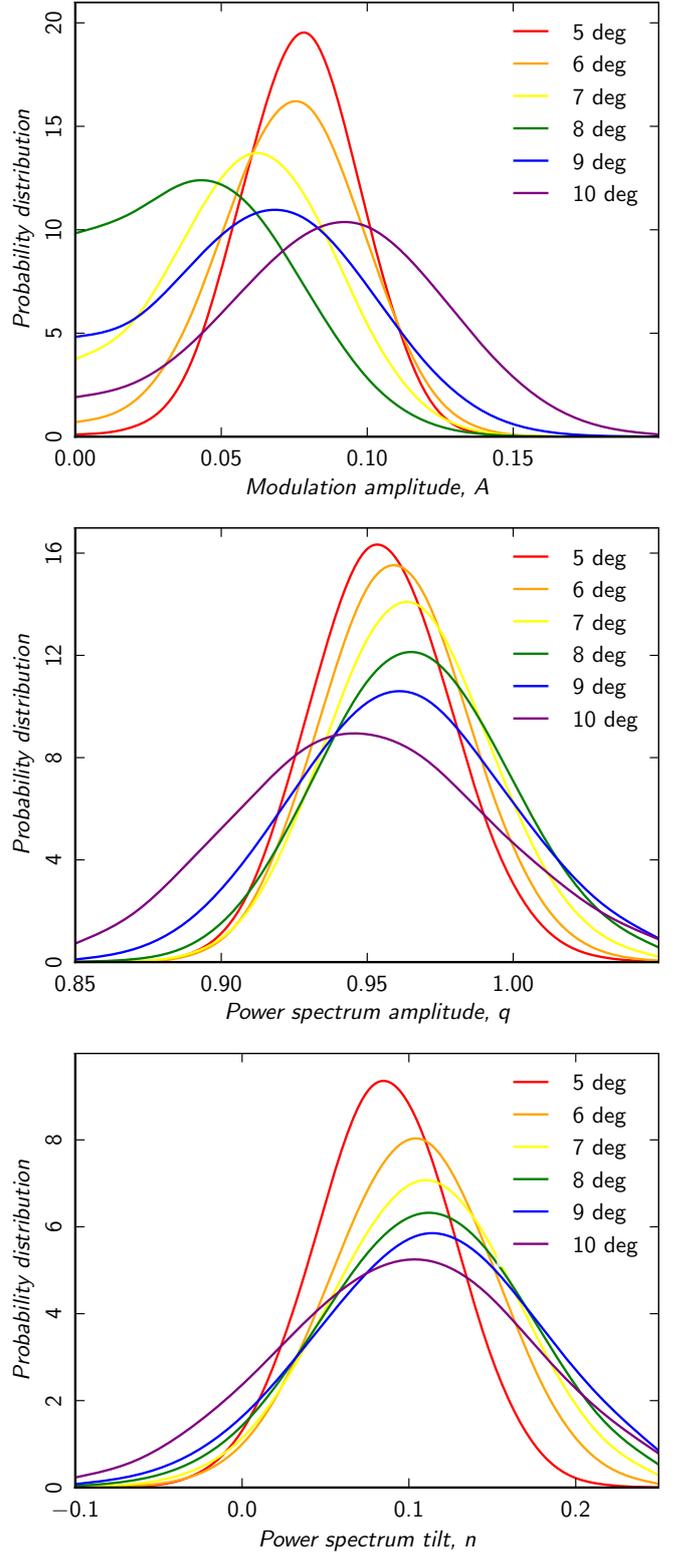

**Fig. 31.** Marginalized dipole modulation amplitude (top), power spectrum amplitude (middle) and power spectrum tilt (bottom) probability distributions as a function of smoothing scale, shown for the `Commander` CMB solution.

**Table 28.** Summary of dipole modulation likelihood results as a function of scale for all four *Planck* CMB solutions.

| Data set | FWHM [°] | $A$ | $(l, b)$ [°] | $\Delta \ln \mathcal{L}$ | Significance |
|---|---|---|---|---|---|
| Commander . . . . . . . . . . . . . . . | 5.0 | $0.078^{+0.020}_{-0.021}$ | $(227, -15) \pm 19$ | 8.8 | $3.5\,\sigma$ |
| NILC . . . . . . . . . . . . . . . . . . . . . | 5.0 | $0.069^{+0.020}_{-0.021}$ | $(226, -16) \pm 22$ | 7.1 | $3.0\,\sigma$ |
| SEVEM . . . . . . . . . . . . . . . . . . . . | 5.0 | $0.066^{+0.021}_{-0.021}$ | $(227, -16) \pm 24$ | 6.7 | $2.9\,\sigma$ |
| SMICA . . . . . . . . . . . . . . . . . . . . | 5.0 | $0.065^{+0.021}_{-0.021}$ | $(226, -17) \pm 24$ | 6.6 | $2.9\,\sigma$ |
| *WMAP*-5 ILC . . . . . . . . . . . . . . | 4.5 | $0.072^{+0.021}_{-0.021}$ | $(224, -22) \pm 24$ | 7.3 | $3.3\,\sigma$ |
| Commander . . . . . . . . . . . . . . . | 6.0 | $0.076^{+0.024}_{-0.025}$ | $(223, -16) \pm 25$ | 6.4 | $2.8\,\sigma$ |
| NILC . . . . . . . . . . . . . . . . . . . . . | 6.0 | $0.062^{+0.025}_{-0.026}$ | $(223, -19) \pm 38$ | 4.7 | $2.3\,\sigma$ |
| SEVEM . . . . . . . . . . . . . . . . . . . . | 6.0 | $0.060^{+0.025}_{-0.026}$ | $(225, -19) \pm 40$ | 4.6 | $2.2\,\sigma$ |
| SMICA . . . . . . . . . . . . . . . . . . . . | 6.0 | $0.058^{+0.025}_{-0.027}$ | $(223, -21) \pm 43$ | 4.2 | $2.1\,\sigma$ |
| Commander . . . . . . . . . . . . . . . | 7.0 | $0.062^{+0.028}_{-0.030}$ | $(223, \quad-8) \pm 45$ | 4.0 | $2.0\,\sigma$ |
| NILC . . . . . . . . . . . . . . . . . . . . . | 7.0 | $0.055^{+0.029}_{-0.030}$ | $(225, -10) \pm 53$ | 3.4 | $1.7\,\sigma$ |
| SEVEM . . . . . . . . . . . . . . . . . . . . | 7.0 | $0.055^{+0.029}_{-0.030}$ | $(226, -10) \pm 54$ | 3.3 | $1.7\,\sigma$ |
| SMICA . . . . . . . . . . . . . . . . . . . . | 7.0 | $0.048^{+0.029}_{-0.029}$ | $(226, -11) \pm 58$ | 2.8 | $1.5\,\sigma$ |
| Commander . . . . . . . . . . . . . . . | 8.0 | $0.043^{+0.032}_{-0.029}$ | $(218, -15) \pm 62$ | 2.1 | $1.2\,\sigma$ |
| NILC . . . . . . . . . . . . . . . . . . . . . | 8.0 | $0.049^{+0.032}_{-0.031}$ | $(223, -16) \pm 59$ | 2.5 | $1.4\,\sigma$ |
| SEVEM . . . . . . . . . . . . . . . . . . . . | 8.0 | $0.050^{+0.032}_{-0.031}$ | $(223, -15) \pm 60$ | 2.5 | $1.4\,\sigma$ |
| SMICA . . . . . . . . . . . . . . . . . . . . | 8.0 | $0.041^{+0.032}_{-0.029}$ | $(225, -16) \pm 63$ | 2.0 | $1.1\,\sigma$ |
| Commander . . . . . . . . . . . . . . . | 9.0 | $0.068^{+0.035}_{-0.037}$ | $(210, -24) \pm 52$ | 3.3 | $1.7\,\sigma$ |
| NILC . . . . . . . . . . . . . . . . . . . . . | 9.0 | $0.076^{+0.035}_{-0.037}$ | $(216, -25) \pm 45$ | 3.9 | $1.9\,\sigma$ |
| SEVEM . . . . . . . . . . . . . . . . . . . . | 9.0 | $0.078^{+0.035}_{-0.037}$ | $(215, -24) \pm 43$ | 4.0 | $2.0\,\sigma$ |
| SMICA . . . . . . . . . . . . . . . . . . . . | 9.0 | $0.070^{+0.035}_{-0.037}$ | $(216, -25) \pm 50$ | 3.4 | $1.8\,\sigma$ |
| *WMAP*-3 ILC . . . . . . . . . . . . . . | 9.0 | 0.114 | $(225, -27)$ | 6.1 | $2.8\,\sigma$ |
| Commander . . . . . . . . . . . . . . . | 10.0 | $0.092^{+0.037}_{-0.040}$ | $(215, -29) \pm 38$ | 4.5 | $2.2\,\sigma$ |
| NILC . . . . . . . . . . . . . . . . . . . . . | 10.0 | $0.098^{+0.037}_{-0.039}$ | $(217, -29) \pm 33$ | 5.0 | $2.3\,\sigma$ |
| SEVEM . . . . . . . . . . . . . . . . . . . . | 10.0 | $0.103^{+0.037}_{-0.039}$ | $(217, -28) \pm 30$ | 5.4 | $2.5\,\sigma$ |
| SMICA . . . . . . . . . . . . . . . . . . . . | 10.0 | $0.094^{+0.037}_{-0.040}$ | $(218, -29) \pm 37$ | 4.6 | $2.2\,\sigma$ |

Taking advantage of the fact that both the signal and noise are assumed Gaussian, the exact likelihood may be written down in a convenient closed form:

$$\mathcal{L}(A, \boldsymbol{p}, q, n) \propto \frac{e^{-\frac{1}{2}\boldsymbol{d}^{\mathrm{T}}\left(\mathsf{M}^{\mathrm{T}}\mathsf{SM}+\mathsf{N}+\alpha\sum_i \boldsymbol{f}_i \boldsymbol{f}_i^{\mathrm{T}}\right)^{-1}\boldsymbol{d}}}{\sqrt{\left|\mathsf{M}^{\mathrm{T}}\mathsf{SM}+\mathsf{N}+\alpha\sum_i \boldsymbol{f}_i \boldsymbol{f}_i^{\mathrm{T}}\right|}}. \tag{33}$$

This expression is the basis of all calculations presented in the rest of this section.

Due to the high computational expense associated with these evaluations, we do not compute the full joint five-parameter model in this analysis, only conditionals of it. However, we iterate once in a Gibbs-sampling like approach, by maximizing each conditional to obtain an approximation to the full maximum-likelihood solution. That is, we first map out the dipole likelihood for the 5° FWHM case, fixing the power spectrum at the fiducial spectrum, $\mathcal{L}(A, \boldsymbol{p}|q = 1, n = 0)$, and locate the maximum-likelihood dipole parameters. Then we map out the corresponding power spectrum conditional, $\mathcal{L}(q, n|A_{m\ell}, \boldsymbol{p}_{m\ell})$. Finally, we update the dipole likelihood with these power spectrum parameters, and evaluate the final results. Note that the power spectrum and dipole modulation parameters are only weakly correlated, and this procedure is therefore close to optimal. Furthermore, the approach is also conservative, in the sense that it will always underestimate the significance of the dipole modulation model; the derived maximum-likelihood value will always lie slightly below the true maximum-likelihood point.

The results from these calculations are summarized in Table 28, listing results for all four *Planck* CMB maps at angular scales between 5 and 10° FWHM. For easy reference, we also list the results from the corresponding 3- and 5-year *WMAP* analyses (Eriksen et al. 2007a; Hoftuft et al. 2009). Note that the former was performed at a `HEALPix` resolution of $N_{\text{side}} = 16$ and the latter at an angular resolution of 4.5° FWHM.

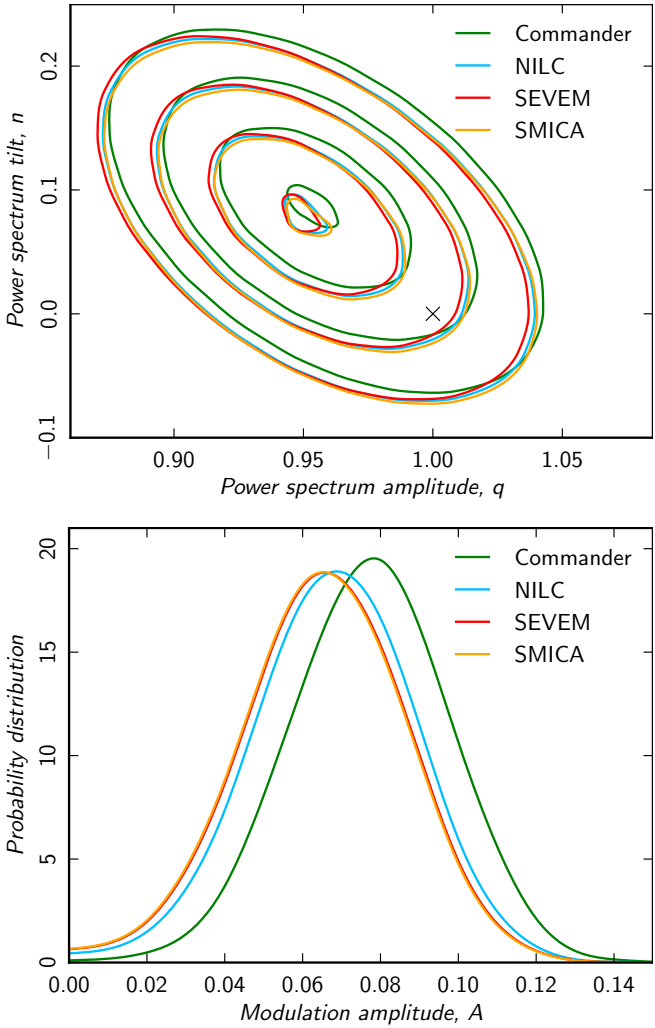

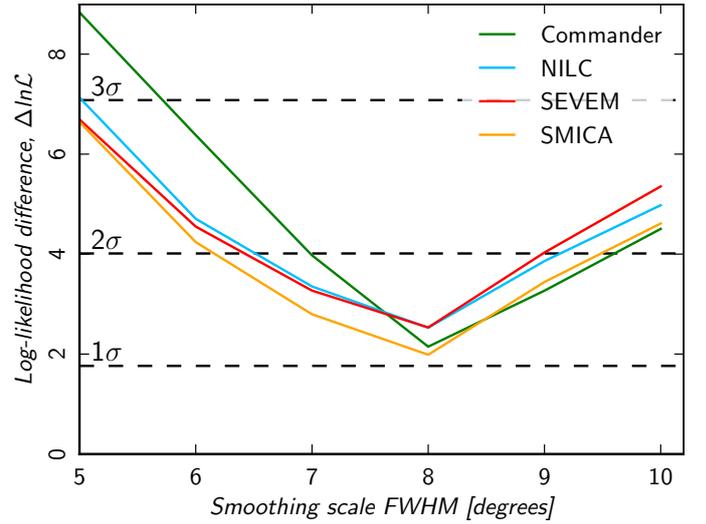

**Fig. 33.** Log-likelihood difference between the best-fit dipole modulation model and the fiducial isotropic model as a function of smoothing scale. Horizontal dashed lines indicate 1, 2, and 3 σ thresholds.

**Fig. 32.** Consistency between component separation algorithms as measured by the dipole modulation likelihood. The top panel shows the marginal power spectrum amplitude for the 5° smoothing scale, the middle panel shows the dipole modulation amplitude, and the bottom panel shows the preferred dipole directions. The coloured area indicates the 95% confidence region for the `Commander` solution, while the dots shows the maximum-posterior directions for the other maps.

Fig. 31 shows marginals for $A$, $q$ and $n$, as derived from the `Commander` CMB solution for all smoothing scales. At least two interesting points can be seen here. First, while there is clearly significant scatter in the derived dipole modulation amplitude for different smoothing scales, as originally pointed out by Hanson & Lewis (2009), all curves appear to be consistent with a single value of $A \sim 0.07$. No other single value fits all scales equally well. Second, it is interesting to note that the low-$\ell$ power spectrum derived here is consistent, but not without some tension, with the fiducial spectrum, $(q, n) = (1, 0)$, around 1.5–2 σ. In particular, there appears to be a slight trend toward a steeper positive spectral index as more weight is put on the larger scales, a result already noted by *COBE*-DMR. The same conclusion is reached using the low-$\ell$ *Planck* likelihood, as described in Planck Collaboration XV (2014).

In Fig. 32 we compare the results from all four CMB solutions for the 5° FWHM smoothing scale. Clearly the results are consistent, despite the use of different algorithms and different treatments of the Galactic plane, demonstrating robustness with respect to the details of the analysis methods. Further, we also note that these results are consistent with those derived from the 5-year *WMAP* ILC map by Eriksen et al. (2007a), demonstrating robustness across experiments. On the other hand, it is notable that a higher dipole amplitude was found for the 3-year *WMAP* ILC map using a larger mask at 9° FWHM.

In Fig. 33 we show the log-likelihood difference between the derived maximum-likelihood point and the isotropic model, $A = 0$, as a function of smoothing scale. The power spectrum parameters are kept fixed at the best-fit values for both points, leaving three additional parameters for the dipole model. The dashed horizontal lines indicate the 1, 2, and 3 σ confidence regions for three degrees of freedom. As has been noted previously in the literature, these significance levels vary with smoothing scale. Taken at face value, the results presented here are suggestive but clearly not decisive, resulting in an unchanged situation with respect to earlier reports. This is of course not unexpected, given that *WMAP* is already strongly cosmic variance limited at these angular scales.

The critical question is whether the trend seen at smaller angular scales in Fig. 33 continues, or if the apparent likelihood peak at 5° FWHM happens to be a local maximum. Hanson & Lewis (2009), and later Bennett et al. (2011), address this question through a computationally cheaper quadratic estimator, al-lowing them to extend a similar analysis to small scales. In doing so, they claim that the apparent likelihood peak at 5° is indeed a local maximum, and the evidence for the modulation model falls off when more data are included.

In this respect, it should be noted that the dipole modulation model was originally proposed by Gordon et al. (2005) as a simple phenomenological characterization of the more general power asymmetry. In particular, it assumes that the modulation amplitude, $A$, is equally strong on all scales. From both the results presented by Hanson & Lewis (2009) and Bennett et al. (2011) and qualitatively shown in Fig. 30, this appears not to hold, as the fractional hemispherical power difference is clearly smaller at $\ell > 300$ than at $\ell < 100$. Furthermore, an alternative power modulation analysis in Planck Collaboration XXIV (2014) finds that the asymmetry on these scales has an amplitude of order 1%. The data, therefore, are not consistent with a simple constant-amplitude dipole modulation of the power. On the other hand, the preferred direction derived from the current low-$\ell$ analysis is remarkably consistent with the high-$\ell$ direction derived in Sect. 5.5. A proper modulation model may therefore need additional spatial structure beyond that proposed by Gordon et al. (2005), as already suggested by Hoftuft et al. (2009) and Moss et al. (2011).

Finally, Doppler boosting as presented in Planck Collaboration XXVII (2014) creates a signature similar to that observed here. However, the present effect is clearly distinct from this, both because the magnitudes of the two effects are very different — $A_{\text{boost}} \sim 0.002$ versus $A_{\text{asym}} \sim 0.07$ — and because the two preferred directions are different — $(l, b)_{\text{boost}} = (264°, 48°)$ versus $(l, b)_{\text{asym}} = (227°, -15°)$. We have analysed the 143 and 217 GHz frequency maps, cleaned for foregrounds using the `SEVEM` algorithm and de-boosted using the best-fit parameters from Planck Collaboration XXVII (2014), and consistently find (very slightly) higher significance levels for these data. The best-fit axis moves by approximately 5° and 2° at 143 and 217 GHz, respectively. The results presented here, therefore, are conservative with respect to boosting corrections.

### 5.7. Generalized modulation

In this section, we study a generalization of the dipolar modulation field analysed in Sect. 5.6 using the Bipolar Spherical Harmonic (BipoSH) formalism. For a statistically isotropic sky, the spherical harmonic space two-point correlation matrix is diagonal, and, given by the angular power spectrum $C_\ell$. The BipoSH representation provides a natural, mathematically complete, generalization of the angular power spectrum that captures statistical isotropy violations via coefficients that are a completely equivalent representation of the spherical harmonic correlation matrix,

$$A_{\ell_1 \ell_2}^{LM} = \sum_{m_1 m_2} \langle a_{\ell_1 m_1} a_{\ell_2 m_2} \rangle C_{\ell_1 m_1 \ell_2 m_2}^{LM}. \quad (34)$$

This relationship combines the off-diagonal spherical harmonic correlations into a bipolar multipole $L, M$ — analogous to the total angular momentum addition of states. The CMB angular power spectrum corresponds to the $L = 0$ BipoSH coefficients $C_\ell = (-1)^\ell A_{\ell\ell}^{00} \delta_{\ell\ell'} / \sqrt{2\ell + 1}$.

A simple model that results in the violation of statistical isotropy arises from the modulation of the of the CMB sky,

$$T(\mathbf{n}) = T_0(\mathbf{n})(1 + M(\mathbf{n})), \quad (35)$$

where $T(\mathbf{n})$ represents the modulated CMB sky, $T_0(\mathbf{n})$ is the underlying (statistically isotropic) random CMB sky and $M(\mathbf{n})$ is a fixed, zero-mean, dimensionless, modulation field. The modulation signal, if any, is expected to be weak and allows quadratic terms in $M$ to be neglected. The BipoSH coefficients for the modulated CMB field ($L > 0$) are then given by

$$A_{\ell_1 \ell_2}^{LM} = \bar{A}_{\ell_1 \ell_2}^{LM} + m_{LM} G_{\ell_1 \ell_2}^{L};$$

$$G_{\ell_1 \ell_2}^{L} = \frac{C_{\ell_1} + C_{\ell_2}}{\sqrt{4\pi}} \sqrt{\frac{(2\ell_1 + 1)(2\ell_2 + 1)}{(2L + 1)}} C_{\ell_1 0 \ell_2 0}^{L0}. \quad (36)$$

**Table 29.** Amplitude ($A$) and direction of the dipole modulation in Galactic coordinates. The measured values of the dipole amplitude and direction are consistent for all maps. The corresponding dipole power for the `SMICA` map is seen at a detection significance of $3.7\,\sigma$, as shown in Fig. 34. For the values in the third column ($\sigma_l = 15.4, \sigma_b = 15.1$).

| Map | $A$ | $(l, b)$ [°] |
|---|---|---|
| `C-R` . . . . . . . | $0.072^{+0.010}_{-0.010}$ | $(218.9, -21.4)$ |
| `NILC` . . . . . . | $0.070^{+0.010}_{-0.010}$ | $(220.3, -20.2)$ |
| `SEVEM` . . . . . | $0.065^{+0.011}_{-0.010}$ | $(221.7, -21.4)$ |
| `SMICA` . . . . . . | $0.073^{+0.010}_{-0.010}$ | $(217.5, -20.2)$ |

Here $\bar{A}_{\ell_1 \ell_2}^{LM}$ corresponds to the BipoSH coefficients of the unknown, but statistically isotropic, unmodulated CMB field, $m_{LM}$ are the spherical harmonic coefficients of the modulating field ($L > 0$), and $C_\ell$ is the best-fit CMB angular power spectrum. The statistically isotropic nature of the unmodulated CMB sky implies that the expectation values of $\bar{A}_{\ell_1 \ell_2}^{LM}$ vanish for $L > 0$, leading to the estimator for the modulation field harmonics,

$$\hat{m}_{LM} = \sum_{\ell_1 \ell_2} w_{\ell_1 \ell_2}^{L} \frac{\hat{A}_{\ell_1 \ell_2}^{LM}}{G_{\ell_1 \ell_2}^{L}}, \quad (37)$$

denoted by the overhat (Hanson & Lewis 2009). The weights $w_{\ell_1 \ell_2}^{L}$ for a minimum variance estimate for the modulation field correspond to

$$w_{\ell_1 \ell_2}^{L} = N \left( \frac{G_{\ell_1 \ell_2}^{L}}{\sigma_{A_{\ell_1 \ell_2}^{LM}}} \right)^2, \quad (38)$$

where $N$ is a normalization chosen such that $\sum_{\ell_1 \ell_2} w_{\ell_1 \ell_2}^{L} = 1$. The BipoSH representation further enables an estimate of the modulation field to be made over specific angular scales by windowing regions in multipole space in the sum over multipoles $\ell_1, \ell_2$ in Eq. 37. This additional information could be very useful in identifying the origin of the statistical isotropy violation, which could be either cosmological or due to systematic artefacts (see Hajian & Souradeep 2003; Hajian & Souradeep 2006).

First, we limit our analysis to the four low resolution $N_{\text{side}} = 32$ CMB maps used in Sect.5.6 and reconstruct the modulation maps for each of them at the same low resolution. The U73 mask is applied to the reconstructed modulation maps before computing $m_{LM}$. The pseudo-power $m_L$ is corrected for the mask applied to the modulation maps. Specifically, for the case of dipole modulation, the pseudo-power $m_L$ is related to the dipole amplitude by $A = 1.5\sqrt{m_1/\pi}$.

A dipole modulation ($L = 1$) signal is detected at $3\,\sigma$ significance in all the maps, as shown in Fig. 34. The amplitude and direction of the dipole modulation match those obtained via a likelihood analysis in Sect. 5.6. The BipoSH representation of modulation confirms the dipole modulation signal found in the low-resolution map. Since this approach enables the reconstruction of any general small amplitude modulation field, the BipoSH representation places constraints on the power in the modulation field at all higher (bipolar) multipoles allowed by the resolution of the CMB maps.

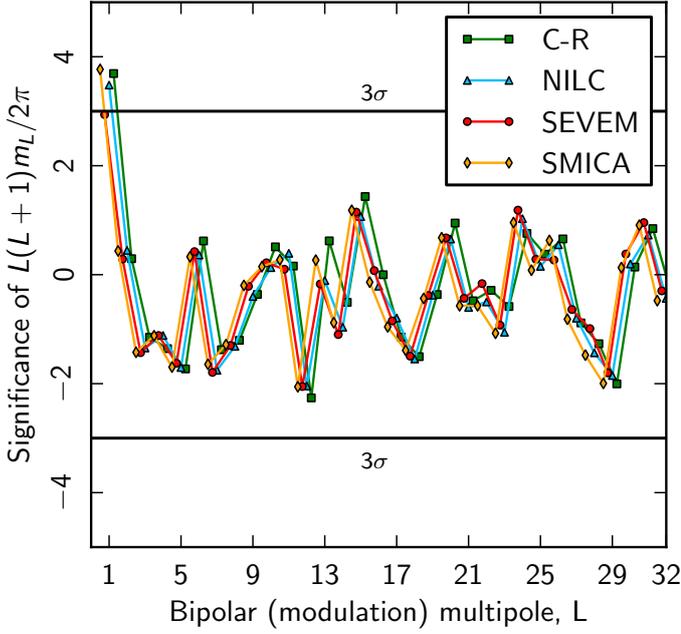

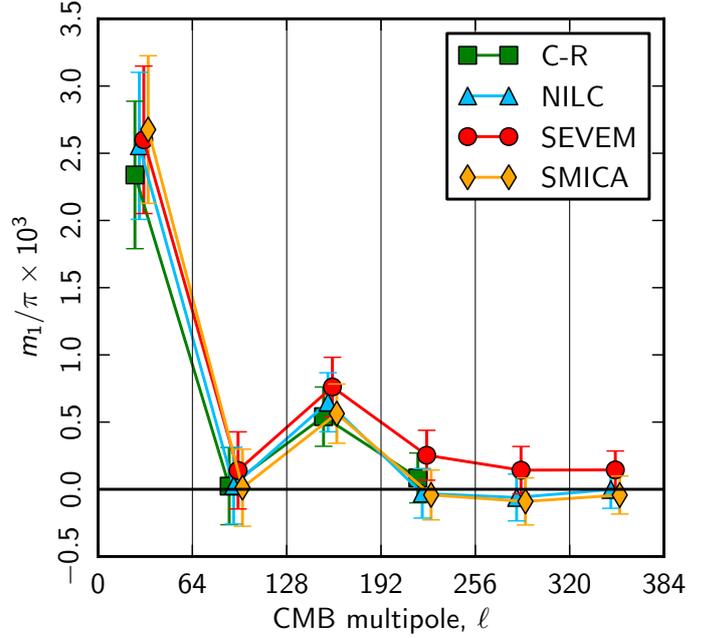

**Fig. 34.** Significance of the modulation power, $L(L + 1)m_L/2\pi$, at bipolar multipoles $L$. The modulation spectra obtained from the four component separation maps (C-R, NILC, SEVEM, and SMICA) are consistent with each other. Dipole ($L = 1$) modulation power is detected in all the spectra at a significance ranging from 2.9 to 3.7 $\sigma$. The solid black lines denote the $3\,\sigma$ significance thresholds. There is no significant power detected at higher multipole of the modulation field, $1 < L \leq 32$.

**Fig. 35.** Measured dipole modulation ($L = 1$) power in CMB multipole bins. This uses the CMB multipole dependence of the BipoSH (modulation) power $L(L + 1)m_L/2\pi$, which can be dissected into bins in $\ell$-space. We establish that significant power in the dipole modulation is limited to $\ell = 2–64$ and does not extend to the higher CMB multipoles considered. The vertical grid lines denote the CMB multipole $\ell$-bins.

We then extend the analysis to higher resolution using maps at $N_{\text{side}} = 256$ for `Commander` and $N_{\text{side}} = 2048$ for `NILC`, `SEVEM`, and `SMICA`, in order to study the above effect in more detail. We repeat the analysis on these higher resolution maps using the U73 mask. Contrary to our expectations based on a scale-independent (i.e., no $\ell$-dependence) model, the significance of the dipole does not increase in the high resolution maps. We then subdivide the $\ell$-range up to $\ell_{\text{max}} = 384$ into uniform bins of size $\Delta\ell = 64$. As seen in Fig. 35, we recover the dipole modulation at over 3 $\sigma$ significance only for the lowest bin ($\ell = 2–64$). This is consistent with the results in Sect. 5.6 and the BipoSH analysis on the corresponding low resolution maps shown in Fig. 34. However, the amplitude of the dipole is consistent with zero within 3 $\sigma$ for all of the higher $\ell$-bins considered. This suggests that the simple modulation model in Eq. 35 is inadequate and should minimally allow for the amplitude, $A(\ell)$, of the dipole to depend on CMB multipole, $\ell$. Although this may appear to be a more complex model, it does not necessarily lack motivation. It is readily conceivable that physical mechanisms that cause a dipolar modulation of the random CMB sky would be scale-dependent and possibly significant only at low wavenumbers. More importantly, such a dipole modulation has also been noted in low resolution *WMAP* data (Eriksen et al. 2007a; Hoftuft et al. 2009). More recently, Bennett et al. (2011) also comment (with-out being quantitative) that the effect is present in the *WMAP* maps, but limited to low $\ell$ and conclude that the $\ell$-dependence rules out a simple modulation explanation. The fact that two independent experiments find this intriguing statistical isotropy violation points to a non-instrumental origin.

The search for modulation power recovered from higher multipoles of the CMB maps yields a null result, as seen in Fig. 35. However, due to our motion with respect to the CMB rest frame, it is expected that the observed CMB map is statistically anisotropic, and this has been demonstrated in Planck Collaboration XXVII (2014). To understand why a Doppler boost induced anisotropy is not detected by the modulation estimator, we first implement an equivalent description in terms of the Doppler boost induced BipoSH coefficients,

$$A^{1M}_{\ell_1\ell_2} = \bar{A}^{1M}_{\ell_1\ell_2} + \beta_{1M} G^1_{\ell_1\ell_2},$$ (39)

$$G^1_{\ell_1\ell_2} = \left\{ b_\nu [G^1_{\ell_1\ell_2}]^{\mathcal{M}} - [G^1_{\ell_1\ell_2}]^\phi \right\} \times$$
$$\sqrt{\frac{(2\ell_1 + 1)(2\ell_2 + 1)}{12\pi}} C^{10}_{\ell_1 0 \ell_2 0},$$

$$[G^1_{\ell_1\ell_2}]^{\mathcal{M}} = [C_{\ell_1} + C_{\ell_2}],$$

$$[G^1_{\ell_1\ell_2}]^\phi = [C_{\ell_1} + C_{\ell_2}]$$
$$+ [C_{\ell_1} - C_{\ell_2}] [\ell_1(\ell_1 + 1) - \ell_2(\ell_2 + 1)]/2,$$

where $\beta_{1M} = \int d\boldsymbol{n} Y_{1M}(\boldsymbol{n}) \boldsymbol{\beta} \cdot \boldsymbol{n}$, $\boldsymbol{\beta} = \boldsymbol{v}/c$ denotes the peculiar velocity of our local rest frame with respect to the CMB, and $b_\nu$ is the frequency dependent boost factor, as discussed in more detail in Planck Collaboration XXVII (2014). Since we perform our analysis on component-separated maps, we use the same approximation, $b_\nu = 2.5$, as advocated there.

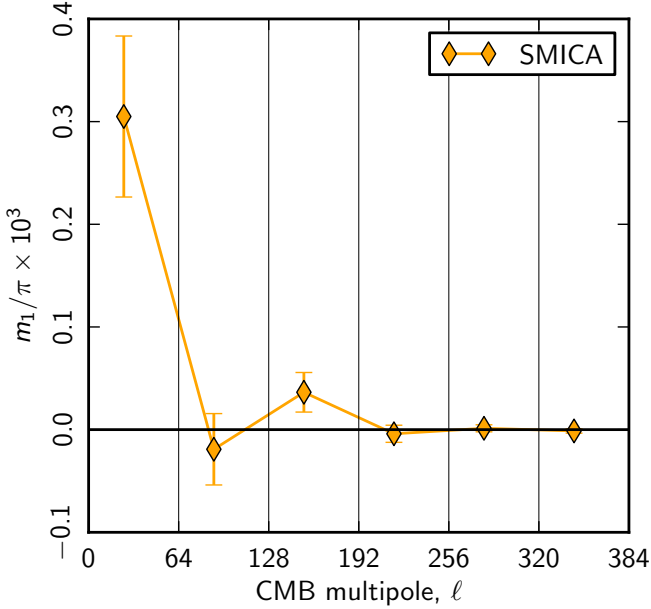

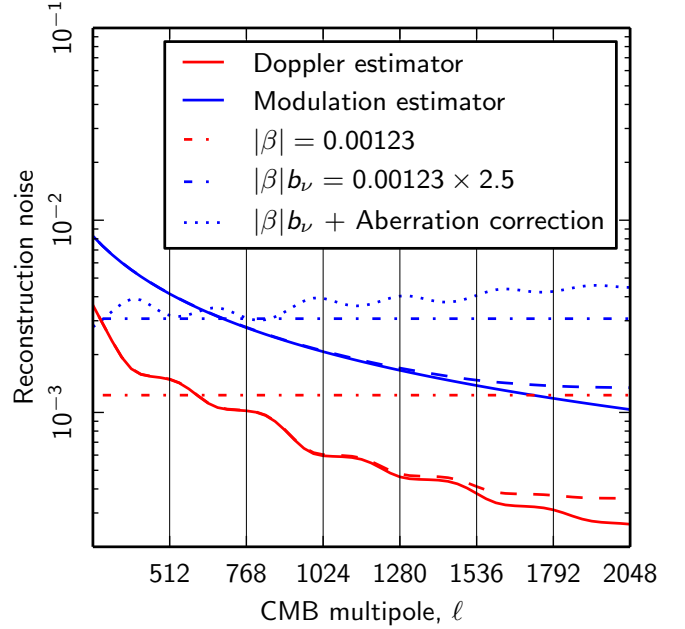

**Fig. 36.** Dipolar ($L = 1$) power measured from different CMB multipole bins, as denoted by the vertical grid lines. Here the Doppler estimator is used to reconstruct the Doppler field ($\boldsymbol{\beta} \cdot \hat{n}$) from SMICA. Significant dipolar power is limited to $\ell = 2$–$64$ and does not extend to the higher CMB multipoles considered.

**Fig. 37.** Reconstruction noise power associated with the modulation (blue) and Doppler (red) estimators as a function of the maximum CMB multipole used. The solid lines correspond to the case of noise-free CMB maps, while the dashed lines correspond to the case where the characteristic *Planck* instrument noise is taken into account. The red dash-dotted horizontal line denotes the expected signal level of the boost due to our local motion, while the blue dash-dotted horizontal line denotes the enhanced boost amplitude as seen by the modulation estimator. The blue dotted line denotes the effective Doppler boost signal as seen by the modulation estimator after accounting for the effects of aberration induced by the boost. Also note that the reconstruction errors shown in red are derived for the total dipole power, rather than the amplitude in a particular direction, and hence are higher by a factor of $\sqrt{3}$ as compared to the errors for individual directions reported in Planck Collaboration XXVII (2014).

A minimum variance estimator for $\beta_{1M}$ is given by the expression in Eq. 37, with the shape function $G^L_{\ell_1\ell_2}$ replaced by the corresponding Doppler boost term given in Eq. 39.

The significant power detected in the reconstructed Doppler field ($\boldsymbol{\beta} \cdot \hat{n}$), from the CMB multipole bin $\ell = 2$ to 64, shown in Fig. 36, corresponds to an effective $|\boldsymbol{\beta}| = 2.6 \times 10^{-2}$. Note that this value of $|\boldsymbol{\beta}|$ is an order of magnitude larger than that of our local motion. Further, we find that the dipole direction is consistent with Fig. 3 of Planck Collaboration XXVII (2014), which depicts the reconstructed direction of the Doppler boost as a function of the CMB multipole.

Figure 37 presents a comparative study of the sensitivity of the two estimators, Doppler and modulation, and depicts the reconstruction noise for the two estimators as a function of the maximum CMB multipole used. The horizontal red dash-dotted line at $|\boldsymbol{\beta}| = 1.23 \times 10^{-3}$ denotes the expected value of the Doppler signal induced by our local motion. The blue dash-dotted line denotes the Doppler boost signal as seen by the modulation estimator and is expected to be enhanced by an effective factor of $b_\nu$ for the component separated maps.

For the case of the modulation estimator, it is seen that the Doppler boost is not expected to significantly contaminate the modulation signal up to $\ell_{\text{max}} \simeq 700$, establishing the fact that the boost effect corresponding to our local motion is not strong enough to affect the modulation signal seen at low CMB multipoles. However, it is expected to significantly add to the modulation signal at higher CMB multipoles.

Fig. 37 also clearly shows that the Doppler estimator is expected to recover the Doppler boost signal at a high significance. The figure also establishes that at $\ell \lesssim 500$, the local motion contribution is not detectable in the Doppler boost search, and hence the signal measured by the Doppler estimator in the range $\ell = 2$–64, depicted in Fig. 36, is expected to be linked to the modulation signal seen at low CMB multipoles.

It is, of course, also possible to extract the BipoSH coefficients $A^{LM}_{\ell_1\ell_2}$, up to the maximum multipole $\ell_{\text{max}}$ allowed by the full resolution *Planck* maps at modest computational expense. This allows us to address a specific indication of statistical isotropy violation previously reported in the literature. Bennett et al. (2011) found non-zero BipoSH power spectra, $A^{20}_{\ell\ell}$ and $A^{20}_{\ell\ell+2}$, at very high statistical significance in the *WMAP* maps as determined in ecliptic coordinates, corresponding to a quadrupolar power asymmetry in the CMB sky. The BipoSH spectra peaked at around $\ell = 250$, and the differences in the BipoSH signal determined from two different frequency bands indicated a non-cosmological origin. Furthermore, the azimuthal symmetry of this BipoSH signal in ecliptic coordinates suggested that it had its origin in some unaccounted-for systematic effect. The findings of Hanson et al. (2010) and Joshi et al. (2012) strongly suggest that the signal arises due to an incomplete treatment of beam asymmetries in the data. Bennett et al. (2012) have subse-quently noted that analysis of the *WMAP*-9 beam-deconvolved maps no longer shows the signal.

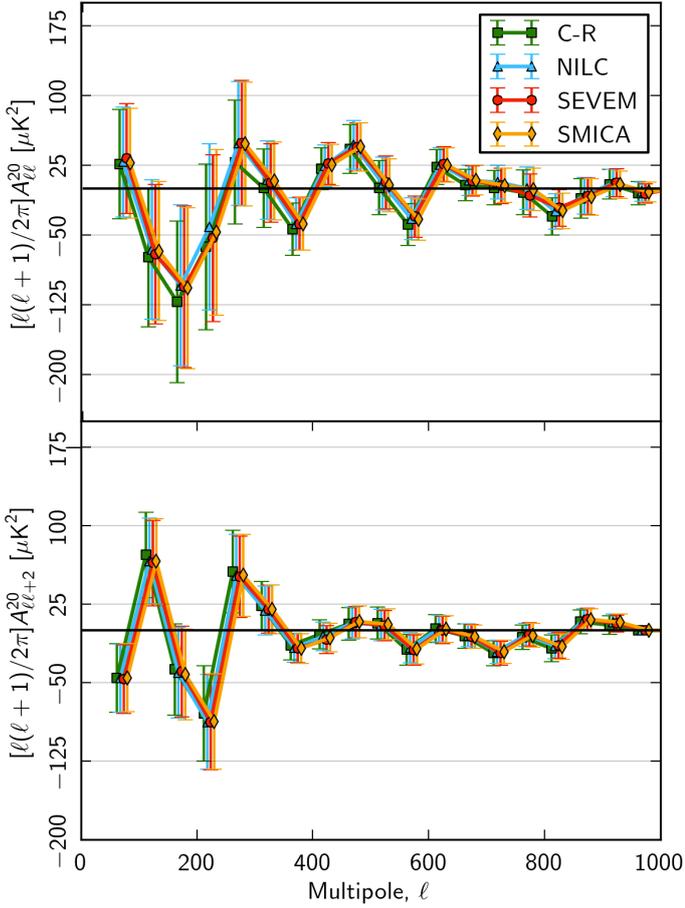

**Fig. 38.** BipoSH power spectra $A_{\ell\ell}^{20}$ and $A_{\ell\ell+2}^{20}$ obtained from the four component separation maps (C-R, NILC, SEVEM, and SMICA) are consistent with each other. Note that no significant ($> 3\,\sigma$) detections are found. This independently establishes the fact that the quadrupolar BipoSH detections made by *WMAP* were due to *WMAP*-specific instrument systematics.

We have computed the $A_{\ell\ell}^{20}$ and $A_{\ell\ell+2}^{20}$ values in ecliptic coordinates for the full resolution *Planck* CMB maps, as shown in Fig. 38. The analysis yields no evidence for BipoSH coefficients that deviate significantly from zero. This provides conclusive observational evidence from independent CMB measurements that the *WMAP* result could only have arisen due to instrumental artefacts in that data set.

In summary, a generalized search for power in the modulation field reveals a statistically significant dipolar modulation of the CMB sky on large angular scales ($\ell \lesssim 64$), with an amplitude of about 0.07. Conversely, no quadrupolar modulation is observed, reconfirming the systematic origin of the corresponding anomaly seen in the *WMAP*-7 data.

### 5.8. Parity asymmetry

#### 5.8.1. Point-parity asymmetry

The CMB sky map may be considered as the sum of even and odd parity functions. Previously, an odd point-parity preference (hereafter parity asymmetry) was observed in the *WMAP* 3-, 5-

, and 7-year data releases (Land & Magueijo 2005b; Kim & Naselsky 2010a,b; Gruppuso et al. 2011; Aluri & Jain 2012; Naselsky et al. 2012). In this section we investigate the parity asymmetry for the *Planck* temperature anisotropy power spectrum derived with a quadratic maximum likelihood (QML) estimator applied to the `Commander-Ruler`, `NILC`, `SEVEM`, and `SMICA` maps at $N_{\mathrm{side}} = 32$, and with a pseudo-$C_\ell$ estimator at $N_{\mathrm{side}} = 64$.

From the CMB anisotropy field defined on the sky, $T(\boldsymbol{n})$, we may construct symmetric and antisymmetric functions using the coordinate inversion $\boldsymbol{n} \to -\boldsymbol{n}$:

$$T^+(\boldsymbol{n}) = \frac{T(\boldsymbol{n}) + T(-\boldsymbol{n})}{2}\,; \qquad T^-(\boldsymbol{n}) = \frac{T(\boldsymbol{n}) - T(-\boldsymbol{n})}{2}. \tag{40}$$

Therefore, $T^+(\boldsymbol{n})$ and $T^-(\boldsymbol{n})$ have even and odd parity, respectively. When combined with the parity property of spherical harmonics, $Y_{\ell m}(\boldsymbol{n}) = (-1)^\ell Y_{\ell m}(-\boldsymbol{n})$, we obtain

$$
\begin{aligned}
T^+(\boldsymbol{n}) &= \sum_{\ell,m} a_{\ell m}\, Y_{\ell m}(\boldsymbol{n})\, \Gamma^+(\ell), \\
T^-(\boldsymbol{n}) &= \sum_{\ell,m} a_{\ell m}\, Y_{\ell m}(\boldsymbol{n})\, \Gamma^-(\ell),
\end{aligned}
\tag{41}
$$

where $n$ is an integer, and $\Gamma^+(\ell) = \cos^2(\ell\pi/2)$, and $\Gamma^-(\ell) = \sin^2(\ell\pi/2)$.

A significant power asymmetry between even and odd multipoles may thus be interpreted as a preference for a particular parity of the anisotropy pattern, connected to the parity asymmetry of the metric perturbations at scales above 1–4 Gpc (Kim & Naselsky 2010a). For investigation of the parity asymmetry we may consider the following quantities (Kim & Naselsky 2010a):

$$
\begin{aligned}
P^+(\ell) &= \sum_{n=2}^{\ell} \Gamma^+(n)\, \frac{n(n+1)}{2\pi}\, C_n\,; \\
P^-(\ell) &= \sum_{n=2}^{\ell} \Gamma^-(n)\, \frac{n(n+1)}{2\pi}\, C_n\,; \\
g(\ell) &= \frac{P^+(\ell)}{P^-(\ell)}.
\end{aligned}
\tag{42}
$$

Here $P^+$ and $P^-$ are the sum of $n(n+1)/2\pi\, C_n$ for even and odd multipoles respectively; the ratio $P^+/P^-$ is associated with the degree of parity asymmetry, where a value of $P^+/P^- < 1$ indicates odd-parity preference, and $P^+/P^- > 1$ indicates even-parity preference.

Following Kim & Naselsky (2010a), we will discuss the range of multipoles $2 \le \ell \le 30$, which belongs to the Sachs-Wolfe plateau of the TT power spectrum, where $\ell(\ell+1) C_\ell \approx$ constant. In order to make a rigorous assessment of the statistical significance of parity asymmetry at low $\ell$, we have compared $g(\ell)$ for the *Planck* power spectra with 1000 simulated CMB maps based on the fiducial *Planck* cosmological model. We compute power spectra using a QML-estimator (Gruppuso et al. 2009) as applied to data at $N_{\mathrm{side}} = 32$ with the U73 mask applied. This yields practically identical power spectrum results for the same $\ell$-range determined with a pseudo-$C_\ell$ estimator applied to maps at $N_{\mathrm{side}} = 64$.

In Fig 39 we show the $g(\ell)$-parameter for the *Planck* power spectra and the corresponding $p$-value$(\ell)$. The $p$-value denotes the fraction of simulations in which the obtained value of $P^+/P^-$ is as low as that observed in the data. Note that the results from the different *Planck* CMB maps yield consistent shapes for the $g(\ell)$ parameter and $p$-value$(\ell)$. The parity asymmetry at $\ell = 22$ is most anomalous, with a corresponding $p$-value in the range 0.002–0.004. Finally, the statistical significance of the parity asymmetry (i.e., low $p$-value) increases when we increase $\ell_{max}$ up to 22–25. Therefore, the odd parity preference cannot simply be attributed to the low quadrupole power. It is plausible the low quadrupole power is not an isolated anomaly, but that it shares an origin with the odd parity preference (see for details Kim & Naselsky 2010a; Naselsky et al. 2012; Kim & Naselsky 2010b).

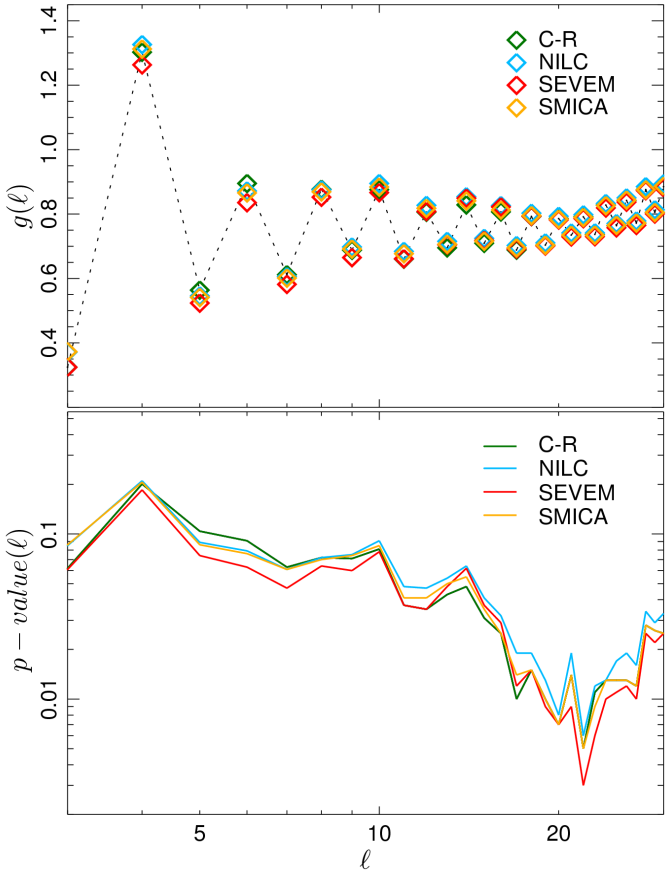

**Fig. 39.** *Upper*: the parity estimator $g(\ell)$ versus $\ell$ for C-R (black diamonds), NILC (red diamonds), SEVEM (blue diamonds), and SMICA (green diamonds). *Lower*: $p$-value for C-R (black solid line), NILC (red line), SEVEM (blue line), and SMICA (green line).

### 5.8.2. Mirror Parity

In this section we investigate the properties of the *Planck* temperature low-resolution maps under reflection with respect to a plane. This search for hidden mirror symmetries and anti-symmetries complements the tests for parity asymmetry, presented in Sect. 5.8.1. Starobinsky (1993) showed how a hidden mirror symmetry might be connected to the non-compact $T^1$ topology, or to a compact $T^3$ topology in which one topological scale is much shorter than the others. The CMB pattern would then exhibit a mirror symmetry with respect to the plane defined by the two large dimensions. Mirror symmetry has been searched for in the *COBE*-DMR data in de Oliveira-Costa et al. (1996), resulting in a lower limit for the scale of the compact dimension as 4 Gpc (see also Gurzadyan et al. 2007; Ben-David et al. 2012 for other more recent analysis). Finelli et al. (2012) analysed hidden mirror symmetry and anti-symmetry properties of the *WMAP* 7-year ILC temperature map, finding a preferred direction that could be considered anomalous at the 93% confidence level with anti-symmetry properties. This direction lies close to the one defining the hemispherical asymmetry.

Following Finelli et al. (2012), we consider the following estimators

$$S^{\pm}(\boldsymbol{n_i}) = \frac{1}{N_{pix}} \sum_{j=1}^{N_{pix}} \left[ \frac{1}{2} \left( \frac{\delta T}{T}(\boldsymbol{n_j}) \pm \frac{\delta T}{T}(\boldsymbol{n_k}) \right) \right]^2 , \quad (43)$$

where the sum is meant over the HEALPix pixels, $N_{pix}$, $\delta T/T(\boldsymbol{n_j})$ is the CMB temperature anisotropy measured at the pixel pointed by the unit vector $\boldsymbol{n_j}$, and $\boldsymbol{n_k}$ is the opposite direction of $\boldsymbol{n_j}$ with respect to the plane defined by $\boldsymbol{n_i}$, i.e.,

$$\boldsymbol{n_k} = \boldsymbol{n_j} - 2(\boldsymbol{n_i} \cdot \boldsymbol{n_j})\boldsymbol{n_i} . \quad (44)$$

The estimator $S^+$ ($S^-$) can be written explicitly in the harmonic domain and its averaged value over isotropic random realization is $\sum_{\ell} \ell (+1) C_{\ell} B_{\ell}^2/(4\pi) (\sum_{\ell} \ell C_{\ell} B_{\ell}^2/(4\pi))$, where $B_{\ell}$ is the Gaussian beam applied to the low-resolution maps.

We compute the quantities $S^{\pm}$ for each of the 3072 directions defined by HEALPix resolution $N_{side} = 16$ map, by allowing the $j$ and $k$ indices to run over the pixels of the low resolution foreground cleaned full-sky maps. We perform the same analysis on 1000 simulated skies and store the minimum and maximum value of $S^{\pm}$ for each of these to correct our statistics for the elsewhere effect.

The minimum value for the $S^+$ estimator is reached for the plane defined by Galactic coordinates $(l, b) = (264°, -17°)$ for all four foreground cleaned maps, with a significance of 0.5% (Commander-Ruler), 1.4% (NILC), 3.1% (SEVEM), and 8.9% (SMICA). The top panel of Fig. 40 shows the minimum value of $S^+$ for each of the four methods and compared to the MC simulations computed for Commander-Ruler, which is considered to be representative.

The minimum value for the $S^-$ estimator is found for a direction which depends slightly on the foreground cleaned CMB map considered and is close to the one associated with the cosmological dipole. This minimum value is not statistically significant for any of the foreground cleaned CMB maps (see bottom panel of Fig. 40).

The anomalous anti-symmetry direction found in the *Planck* CMB data is close to that found for the dipolar modulation in Sect. 5.6, suggesting possible connections between them. The direction that minimizes $S^+$ ($S^-$) for *Planck* is almost exactly the same as the one found for the *WMAP* 7-year ILC map in Finelli et al. (2012).

### 5.9. The Cold Spot

The Cold Spot was identified in the *WMAP* first year data (Vielva et al. 2004) through the estimation of the kurtosis of the Spherical Mexican Hat Wavelet (SMHW, e.g., Martínez-González et al. 2002) coefficients, and confirmed (Cruz et al. 2005) by analysing the area of the SMHW coefficients above/below a given threshold. Since its detection, the Cold Spot has been extensively studied and verified with a large battery of statistical probes (e.g., Mukherjee & Wang 2004; Cayón et al. 2005; McEwen et al. 2005; Cruz et al. 2007a; Räth et al. 2007b; Vielva et al. 2007; Pietrobon et al. 2008; Gurzadyan et al. 2009; Rossmanith et al. 2009b). A complete review of the Cold Spot can be found in Vielva (2010), including a discussion on possible explanations of its nature.

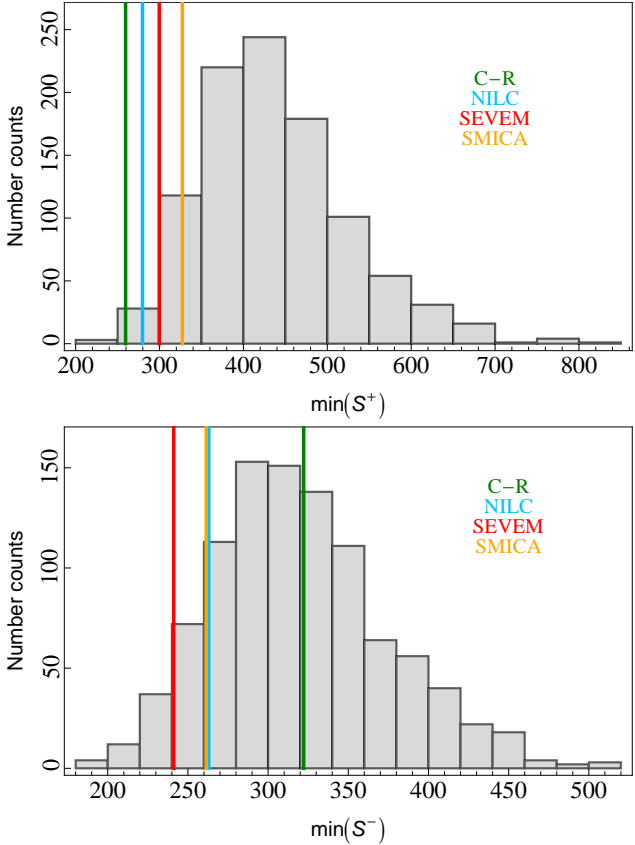

**Fig. 40.** Top panel: $S^+$ statistic. The vertical lines show the minimum value for the estimator as computed on low resolution `C-R`, `NILC`, `SEVEM`, and `SMICA` maps. The grey histogram shows the same quantity computed from 1000 simulated maps processed by `C-R`. Bottom panel: as above for $S^-$.

The analysis of the kurtosis of the SMHW coefficients has already been addressed in Sect. 4.5. We have checked that the kurtosis of the coefficients corresponding to the four *Planck* cleaned frequency maps is larger than the expected value obtained from simulations, with a modified upper tail probability of around 0.01. This is compatible with the value obtained from *WMAP*.

Nevertheless, the Cold Spot is more robustly described in terms of a morphological quantity: the area of the SMHW coefficients above/below a given threshold. At a given scale $R$ and threshold $\nu$, the cold ($A_R^{-\nu}$) and hot ($A_R^{+\nu}$) areas of the SMHW coefficients are defined as:

$$A_R^{-\nu} \equiv \#\{\omega_T(R, p) < -\nu\};$$
$$A_R^{+\nu} \equiv \#\{\omega_T(R, p) < +\nu\}. \tag{45}$$

Here # represents the number operator, i.e, it indicates for how many pixels $p$, the specific condition defined between the braces is satisfied.

Table 30 summarizes the results for the hot and cold areas determined for the four CMB maps analysed with the U73 mask (and its associated exclusions masks). The cold area is anomalous at scales between $R = 200$ and $R = 300'$, similar to the sizes already highlighted with the kurtosis analysis. We see that the higher the threshold, the smaller the upper tail probability associated with the *Planck* CMB map. In particular, the cold area has a upper tail probability of 0.003 at $\nu > 4\,\sigma_R$ and for $R = 300'$.

**Table 30.** Upper tail probability (UTP, in %) associated to the cold (top) and hot (bottom) areas. Results are given for a $\nu > 4\,\sigma_R$ threshold and for the four *Planck* CMB maps. The three most significant scales associated to the Cold Spot are shown. Analysis performed on the exclusions masks associated with the U73 mask.

| Area | Scale ['] | UTP | | | |
|------|-----------|-----|------|-------|-------|
| | | C-R | NILC | SEVEM | SMICA |
| Cold . . . . . . | 200 | 1.6 | 1.1 | 1.2 | 1.1 |
| | 250 | 0.3 | 0.3 | 0.3 | 0.3 |
| | 300 | 0.3 | 0.3 | 0.3 | 0.3 |
| Hot . . . . . . . | 200 | 2.3 | 1.6 | 1.8 | 1.6 |
| | 250 | 2.7 | 2.2 | 2.4 | 2.2 |
| | 300 | 4.9 | 3.7 | 4.1 | 3.8 |

**Table 31.** Upper tail probability (in %) associated to the cold (top) and hot (bottom) areas. Results are given for a $\nu > 4\,\sigma_R$ threshold and for the four *Planck* CMB maps. The three most significant scales associated to the Cold Spot are shown. Analysis performed on the exclusions masks associated with the CG70 mask. An ellipsis (...) indicates that no area above that threshold was found on the data.

| Area | Scale ['] | UTP | | | |
|------|-----------|-----|------|-------|-------|
| | | C-R | NILC | SEVEM | SMICA |
| Cold . . . . . . | 200 | 1.0 | 1.0 | 0.9 | 1.0 |
| | 250 | 0.3 | 0.3 | 0.3 | 0.3 |
| | 300 | 0.2 | 0.2 | 0.2 | 0.2 |
| Hot . . . . . . . | 200 | 15.1 | 14.5 | 14.6 | 14.5 |
| | 250 | ... | ... | ... | ... |
| | 300 | ... | ... | ... | ... |

**Table 32.** Upper tail probabilities (in %) associated with the cold (top) and hot (bottom) areas. Results are given for a $\nu > 4\,\sigma_R$ threshold and for the four *Planck* CMB maps. The three most significant scales associated to the Cold Spot are shown. Analysis performed on the exclusions masks associated with the CG60 mask. An ellipsis (...) indicates that no area above that threshold was found on the data.

| Area | Scale ['] | UTP | | | |
|------|-----------|-----|------|-------|-------|
| | | C-R | NILC | SEVEM | SMICA |
| Cold . . . . . . | 200 | 1.1 | 0.9 | 0.8 | 0.9 |
| | 250 | 0.1 | 0.1 | 0.1 | 0.1 |
| | 300 | 0.1 | 0.1 | 0.1 | 0.1 |
| Hot . . . . . . . | 200 | ... | ... | ... | ... |
| | 250 | ... | ... | ... | ... |
| | 300 | ... | ... | ... | ... |

Notice that the most significant deviation comes from the cold area, although the hot area is marginally compatible. However, the cold area represents the most robust detection of an anomaly, since it is robust to the mask employed (see Tables 31 and 32).

The information provided in the previous Tables is also represented (for the $R = 300'$ scale) in Fig. 41. In these panels we show the anomalous cold (in blue) and hot (in red) areas for thresholds $\nu > 3.0\sigma_R$ as determined from the `SMICA` map. The most extreme value (in terms of $\sigma_R$) of each area is indicated.

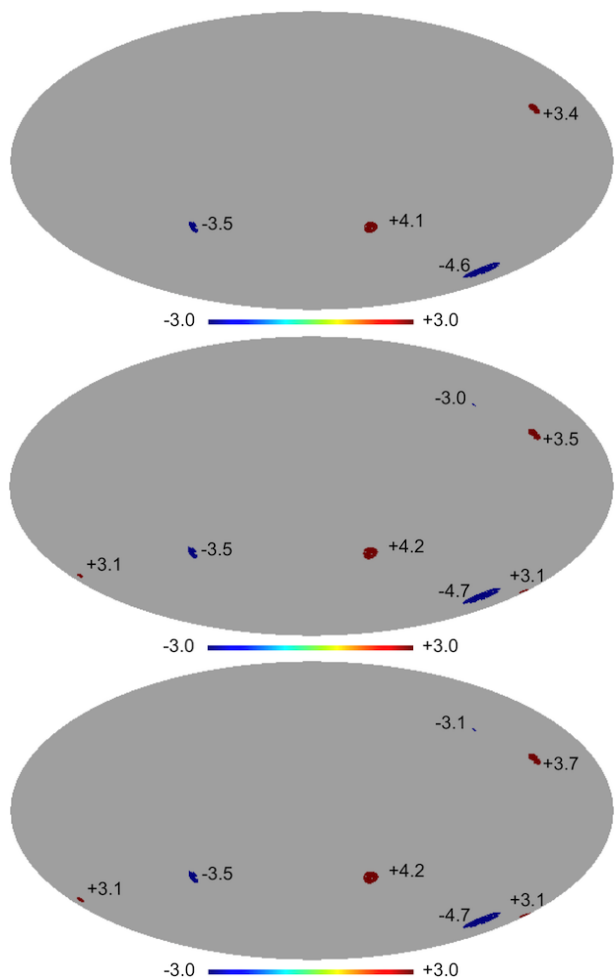

**Fig. 41.** SMHW coefficients at $R = 300$ arc minutes, above (and below) a $+3.0\,\sigma$ ($-3.0\,\sigma$) threshold. Results for the three masks considered in the analysis are shown: U73 mask (top); CG70 (middle); and CG60 (bottom).

The coldest area corresponds to the Cold Spot with the minimum value of the wavelet coeficient at the position $(209°, -57°)$ in Galactic coordinates, whereas the hotest area has already been identified in the *WMAP* data (e.g., Vielva et al. 2007) as an anomalous hot spot. This does not depends significantly on the mask considered in the analysis. From these results it is clear that the Cold Spot anomaly is present in both *WMAP* and *Planck* data.

### 5.10. Interpretation of anomalies

The results presented here in Sect. 5 demonstrate that many features previously observed in the *WMAP* data are present also in the *Planck* sky. This agreement between two independent experiments effectively rules out the possibility that their origin lies in systematic artefacts present in either data set. In particular, there is evidence for a violation of statistical isotropy at least on large angular scales in the context of the *Planck* fiducial sky model. Moreover, a power asymmetry extends to scales corresponding to $\ell \simeq 600$, whilst fits at low $\ell$ to a model containing a dipole modulation field yield results in excess of $3\,\sigma$ significance. In addition, there is evidence from such fits that the low-$\ell$ spectrum of the *Planck* data departs from the fiducial spectrum in both amplitude and slope. These results could have profound implications for cosmology. It is therefore pertinent to consider whether a model can be proposed to provide a common origin for the anomalies.

The microwave sky is manifestly non-Gaussian and anisotropic, with known contributions from Galactic astrophysical foregrounds, lensing of CMB anistropies by the intervening matter distribution, and the ISW. However, the excellent performance of the component separation algorithms used here in rejecting diffuse foregrounds argues strongly against known Galactic emission as the source of the anomalies.

Schwarz et al. (2004), Copi et al. (2007), Maris et al. (2011) and Hansen et al. (2012) suggested that diffuse Solar System emission could contribute to the observed structure on large angular scales, although it is not expected that the classical Zodiacal Light Emission or Kuiper Belt objects are responsible. Planck Collaboration XIV (2014) presents the current *Planck* contribution to the modelling of the Zodiacal cloud. Moreover, the association of the anomalies with the ecliptic reference frame seems, at best, to be confined to the very largest angular scales, since the preferred direction determined for the dipole modulation field analysis is separated from the south ecliptic pole by more than $45°$, with even larger separations found for the power asymmetry direction.

Another possibility is that the anomalies have their origin in the local Universe. According to Francis & Peacock (2009), the removal of the ISW signal originating within the volume at $z < 0.3$ from *WMAP* data reduces the significance of the apparent alignment between the CMB quadrupole and octopole and the Cold Spot. Efstathiou et al. (2010) have used the same correction to yield an increase in the structure of the two-point correlation function for angular separations less than $60°$, that had been noted as apparently anomalous since the first *WMAP* data release. A future possibility is that *Planck* itself will be able to reconstruct the ISW signal and test its impact on issues related to isotropy and non-Gaussianity. Planck Collaboration XIX (2014) presents maps of the effect based on the current data release.

Of more interest to us is that the anomalies are genuinely cosmological in origin. In that context, obvious candidate models include those with simply or multi-connected topology. In a companion paper (Planck Collaboration XXVI 2014), a subset of such models are considered and the signatures of their specific correlation structures on the sky are searched for. However, no detections are found, but rather the scale of topology is limited to be of order the diameter of the last-scattering surface or greater. More interestingly, they reconsider Bianchi VII$_h$ models that were previously demonstrated to show statistical correlation with the *WMAP* data (Jaffe et al. 2005, 2006; Bridges et al. 2007; McEwen et al. 2013), albeit with parameters inconsistent with standard cosmological parameters. In this new analysis, the Bianchi parameters are physically coupled to the cosmological ones, yielding no evidence for a Bianchi VII$_h$ cosmology. However, as before, when treated simply as a template for structure contained in the CMB sky, a best-fit pattern is found to be in good agreement with the old results. Previous analyses (Jaffe et al. 2005; Cayón et al. 2006; McEwen et al. 2006) have shown that when the CMB sky is corrected for such a template, many of the large-scale anomalies are no longer present at a statistically significant level. It is likely that such an effect will persist for Bianchi-corrected *Planck* data, and we have made an explicit test as to whether the best-fit Bianchi template can also explain the presence of phase correlations. We therefore repeated the surrogate analysis from Sect. 5.4 for the appropriately corrected `SMICA` map. Figure 42 presents the result for the corresponding significance map. It is clear that the signature for hemispherical asymmetry is drastically reduced, thereby rendering the signal formally statistically insignificant. Thus, the best-fit Bianchi model can also account for the asymmetries induced by higher order phase correlations. It should also be noted that subtracting the best-fit Bianchi template from the data, outside the U73 mask, explains the anomalous skewness and kurtosis values but not the variance, for which the corresponding lower tail probabilities are 0.008, 0.166, and 0.306, respectively. Given the lack of consistency of the physical parameters of the model with the *Planck* cosmological model, the results obtained using Bianchi-subtracted input maps might be considered moot, however, the morphology of the maps may provide insight into the type of underlying structures associated with the anomalies.

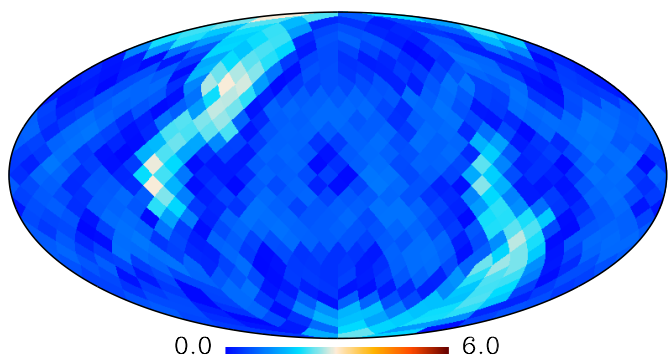

**Fig. 42.** Same as Fig. 23 but with the best fit Bianchi template subtracted from the `SMICA` map.

Although the Cold Spot is also rendered statistically insignificant by the Bianchi template, other possible explanations about its nature have been considered, including the late evolution of the large-scale structure (e.g., Inoue & Silk 2006, 2007), the Sunyaev-Zeldovich effect (e.g., Cruz et al. 2008), residual foregrounds (Cruz et al. 2006), or a cosmic texture (e.g., Cruz et al. 2007b). Further tests based on gravitational lensing (Das & Spergel 2009) or CMB polarization (Vielva et al. 2011) have been suggested to confirm or reject some of the proposed explanations.

The presence of primordial magnetic fields (PMFs) due to either pre- or post-recombination mechanisms could also provide a physical basis for some of the anomalies discussed in this paper. Specifically, PMFs with coherence scales comparable to the present day horizon could result in Alfvén waves in the early Universe that generate specific signatures on the sky via the Doppler and integrated Sachs-Wolfe effects. In particular, a preferred angular direction in the CMB anisotropy can be induced (Durrer et al. 1998; Kim & Naselsky 2009) that leads to structure in the spherical harmonic mode correlation matrix (Kahniashvili et al. 2008). Appendix A presents a search for the predicted correlations between harmonic modes separated by $\Delta\ell = 0, \pm2$, and $\Delta m = 0, \pm1, \pm2$, allowing constraints to be placed on the Alfvén wave amplitude. Further constraints on PMFs based on the power spectrum and bispectrum have been provided in Planck Collaboration XVI (2014) and Planck Collaboration XXIV (2014), respectively.

To conclude, when analysing a data set as complex and rich as that provided by *Planck*, some statistical outliers will be expected. However, it should be clear that the evidence for some of the large-angular scale anomalies is significant indeed, yet few physically compelling models have been proposed to account for them, and none so far that provide a common origin. The dipole modulation model that was analysed here was phenomenologically motivated, and is detected in the data at relatively high significance. Whether it can resolve the anomalous nature of other observed features remains to be evaluated.

## 6. Implications for $C_\ell$ and cosmological parameter estimation

The approach to $C_\ell$ estimation, the construction of the *Planck* likelihood and subsequent inference of cosmological parameters are described in the accompanying papers Planck Collaboration XV (2014) and Planck Collaboration XVI (2014). For these studies, specific assumptions are made about the isotropy and Gaussianity of the primordial fluctuations observed in the CMB. The latter in particular seems to be well-supported by the comprehensive set of tests applied to the *Planck* data in Sect. 4. The most significant discrepancies are seen in association with the Cold Spot (Sect. 5.9), which constitutes a localized region of sky of about 10° in size. Its impact on cosmological parameters is then likely to be relatively insignificant, and masking of the region could easily test this assertion.

It is well-known that the quadrupole and octopole have low amplitudes relative to the best-fit cosmological power-spectrum. The contribution of those multipoles to cosmological parameter estimation is very small due to the associated cosmic variance on these scales, and thus the direct impact of their alignment (as discussed in Sect. 5.1) is also likely to be small. Remarkably, however, Planck Collaboration XV (2014) presents evidence that the low-$\ell$ multipole range from 2 to 30 is coherently low, and is not well accounted for by the standard ΛCDM model. Moreover, this conclusion is a consequence of the fact that the cosmological parameters are strongly influenced by the $\ell = 1000$–1500 range, previously inaccessible to *WMAP*. Consistent findings have been presented here in the form of the low variance of the data in Sects. 4.1 and 5.2, although this is largely driven by the quadupole and octopole alone.

The question therefore remains as to whether there is a deeper connection with the cosmological anomalies seen in both the *WMAP* and *Planck* data sets particularly on large angular scales. Indeed, the hemispherical asymmetry and dipolar power modulation discussed in Sect. 5 could have a more important impact in that they directly address whether a broader class of cosmological models should be considered. Indeed, the low-$\ell$ signature seen in the data has previously been associated with missing power in a universe with simply- or multiply-connected topology. However, there are specific morphological signatures of such topologies that have not been detected in the *Planck* data (Planck Collaboration XXVI 2014).

However, the phenomenologically motivated dipole modulation model due to Gordon et al. (2005) yields a significant fit to the data, as seen in Sects. 5.6 and 5.7. The former also shows some evidence for a departure from the *Planck* fiducial power spectrum in both amplitude and slope. Both of these analyses are in good agreement in terms of the direction of the dipolar modulation field with the model independent power asymmetry analysis of Sect. 5.5.

A qualitative exploration as to how these features are reflected in the low-$\ell$ power spectrum is provided in Fig. 43. Specifically, the plot presents the angular power spectra computed using a quadratic maximum-likelihood (QML) estimator (Paci et al. 2010, 2013) from the $N_{\text{side}} = 16$ `SMICA` map after application of the U73 mask used in this paper. The *Planck* fiducial power spectrum is also shown for comparison. Clearly, there is a deficit of power as expected when no further partitioning of the sky is applied. However, further interesting properties of the data are revealed when spectra are computed for the two opposing hemispheres defined by the preferred direction from the $\ell_{max} = 1500$ analysis in Sect. 5.5. In the positive direction, there is improved agreement between the derived spectrum and the *Planck* fiducial sky, but with an interesting oscillation between odd and even modes. For the negative direction, an overall suppression of power is again seen. It would be interesting to test the connection between these spectral features and the phase correlations detected in Sect. 5.4 or the evidence for parity violation presented in Sect. 5.8. The observations may, in part, reflect the presence of visually striking features noted by Bennett et al. (2011) — the four elongated cold fingers stretching from near the Galactic equator to the south Galactic pole and a prominent cold spot near the centre of the map. Very similar results are found when the preferred direction is specified either by the $\ell_{max} = 600$ power asymmetry analysis, or the dipolar modulation direction in Sect. 5.6.

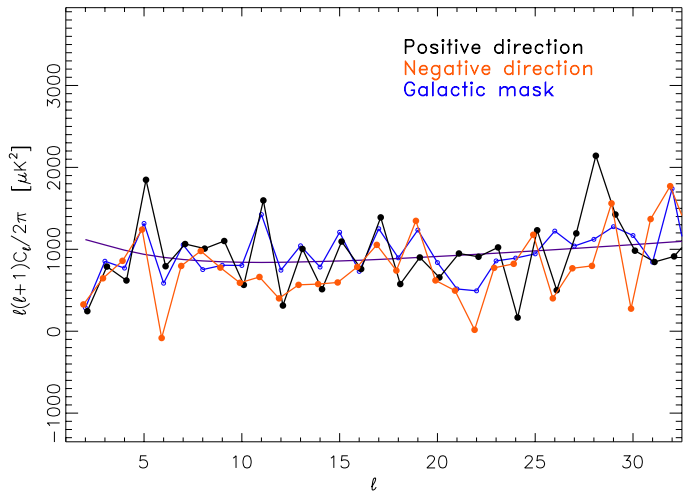

**Fig. 43.** The angular power spectrum on large angular scales computed from the `SMICA` map at $N_{side} = 16$ on the opposing hemispheres defined by the maximal asymmetry direction $(l, b) = (220°, 4°)$ determined in Sect. 5.5.

However, Fig. 30 and the corresponding analysis suggest that the asymmetry in power between hemispheres extends to much smaller angular scales, whilst Planck Collaboration XXVII (2014) demonstrates the presence of a modulation due to Doppler boosting up to $\ell \simeq 2000$. Whether such properties of the data would have implications for parameter estimation may need further exploration.

# 7. Conclusions

In this paper, we have tested the statistical isotropy and Gaussianity of the CMB using data from the *Planck* satellite. We have demonstrated that little evidence is seen for non-Gaussianity, although some deviations from isotropy are found.

Most of the tests performed in Sect. 4 showed an overall consistency with the null hypothesis, as represented by a set of realistic simulations based on a *Planck* fiducial sky model and including the secondary ISW-lensing effect (as detected for the first time with the *Planck* data, see Planck Collaboration XIX

2014). However, two important exceptions were seen. The variance of the CMB signal was found to be significantly lower than expected, with the anomalously low signal seemingly localized in the northern ecliptic hemisphere (Sect. 4.1). This result was also confirmed with the low variance of the wavelet coefficients that was seen on scales above a few degrees (see Sect. 4.5). Moreover, a significant deviation from Gaussianity was found in the form of a positive kurtosis of the wavelet coefficients.

These results correspond to statistical features on large angular scales where numerous anomalies were previously observed in the *WMAP* data. In Sect. 5, we revisited these in the light of the *Planck* data and found results in excellent agreement with those for *WMAP*. In particular, the most significant anomalies, namely the quadrupole-octopole alignment (Sect. 5.1), the low variance (Sect. 5.2), hemispherical asymmetry (Sect. 5.3), phase correlations (Sect. 5.4), power asymmetry (Sect. 5.5), dipolar modulation (Sect. 5.6), generalized power modulation (Sect. 5.7), parity asymmetry (Sect. 5.8), and the Cold Spot (Sect. 5.9) have been confirmed with the *Planck* data. Attempts to explain the observed features in terms of systematic artefacts, local astrophysical sources of emission, or structure in the local Universe have not been successful. It is clear that these anomalies represent real features of the CMB sky.

However, it is difficult to make a detailed interpretation of the anomalies in the absence of theoretical models, in particular with regard to the role of a posterior choices. Nevertheless, *Planck* does offer new possibilities to check the a posteriori claims in this context as a consequence of its superior multipole content that cannot easily been probed by *WMAP*.

Phenomenological models have been suggested to account for the observations. The dipolar power modulation approach due to Gordon et al. (2005) was explicitly tested in Sect. 5.6 and found to represent a good fit to the large scale asymmetry, corresponding to a detection at about $3\,\sigma$ significance. This result was confirmed by the more generalized modulation study in Sect. 5.7, which also ruled out the presence of modulation fields of higher order. Alternatively, a Bianchi template fit to the data performed in Planck Collaboration XXVI (2014) can provide a good fit to the hemispherical asymmetry, the Cold Spot and the phase correlations, but corresponds to values of the cosmological parameters incompatible with those derived in Planck Collaboration XVI (2014). Clearly, these do not provide complete and satisfactory explanations for the observations, and more physically motivated models should be sought.

This may also be indicated by the cosmological parameter studies presented in Planck Collaboration XV (2014) and Planck Collaboration XVI (2014). Here, it was demonstrated that while the power spectrum determined from the *Planck* temperature data is extremely consistent with a basic 6-parameter $\Lambda$CDM model, the low-$\ell$ multipoles ($\ell \leq 30$) deviate from the best-fit model although at a significance that does not appear to exceed $2.7\,\sigma$. However, this is precisely the regime where many of the anomalies presented in this paper seem to manifest themselves, and where qualitatively interesting differences are observed in the power-spectra for two hemispheres defined by the preferred direction for the dipolar power modulation.

Finally, it is expected that the polarization data that will become available with the 2014 data release should provide valuable information on the nature of the CMB anomalies. Then, the presence, or even absence, of a specific signature in the data should help to elucidate the physical mechanism that is causing the anomaly (see Vielva et al. 2011, Frommert & Enßlin 2010 and Dvorkin et al. 2008 for examples related to the Cold Spot, mode alignment, and dipolar modulation, respectively). In par-ticular, a deviation of isotropy present at recombination should be reflected in both the temperature and polarization data, with a correlated signal. It may be that the statistical anomalies currently described in this paper are a hint of more profound physical phenomena that are yet to be revealed.

*Acknowledgements.* The development of Planck has been supported by: ESA; CNES and CNRS/INSU-IN2P3-INP (France); ASI, CNR, and INAF (Italy); NASA and DoE (USA); STFC and UKSA (UK); CSIC, MICINN, JA and RES (Spain); Tekes, AoF and CSC (Finland); DLR and MPG (Germany); CSA (Canada); DTU Space (Denmark); SER/SSO (Switzerland); RCN (Norway); SFI (Ireland); FCT/MCTES (Portugal); and PRACE (EU). A description of the Planck Collaboration and a list of its members, including the technical or scientific activities in which they have been involved, can be found at `http://www.sciops.esa.int/index.php?project=planck&page=Planck_Collaboration`. We acknowledge the use of resources from the Norewegian national super computing facilities NOTUR. The modal and KSW bispectrum estimator analysis was performed on the COSMOS supercomputer, part of the STFC DiRAC HPC Facility. We further acknowledge the computer resources and technical assistance provided by the Spanish Supercomputing Network nodes at Universidad de Cantabria and Universidad Politécnica de Madrid as well as by the Advanced Computing and e-Science team at IFCA. Some of the results in this paper have been derived using the `HEALPix` package.

# Appendix A: Constraints on Alfvén waves

Observations of synchrotron emission and Faraday rotation provide increasing evidence that large-scale astrophysical systems in the Universe are pervaded by magnetic fields. These huge systems include the Ly-$\alpha$ forest and intercluster regions (see Kronberg 2009, for a review). Both pre- and post-recombination mechanisms could result in a background of nano-gauss fields that might be detectable in large-scale structures or the CMB, although at present no imprints of these Primordial Magnetic Fields (PMFs) have been detected therein.

Here, we report our findings based on an analysis of the *Planck* data to search for the predicted signature of statistical anisotropy due to PMFs. Specifically, PMFs with coherence scales comparable to the present day horizon may induce and sustain Alfvén waves in the early Universe that can leave observable imprints on the CMB via the Doppler and integrated Sachs-Wolfe effects. In particular, this results in a preferred angular direction in the CMB anisotropy, therefore breaking statistical isotropy (Durrer et al. 1998, Kim & Naselsky 2009).

Durrer et al. (1998) showed that cosmological Alfvén waves generate a fractional CMB anisotropy for a Fourier mode $\boldsymbol{k}$:

$$\frac{\delta T}{T_0}(\boldsymbol{n}, \boldsymbol{k}) \approx \boldsymbol{n} \cdot \boldsymbol{\Omega}(k, \eta_{\text{last}}) = \boldsymbol{n} \cdot \boldsymbol{\Omega_0} \, v_{\text{A}} k \, \eta_{\text{last}} \, \boldsymbol{B} \cdot \boldsymbol{k} \,, \qquad (A.1)$$

where $\boldsymbol{n}$ denotes sky direction, $\boldsymbol{B}$ is a unit vector in the direction of the coherent PMF, $\boldsymbol{\Omega}(\boldsymbol{k}, \eta_{\text{last}})$ is the Gauge invariant linear combination associated with vector perturbations, $\eta_{\text{last}}$ denotes the conformal time at the moment of baryon-photon decoupling, and $T_0$ is the CMB monopole temperature 2.7255 K (Fixsen 2009). Durrer et al. (1998) assumed that the vector perturbations are initially created by some random stochastic PMF and have the following statistical properties over an ensemble of universes:

$$\langle \Omega_0^i(\boldsymbol{k})\,\Omega_0^j(\boldsymbol{k}) \rangle = (\delta_{ij} - k_i\,k_j)P(k). \tag{A.2}$$

Here, $P(k)$ is the power spectrum assumed to follow a simple power law (Durrer et al. 1998):

$$P(k) = A_v \frac{k^{n_v}}{k_0^{n_v+3}}, \tag{A.3}$$

where $k_0$ is a pivot wavenumber set to 0.05/Mpc in this analysis. The Alfvén wave velocity is given by (Durrer et al. 1998):

$$v_{\text{A}} = \frac{B}{2\sqrt{\pi(\rho_r + p_r)}} \approx 4 \times 10^{-4} \frac{B}{10^{-9}\text{Gauss}}, \tag{A.4}$$

where $\rho_r$ and $p_r$ are the density and the pressure of photons, and the speed of light $c$ is set to 1.

Kahniashvili et al. (2008) showed that the presence of Alfvén waves in the early Universe leads to specific correlations of the CMB in harmonic space:

$$\langle a_{\ell m}^* a_{\ell m} \rangle = C_\ell + \frac{\ell(\ell+1)}{(2\ell-1)(2\ell+3)}\Bigg\{ (\ell^2 + \ell - 3)\cos^2\theta_{\text{B}}$$
$$+ \ell(\ell+1) - (3\cos^2\theta_{\text{B}} - 1)m^2\left[1 - \frac{3}{\ell(\ell+1)}\right]\Bigg\}I_{\text{d}}^{\ell,\ell};$$

$$\langle a_{\ell,m+1}^* a_{\ell,m} \rangle = -\sin 2\theta_{\text{B}}\exp[\pm\imath\phi_{\text{B}}]\,\frac{\ell^2+\ell-3}{(2\ell-1)(2\ell+3)}\left(m\pm\frac{1}{2}\right)$$
$$\times \sqrt{(\ell \mp m)(\ell \pm m+1)}\,I_{\text{d}}^{\ell,\ell};$$

$$\langle a_{\ell,m+2}^* a_{\ell,m} \rangle = -\frac{1}{2}\sin^2\theta_{\text{B}}\exp[\pm\imath 2\phi_{\text{B}}]\,\frac{\ell^2+\ell-3}{(2\ell-1)(2\ell+3)}$$
$$\times \sqrt{(\ell \mp m)(\ell \mp m-1)(\ell \pm m+1)(\ell \pm m+2)}\,I_{\text{d}}^{\ell,\ell};$$

$$\langle a_{\ell+2,m}^* a_{\ell,m} \rangle = -(3\cos^2\theta_{\text{B}} - 1)\frac{(\ell+3)\ell}{2(2\ell+3)\sqrt{(2\ell+1)(2\ell+5)}}$$
$$\times \sqrt{((\ell+1)^2 - m^2)((\ell+2)^2 - m^2)}\,I_{\text{d}}^{\ell,\ell+2};$$

$$\langle a_{\ell-2,m}^* a_{\ell,m} \rangle = -(3\cos^2\theta_{\text{B}} - 1)\frac{(\ell+1)(\ell-2)}{2(2\ell-1)\sqrt{(2\ell-3)(2\ell+1)}}$$
$$\times \sqrt{((\ell-1)^2 - m^2)(\ell^2 - m^2)}\,I_{\text{d}}^{\ell,\ell-2};$$

$$\langle a_{\ell+2,m+1}^* a_{\ell,m} \rangle = \sin 2\theta_{\text{B}}\exp[\pm\imath\phi_{\text{B}}]\frac{(\ell+3)l}{2(2\ell+3)\sqrt{(2\ell+1)(2\ell+5)}}$$
$$\times \sqrt{((\ell+1)^2 - m^2)(\ell \pm m+2)(\ell \pm m+3)}\,I_{\text{d}}^{\ell,\ell+2};$$

$$\langle a_{\ell-2,m+1}^* a_{\ell,m} \rangle =$$
$$-\sin 2\theta_{\text{B}}\exp[\pm\imath\phi_{\text{B}}]\frac{(\ell+1)(\ell-2)}{2(2\ell-1)\sqrt{(2\ell-3)(2\ell+1)}}$$
$$\times \sqrt{(\ell^2 - m^2)(\ell \mp m-1)(\ell \mp m-2)}\,I_{\text{d}}^{\ell,\ell-2};$$

$$\langle a_{\ell+2,m+2}^* a_{\ell,m} \rangle =$$
$$-\frac{1}{2}\sin^2\theta_{\text{B}}\exp[\pm\imath 2\phi_{\text{B}}]\frac{(\ell+3)\ell}{2(2\ell+3)\sqrt{(2\ell+1)(2\ell+5)}}$$
$$\times \sqrt{((\ell \pm m+1)(\ell \pm m+2)(\ell \pm m+3)(\ell \pm m+4)}\,I_{\text{d}}^{\ell,\ell+2};$$

$$\langle a_{\ell-2,m+2}^* a_{\ell,m} \rangle =$$
$$-\frac{1}{2}\sin^2\theta_{\text{B}}\exp[\pm\imath 2\phi_{\text{B}}]\frac{(\ell+1)(\ell-2)}{2(2\ell-1)\sqrt{(2\ell-3)(2\ell+1)}}$$
$$\times \sqrt{((\ell \mp m-3)(\ell \mp m-2)(\ell \mp m-1)(\ell \mp m)}\,I_{\text{d}}^{\ell,\ell-2}.$$

Here $C_\ell$ is the power spectrum in the absence of Alfvén waves, $\theta_{\text{B}}$ and $\phi_{\text{B}}$ are the spherical angles of a PMF direction $\boldsymbol{B}$, and $I_{\text{d}}^{\ell\ell'}$ is given by

$$I_{\text{d}}^{\ell\ell'} = \frac{2T_0^2}{\pi}\left(\frac{\eta_{\text{last}}}{\eta_0}\right)^2 A_v v_{\text{A}}^2$$
$$\times \int d\ln k\left(\frac{k}{k_0}\right)^{n_v+3}\exp\left(-2\frac{k^2}{k_{\text{D}}^2}\right)j_\ell(k\eta_0)\,j_{\ell'}(k\eta_0),$$

where $\eta_0$ is the present conformal time, and $k_D$ denotes the comoving wavenumber of the dissipation scale, due to photon viscosity and given by approximately $10/\eta_{\text{last}}$ (Durrer et al. 1998). The damping effect becomes significant on multipoles $\ell \gtrsim 500$ (Durrer et al. 1998). As shown above, Alfvén waves in the early Universe produce correlations between harmonic modes separated by $\Delta\ell = 0, \pm 2$, and $\Delta m = 0, \pm 1, \pm 2$. Investigating these imprints, we may impose a constraint on the Alfvén waves. In the weak Alfvén wave limit, the CMB data log-likelihood can be expanded as follows:

$$\mathcal{L} \approx \mathcal{L}|_{A_v v_{\text{A}}^2 = 0} + \frac{\partial\mathcal{L}}{\partial(A_v v_{\text{A}}^2)}\bigg|_{A_v v_{\text{A}}^2 = 0} A_v v_{\text{A}}^2 + \frac{1}{2}\frac{\partial^2\mathcal{L}}{\partial(A_v v_{\text{A}}^2)^2}\bigg|_{A_v v_{\text{A}}^2 = 0}(A_v v_{\text{A}}^2)^2$$
$$+ O((A_v v_{\text{A}}^2)^3). \tag{A.5}$$

Since all correlations produced by Alfvén waves are proportional to $A_v v_{\text{A}}^2$, the first term in Eq. A.5 is simply equal to the likelihood of the standard cosmological model. The first and second derivative of the likelihood are obtained by

$$\frac{\partial\mathcal{L}}{\partial\lambda} = \mathcal{H} - \langle\mathcal{H}\rangle, \qquad \frac{\partial^2\mathcal{L}}{\partial\lambda^2} = -\langle\mathcal{H}^2\rangle + \langle\mathcal{H}\rangle\langle\mathcal{H}\rangle, \tag{A.6}$$

where

$$\mathcal{H} = \frac{1}{2}\left[\mathsf{C}^{-1}\boldsymbol{a}\right]^\dagger \frac{\partial\mathsf{C}}{\partial\lambda}\left[\mathsf{C}^{-1}\boldsymbol{a}\right]. \tag{A.7}$$

Here $\boldsymbol{a}$ is the data vector, consisting of the spherical harmonic coefficients, $a_{\ell m}$, of the CMB anisotropy data, and $\mathsf{C}$ is their covariance matrix.

In our analysis, we consider the four foreground cleaned CMB maps `Commander-Ruler`, `NILC`, `SEVEM`, and `SMICA` and apply the common mask. We assume the fiducial *Planck* cosmological model, and use MC simulations to estimate the ensemble average values for signal and noise, as required in Eq. A.6. The quantity $\mathsf{C}^{-1}\boldsymbol{a}$ from Eq. A.7 was then determined by the messenger field method (Elsner & Wandelt 2013). The `CosmoMC` package (Lewis & Bridle 2002) is then used as a generic sampler in order to obtain the posterior probability for the Alfvén wave parameters $\{A_v v_{\text{A}}^2, n_v, \theta_{\text{B}}, \phi_{\text{B}}\}$.

In Table A.1, we show the upper bounds on the Alfvén wave amplitude $A_v v_{\text{A}}^2$ at various confidence levels, after marginalizing over the spectral index $n_v$ and the direction $\theta_{\text{B}}, \phi_{\text{B}}$. From the analysis of the *Planck* data, we impose an upper bound on the Alfvén wave amplitude that is tighter than that from the *WMAP* data by more than one order of magnitude.

**Table A.1.** *Planck* constraints on the Alfvén wave amplitude $A_v v_A^2$.

| | Confidence level | | |
|---|---|---|---|
| Method | 68% | 95% | 99.7% |
| C-R . . . . . . . . . . | $< 0.52 \times 10^{-11}$ | $< 1.45 \times 10^{-11}$ | $< 2.66 \times 10^{-11}$ |
| NILC . . . . . . . . . | $< 0.58 \times 10^{-11}$ | $< 1.79 \times 10^{-11}$ | $< 3.44 \times 10^{-11}$ |
| SEVEM . . . . . . . . | $< 0.65 \times 10^{-11}$ | $< 1.92 \times 10^{-11}$ | $< 4.43 \times 10^{-11}$ |
| SMICA . . . . . . . . | $< 0.67 \times 10^{-11}$ | $< 2.18 \times 10^{-11}$ | $< 4.14 \times 10^{-11}$ |

[1] APC, AstroParticle et Cosmologie, Université Paris Diderot, CNRS/IN2P3, CEA/Irfu, Observatoire de Paris, Sorbonne Paris Cité, 10, rue Alice Domon et Léonie Duquet, 75205 Paris Cedex 13, France

[2] Aalto University Metsähovi Radio Observatory and Dept of Radio Science and Engineering, P.O. Box 13000, FI-00076 AALTO, Finland

[3] African Institute for Mathematical Sciences, 6-8 Melrose Road, Muizenberg, Cape Town, South Africa

[4] Agenzia Spaziale Italiana Science Data Center, Via del Politecnico snc, 00133, Roma, Italy

[5] Agenzia Spaziale Italiana, Viale Liegi 26, Roma, Italy

[6] Astrophysics Group, Cavendish Laboratory, University of Cambridge, J J Thomson Avenue, Cambridge CB3 0HE, U.K.

[7] Astrophysics & Cosmology Research Unit, School of Mathematics, Statistics & Computer Science, University of KwaZulu-Natal, Westville Campus, Private Bag X54001, Durban 4000, South Africa

[8] CITA, University of Toronto, 60 St. George St., Toronto, ON M5S 3H8, Canada

[9] CNRS, IRAP, 9 Av. colonel Roche, BP 44346, F-31028 Toulouse cedex 4, France

[10] California Institute of Technology, Pasadena, California, U.S.A.

[11] Centre for Theoretical Cosmology, DAMTP, University of Cambridge, Wilberforce Road, Cambridge CB3 0WA, U.K.

[12] Centro de Estudios de Física del Cosmos de Aragón (CEFCA), Plaza San Juan, 1, planta 2, E-44001, Teruel, Spain

[13] Computational Cosmology Center, Lawrence Berkeley National Laboratory, Berkeley, California, U.S.A.

[14] Consejo Superior de Investigaciones Científicas (CSIC), Madrid, Spain

[15] DSM/Irfu/SPP, CEA-Saclay, F-91191 Gif-sur-Yvette Cedex, France

[16] DTU Space, National Space Institute, Technical University of Denmark, Elektrovej 327, DK-2800 Kgs. Lyngby, Denmark

[17] Département de Physique Théorique, Université de Genève, 24, Quai E. Ansermet,1211 Genève 4, Switzerland

[18] Departamento de Física Fundamental, Facultad de Ciencias, Universidad de Salamanca, 37008 Salamanca, Spain

[19] Departamento de Física, Universidad de Oviedo, Avda. Calvo Sotelo s/n, Oviedo, Spain

[20] Departamento de Matemáticas, Estadística y Computación, Universidad de Cantabria, Avda. de los Castros s/n, Santander, Spain

[21] Department of Astronomy and Astrophysics, University of Toronto, 50 Saint George Street, Toronto, Ontario, Canada

[22] Department of Astrophysics/IMAPP, Radboud University Nijmegen, P.O. Box 9010, 6500 GL Nijmegen, The Netherlands

[23] Department of Electrical Engineering and Computer Sciences, University of California, Berkeley, California, U.S.A.

[24] Department of Physics & Astronomy, University of British Columbia, 6224 Agricultural Road, Vancouver, British Columbia, Canada

[25] Department of Physics and Astronomy, Dana and David Dornsife College of Letter, Arts and Sciences, University of Southern California, Los Angeles, CA 90089, U.S.A.

[26] Department of Physics and Astronomy, University College London, London WC1E 6BT, U.K.

[27] Department of Physics, Florida State University, Keen Physics Building, 77 Chieftan Way, Tallahassee, Florida, U.S.A.

[28] Department of Physics, Gustaf Hällströmin katu 2a, University of Helsinki, Helsinki, Finland

[29] Department of Physics, Princeton University, Princeton, New Jersey, U.S.A.

[30] Department of Physics, University of Alberta, 11322-89 Avenue, Edmonton, Alberta, T6G 2G7, Canada

[31] Department of Physics, University of California, Berkeley, California, U.S.A.

[32] Department of Physics, University of California, One Shields Avenue, Davis, California, U.S.A.

[33] Department of Physics, University of California, Santa Barbara, California, U.S.A.

[34] Department of Physics, University of Illinois at Urbana-Champaign, 1110 West Green Street, Urbana, Illinois, U.S.A.

[35] Dipartimento di Fisica e Astronomia G. Galilei, Università degli Studi di Padova, via Marzolo 8, 35131 Padova, Italy

[36] Dipartimento di Fisica e Astronomia, Università degli Studi di Bologna, viale Berti Pichat 6/2, I-40127, Bologna, Italy

[37] Dipartimento di Fisica e Scienze della Terra, Università di Ferrara, Via Saragat 1, 44122 Ferrara, Italy

[38] Dipartimento di Fisica, Università La Sapienza, P. le A. Moro 2, Roma, Italy

[39] Dipartimento di Fisica, Università degli Studi di Milano, Via Celoria, 16, Milano, Italy

[40] Dipartimento di Fisica, Università degli Studi di Trieste, via A. Valerio 2, Trieste, Italy

[41] Dipartimento di Fisica, Università di Roma Tor Vergata, Via della Ricerca Scientifica, 1, Roma, Italy

[42] Dipartimento di Matematica, Università di Roma Tor Vergata, Via della Ricerca Scientifica, 1, Roma, Italy

[43] Discovery Center, Niels Bohr Institute, Blegdamsvej 17, Copenhagen, Denmark

[44] Dpto. Astrofísica, Universidad de La Laguna (ULL), E-38206 La Laguna, Tenerife, Spain

[45] European Space Agency, ESAC, Planck Science Office, Camino bajo del Castillo, s/n, Urbanización Villafranca del Castillo, Villanueva de la Cañada, Madrid, Spain

[46] European Space Agency, ESTEC, Keplerlaan 1, 2201 AZ Noordwijk, The Netherlands

[47] Helsinki Institute of Physics, Gustaf Hällströmin katu 2, University of Helsinki, Helsinki, Finland

[48] INAF - Osservatorio Astronomico di Padova, Vicolo dell'Osservatorio 5, Padova, Italy

[49] INAF - Osservatorio Astronomico di Roma, via di Frascati 33, Monte Porzio Catone, Italy

[50] INAF - Osservatorio Astronomico di Trieste, Via G.B. Tiepolo 11, Trieste, Italy

[51] INAF Istituto di Radioastronomia, Via P. Gobetti 101, 40129 Bologna, Italy

[52] INAF/IASF Bologna, Via Gobetti 101, Bologna, Italy

[53] INAF/IASF Milano, Via E. Bassini 15, Milano, Italy

[54] INFN, Sezione di Bologna, Via Irnerio 46, I-40126, Bologna, Italy

[55] INFN, Sezione di Roma 1, Università di Roma Sapienza, Piazzale Aldo Moro 2, 00185, Roma, Italy

[56] IPAG: Institut de Planétologie et d'Astrophysique de Grenoble, Université Joseph Fourier, Grenoble 1 / CNRS-INSU, UMR 5274, Grenoble, F-38041, France

[57] ISDC Data Centre for Astrophysics, University of Geneva, ch. d'Ecogia 16, Versoix, Switzerland

[58] IUCAA, Post Bag 4, Ganeshkhind, Pune University Campus, Pune 411 007, India

[59] Imperial College London, Astrophysics group, Blackett Laboratory, Prince Consort Road, London, SW7 2AZ, U.K.

[60] Infrared Processing and Analysis Center, California Institute of Technology, Pasadena, CA 91125, U.S.A.

[61] Institut Néel, CNRS, Université Joseph Fourier Grenoble I, 25 rue des Martyrs, Grenoble, France

[62] Institut Universitaire de France, 103, bd Saint-Michel, 75005, Paris, France

[63] Institut d'Astrophysique Spatiale, CNRS (UMR8617) Université Paris-Sud 11, Bâtiment 121, Orsay, France

[64] Institut d'Astrophysique de Paris, CNRS (UMR7095), 98 bis Boulevard Arago, F-75014, Paris, France

[65] Institute for Space Sciences, Bucharest-Magurale, Romania

[66] Institute of Astronomy and Astrophysics, Academia Sinica, Taipei, Taiwan

[67] Institute of Astronomy, University of Cambridge, Madingley Road, Cambridge CB3 0HA, U.K.

[68] Institute of Theoretical Astrophysics, University of Oslo, Blindern, Oslo, Norway

[69] Instituto de Astrofísica de Canarias, C/Vía Láctea s/n, La Laguna, Tenerife, Spain

[70] Instituto de Física de Cantabria (CSIC-Universidad de Cantabria), Avda. de los Castros s/n, Santander, Spain

[71] Jet Propulsion Laboratory, California Institute of Technology, 4800 Oak Grove Drive, Pasadena, California, U.S.A.

[72] Jodrell Bank Centre for Astrophysics, Alan Turing Building, School of Physics and Astronomy, The University of Manchester, Oxford Road, Manchester, M13 9PL, U.K.

[73] Kavli Institute for Cosmology Cambridge, Madingley Road, Cambridge, CB3 0HA, U.K.

[74] LAL, Université Paris-Sud, CNRS/IN2P3, Orsay, France

[75] LERMA, CNRS, Observatoire de Paris, 61 Avenue de l'Observatoire, Paris, France

[76] Laboratoire AIM, IRFU/Service d'Astrophysique - CEA/DSM - CNRS - Université Paris Diderot, Bât. 709, CEA-Saclay, F-91191 Gif-sur-Yvette Cedex, France

[77] Laboratoire Traitement et Communication de l'Information, CNRS (UMR 5141) and Télécom ParisTech, 46 rue Barrault F-75634 Paris Cedex 13, France

[78] Laboratoire de Physique Subatomique et de Cosmologie, Université Joseph Fourier Grenoble I, CNRS/IN2P3, Institut National Polytechnique de Grenoble, 53 rue des Martyrs, 38026 Grenoble cedex, France

[79] Laboratoire de Physique Théorique, Université Paris-Sud 11 & CNRS, Bâtiment 210, 91405 Orsay, France

[80] Lawrence Berkeley National Laboratory, Berkeley, California, U.S.A.

[81] Max-Planck-Institut für Astrophysik, Karl-Schwarzschild-Str. 1, 85741 Garching, Germany

[82] Max-Planck-Institut für Extraterrestrische Physik, Giessenbachstraße, 85748 Garching, Germany

[83] McGill Physics, Ernest Rutherford Physics Building, McGill University, 3600 rue University, Montréal, QC, H3A 2T8, Canada

[84] MilliLab, VTT Technical Research Centre of Finland, Tietotie 3, Espoo, Finland

[85] Mullard Space Science Laboratory, University College London, Surrey RH5 6NT, U.K.

[86] National University of Ireland, Department of Experimental Physics, Maynooth, Co. Kildare, Ireland

[87] Niels Bohr Institute, Blegdamsvej 17, Copenhagen, Denmark

[88] Observational Cosmology, Mail Stop 367-17, California Institute of Technology, Pasadena, CA, 91125, U.S.A.

[89] Optical Science Laboratory, University College London, Gower Street, London, U.K.

[90] SB-ITP-LPPC, EPFL, CH-1015, Lausanne, Switzerland

[91] SISSA, Astrophysics Sector, via Bonomea 265, 34136, Trieste, Italy

[92] School of Physics and Astronomy, Cardiff University, Queens Buildings, The Parade, Cardiff, CF24 3AA, U.K.

[93] School of Physics and Astronomy, University of Nottingham, Nottingham NG7 2RD, U.K.

[94] Space Sciences Laboratory, University of California, Berkeley, California, U.S.A.

[95] Special Astrophysical Observatory, Russian Academy of Sciences, Nizhnij Arkhyz, Zelenchukskiy region, Karachai-Cherkessian Republic, 369167, Russia

[96] Stanford University, Dept of Physics, Varian Physics Bldg, 382 Via Pueblo Mall, Stanford, California, U.S.A.

[97] Sub-Department of Astrophysics, University of Oxford, Keble Road, Oxford OX1 3RH, U.K.

[98] Theory Division, PH-TH, CERN, CH-1211, Geneva 23, Switzerland

[99] UPMC Univ Paris 06, UMR7095, 98 bis Boulevard Arago, F-75014, Paris, France

[100] Université de Toulouse, UPS-OMP, IRAP, F-31028 Toulouse cedex 4, France

[101] University of Granada, Departamento de Física Teórica y del Cosmos, Facultad de Ciencias, Granada, Spain

[102] Warsaw University Observatory, Aleje Ujazdowskie 4, 00-478 Warszawa, Poland